\pdfoutput=1
\documentclass[cernpreprint, texlive=2023, UKenglish, texmf, orcidlogo]{atlasdoc}
 
\usepackage[subfig]{atlaspackage}
 
\usepackage{atlasbiblatex}
 
\usepackage[hepprocess,hepparticle]{atlasphysics}

\graphicspath{{logos/}{figures/}}
\usepackage{multirow}
\usepackage{algorithm}
\usepackage{algpseudocode}
\usepackage{heppennames}
\usepackage{hepnames}

\usepackage{ANA-HDBS-2020-19-PAPER-defs}

\AtlasTitle{Search for heavy resonances decaying into a $Z$ or $W$ boson and a Higgs boson in final states
with leptons and $b$-jets in $\SI{139}{\per\femto\barn}$ of $pp$ collisions at
$\sqrt{s}=\SI{13}{TeV}$ with the ATLAS detector}
 
\AtlasAbstract{
This article presents a search for new resonances decaying into a $Z$ or $W$ boson and a $125$~GeV
Higgs boson $h$, and it targets the $\nu\bar{\nu}b\bar{b}$, $\ell^+\ell^-b\bar{b}$, or $\ell^{\pm}{\nu}b\bar{b}$
final states, where $\ell=e$ or $\mu$, in proton--proton collisions at $\sqrt{s}=$13~TeV.
The data used correspond to a total integrated luminosity of $\SI{139}{\per\femto\barn}$ collected by the ATLAS detector during
Run~2 of the LHC at CERN.
The search is conducted by examining the reconstructed invariant or
transverse mass distributions of $Zh$ or $Wh$ candidates for evidence of a localised excess in
the mass range from $220$~GeV to 5~TeV. No significant excess is observed and $95\%$
confidence-level upper limits between 1.3~pb and 0.3~fb are placed on the production
cross section times branching fraction of neutral and charged spin-1 resonances and
CP-odd scalar bosons. These limits are converted into constraints on the parameter space of
the Heavy Vector Triplet model and the two-Higgs-doublet model.}
 
\author{The ATLAS Collaboration}
 
\AtlasRefCode{HDBS-2020-19}
\PreprintIdNumber{CERN-EP-2022-115}

\AtlasJournalRef{JHEP 06 (2023) 016}
\AtlasDOI{10.1007/JHEP06(2023)016}

\hypersetup{pdftitle={ATLAS document},pdfauthor={The ATLAS Collaboration}}
 
\begin{document}
 
\maketitle

% The next lines are included from the .//texfiles/Introduction.tex input file
\section{Introduction}
\label{sec:Introduction}
 
Following the discovery of the Higgs boson~\cite{HIGG-2012-27,CMS-HIG-12-028} by the ATLAS and CMS collaborations
at the Large Hadron Collider (LHC), measurements of its properties
so far indicate consistency with the Standard Model (SM) predictions. Nevertheless, several questions related
to electroweak symmetry breaking remain open, in particular how the Higgs boson mass is protected against
large radiative corrections (the naturalness problem~\cite{naturalness1,naturalness2,naturalness3}) and whether
the Higgs boson is part of an extended scalar sector. \par
 
Various models with dynamical electroweak symmetry breaking scenarios attempt to solve the
naturalness problem by assuming new strong interactions at a higher energy scale. These models
generally predict the existence of new vector resonances that have significant branching fractions for
decays into a vector boson and a Higgs boson. Examples are Minimal Walking Technicolour~\cite{Sannino:2004qp,Foadi:2007ue,Belyaev:2008yj},
Little Higgs~\cite{Schmaltz:2005ky}, or composite Higgs models~\cite{Dugan:1984hq,Agashe:2004rs}.
Resonance searches are typically not sensitive to all free parameters of the underlying theory,
so simplified models are generally used, such as the Heavy Vector Triplet (HVT) parameterised
Lagrangian~\cite{Pappadopulo:2014qza, deBlas:2012qp}, which adds an additional SU(2) field to the
SM and provides a restricted number of new couplings. \par
 
A second class of models extend the scalar sector by including additional Higgs singlets or doublets~\cite{2HDM}.
Examples of models with an extended scalar sector are the minimal supersymmetric SM~\cite{MSSM_I,MSSM_II,MSSM_III,MSSM_IV,MSSM_V},
axion models~\cite{Axion}, or baryogenesis models~\cite{baryogenesis}. Again, instead of targeting any of these specific theories,
a generic two-Higgs-doublet model (2HDM) with a CP-conserving Higgs potential is probed. The scalar sector of the theory consists
of five Higgs bosons: two charged ($H^{\pm}$), two neutral CP-even (\h, $H$), and one neutral CP-odd (\A). \par
 
In this article, ATLAS presents a search for new heavy resonances decaying into a $125$~GeV
Higgs boson \h\ and a \Z\ or \W\ boson in 139~\ifb of proton--proton ($pp$) collision data at $\sqrt{s}=$13~\TeV.
These resonances are assumed to be either a new heavy vector boson, denoted hereafter by \Zprime\ and \Wprime, or a heavy CP-odd scalar boson \A.
The article targets leptonic decays of the \Z\ and \W\ bosons, and requires $b$-quark pair decays for the Higgs boson.
Therefore, the search is performed in the following three final states: \vvbb, \lvbb\ and \llbb.
In this article, $\ell$ represents $e$ or $\mu$ unless otherwise stated. \par
 
Searches in the same final states have been performed at $\sqrt{s}=$13~\TeV\ with an integrated luminosity of 36.1~\ifb by ATLAS~\cite{EXOT-2016-10} and of 137~\ifb by CMS~\cite{CMS:2021lyi,CMS:2021fyk}.
Results in other final states have also been published: ATLAS has performed a
search in the fully hadronic final state~\cite{HDBS-2018-11}, based on an integrated luminosity of 139~\ifb, and CMS has published results in the fully hadronic final state~\cite{cmsVHqqbb2017} and the $\tau^+\tau^-\ell^+\ell^-$ final state~\cite{cmstautaull:2020}, based on a 35.9~\ifb dataset. \par
 
Apart from the significantly larger dataset used here, several improvements have been implemented since the previous ATLAS publication~\cite{EXOT-2016-10}, including
improved $b$-tagging, lepton isolation and jet reconstruction, a reoptimisation of the lepton selection in the \llbb\ channel and a new selection based on
the missing transverse momentum (\met) significance in the \vvbb\ channel (see Sections~\ref{sec:Reconstruction} and~\ref{sec:AnalysisStrategy}).
The refined $b$-tagging techniques lead to a significantly higher signal acceptance for resonances with masses of multiple \TeV, while the
other changes allow for an improved background rejection.
 
The search presented in this article is performed by seeking a localised excess in the distribution of a channel-dependent proxy for the
resonance mass of the \vvbb, \lvbb\ and \llbb\ systems. Signal mass hypotheses in a range from $220~\GeV$ to $5~\TeV$ are tested. Due to the wide range of probed
Higgs boson transverse momenta, two methods are used to reconstruct Higgs boson candidates. At low transverse momenta, the decay products of
the Higgs boson are reconstructed as individual jets, while at high transverse momenta they are reconstructed as a single large-radius jet.
 
The search selects events in signal regions (SR) and background-dominated control regions (CR) based on requirements placed on kinematic
properties of final-state particles and on event-level quantities. The statistical interpretation is performed with a binned maximum-likelihood
fit to data based on the reconstructed resonance mass, using the signal and control regions. The major backgrounds are modelled using simulation
and their normalisations are determined by the fit. Different fit models are used depending on the targeted signal hypothesis, production mode, or final state.
 
The results of the HVT \Zprime\ and \Wprime\ searches are interpreted in two benchmark models. In the first model, referred to as \textit{Model A}, the branching
fractions to fermions and gauge bosons are comparable, as in some models with an extended gauge symmetry~\cite{Barger:1980ix}. For \textit{Model B}, fermionic
couplings are suppressed, as in strong dynamical models such as the minimal composite Higgs model~\cite{Contino:2011np}.
Constraints are also set on the coupling-strength scale factors $g_\text{F}$ and $g_\text{H}$, which modulate the coupling of the \Zprime\ and \Wprime\ bosons to fermions
and the Higgs and gauge bosons, respectively. The search focuses on high resonance masses, ranging from 300~\GeV\ to 5~\TeV.
 
The search for \AZh\ focuses on smaller masses, ranging from 220~\GeV\ to  2~\TeV, because the Higgs potential becomes unstable for sufficiently large
$A$ boson masses in the class of models targeted by this article. The results of the \AZh\ search are interpreted as constraints on the
ratio of the vacuum expectation values of the two Higgs doublets, $\tan\beta$, and  on $\cos(\beta-\alpha)$, where $\alpha$ is the
mixing angle between the two CP-even Higgs bosons, for  Type-I, Type-II, Lepton-specific, and Flipped 2HDMs. These models differ with
respect to which doublets couple to the up-type and down-type quarks and to the charged leptons~\cite{2HDM}.
Production via gluon--gluon fusion (\ggA) and production with associated $b$-quarks (\bbA) are both considered in this search.
 
Representative lowest-order Feynman diagrams of the relevant production modes\footnote{In the HVT benchmark \textit{Models A} and \textit{B}, the \Zprime\ and \Wprime\ bosons are
produced mainly via quark--antiquark annihilation, while production via vector-boson fusion is relevant only for the
fermiophobic scenario (i.e.\ \textit{Model C}).} for \Zprime, \Wprime, and $A$ bosons in $pp$
collisions and the subsequent decays into the \vvbb, \llbb, and \lvbb\ final states are depicted in Figure~\ref{fig:feynman-diagrams}.
 
This article is structured as follows. Sections~\ref{sec:ATLASDet} and~\ref{sec:DataMCSamples} provide a brief description of the
ATLAS experiment and of the data and simulated event samples, respectively. The event reconstruction and selections are discussed
in Sections~\ref{sec:Reconstruction} and~\ref{sec:AnalysisStrategy}.  The background estimation and systematic uncertainties are
described in Sections~\ref{sec:Background} and~\ref{sec:Systematics}, respectively. Finally, Sections~\ref{sec:Results}
and~\ref{sec:Conclusion} detail the statistical analysis and then provide a discussion of the results and concluding remarks.
 
\begin{figure*}[h!]
\centering
\subfloat[]{
\includegraphics[width=0.4\textwidth]{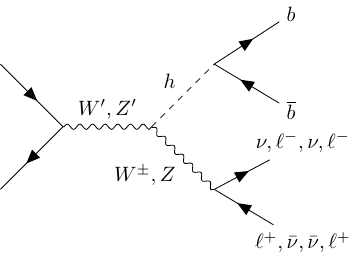}
}\\
\subfloat[]{
\includegraphics[width=0.45\textwidth]{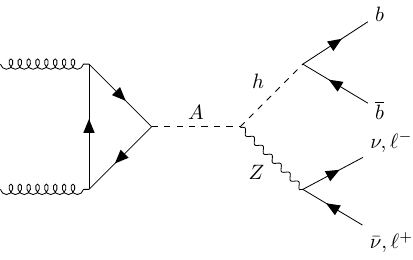}
}
\subfloat[]{
\includegraphics[width=0.45\textwidth]{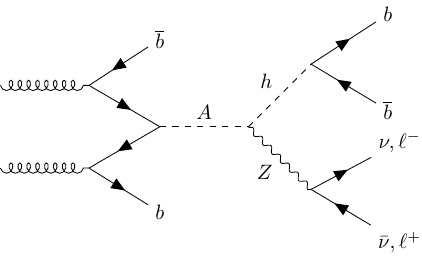}
}
\caption{Representative lowest-order Feynman diagrams of the resonant production of a $W$ or $Z$ boson and a Higgs boson via (a)
quark--antiquark annihilation, (b) gluon--gluon fusion and (c) $b$-associated production. The subsequent decays into the
\vvbb, \lvbb\ and \llbb\ final states are also depicted, where $\ell = e, \mu, \mathrm{and}\,\tau$.}
\label{fig:feynman-diagrams}
\end{figure*}

% End of text imported from the .//texfiles/Introduction.tex input file

% The next lines are included from the .//texfiles/ATLASDet.tex input file
\section{ATLAS detector} \label{sec:ATLASDet}
 
The ATLAS experiment~\cite{PERF-2007-01} at the LHC is a multipurpose particle detector
with a forward--backward symmetric cylindrical geometry and a near \(4\pi\) coverage in
solid angle.\footnote{ATLAS uses a right-handed coordinate system with its origin at the nominal interaction point (IP)
in the centre of the detector and the \(z\)-axis along the beam pipe.
The \(x\)-axis points from the IP to the centre of the LHC ring, and the \(y\)-axis points upwards.
Cylindrical coordinates \((r,\phi)\) are used in the transverse plane, \(\phi\) being the azimuthal angle around the \(z\)-axis. The pseudorapidity is defined in terms of the polar angle \(\theta\) as \(\eta = -\ln \tan(\theta/2)\).
Angular distance is measured in units of \(\Delta R \equiv \sqrt{(\Delta\eta)^{2} + (\Delta\phi)^{2}}\).
}
It consists of an inner tracking detector (ID) surrounded by a thin superconducting solenoid
providing a \SI{2}{\tesla} axial magnetic field, electromagnetic and hadron calorimeters, and a muon spectrometer.
The inner tracking detector covers the pseudorapidity range \(|\eta| < 2.5\).
It consists of silicon pixel, silicon microstrip, and transition radiation tracking detectors.
Lead/liquid-argon (LAr) sampling calorimeters provide electromagnetic energy measurements
with high granularity.
A steel/scintillator-tile hadron calorimeter covers the central pseudorapidity range (\(|\eta| < 1.7\)).
The endcap and forward regions are instrumented with LAr calorimeters
for both the electromagnetic and hadronic energy measurements up to \(|\eta| = 4.9\).
The muon spectrometer surrounds the calorimeters and is based on
three large superconducting air-core toroidal magnets with eight coils each.
The field integral of the toroids ranges between \num{2.0} and \SI{6.0}{\tesla\metre}
across most of the detector.
The muon spectrometer includes a system of precision tracking chambers and fast detectors for triggering.
A two-level trigger system is used to select events.
The first-level trigger is implemented in hardware and uses a subset of the detector information
to accept events at a rate below \SI{100}{\kHz}.
This is followed by a software-based trigger that
reduces the accepted event rate to \SI{1}{\kHz} on average
depending on the data-taking conditions.
An extensive software suite~\cite{ATL-SOFT-PUB-2021-001} is used in the reconstruction
and analysis of real and simulated data, in detector operations, and in the trigger
and data acquisition systems of the experiment.

% End of text imported from the .//texfiles/ATLASDet.tex input file

% The next lines are included from the .//texfiles/DataMCSamples.tex input file
\section{Data and Monte Carlo samples}\label{sec:DataMCSamples}

The data used in this analysis were recorded with the ATLAS detector between 2015 and 2018 in $pp$ collisions at $\sqrt{s}= 13$~\TeV\ and correspond to a total
integrated luminosity of 139~\ifb~\cite{LumiDetermination}. The data are required to satisfy criteria that ensure that the detector was in good operating condition \cite{DAPR-2018-01}.
Monte Carlo (MC) simulation samples are used to model all the major backgrounds and the signal processes.
 
For the HVT interpretations of the ATLAS data, the production of \Zprime\ and \Wprime\ bosons via quark--antiquark annihilation
was modelled at leading-order (LO) accuracy by $\MADGRAPH5$\,2.3.3 \cite{Alwall:2014hca}
interfaced with \Pythia[8.186]~\cite{Sjostrand:2007gsChanged}, which used the A14 set of tuned parameters~\cite{ATL-PHYS-PUB-2014-021} and the \NNPDF[2.3lo] parton distribution function (PDF) set~\cite{Ball:2011mu}.
Events were generated for a range of resonance masses from 300\,\GeV\ to 5\,\TeV\ using the benchmark HVT \textit{Model A}.
Separate generation of signal events for HVT \textit{Model A} and \textit{Model B} is not necessary as both give rise
to mass peaks in the $Zh$ and $Wh$ system with a width that is dominated by the experimental resolution.
Higgs boson decays into $b\bar{b}$ or $c\bar{c}$ are considered, assuming the branching fractions $B\left( h\to b\bar{b}/c\bar{c}\right) = 0.569/0.0287$
and a Higgs boson mass of $m_{h} = 125\,\GeV$. Even though this search does not target the decay $h\rightarrow c\bar{c}$
directly, it is sensitive to such events. The $Z$ and $W$ bosons were required to decay via $Z\rightarrow \nu\bar{\nu}$,
$Z\rightarrow \ell^+\ell^-$, or $W\rightarrow \ell^\pm\nu$, with $\ell = e, \mu, \mathrm{and}\,\tau$. Contributions from $Z'\rightarrow WW\rightarrow \ell^\pm \nu qq$
or $W’ \rightarrow WZ\rightarrow \ell^\pm \nu qq, \ell^+\ell^- qq$ decays are negligible to these searches.
 
For the 2HDM interpretations of the ATLAS data, events for the gluon--gluon fusion production of \A\ bosons (\ggA) were generated in the narrow-width
approximation with $\MADGRAPH5$\,2.3.3 at LO accuracy and interfaced to \Pythia[8.186], which used the A14 tune
and the \NNPDF[2.3lo] PDF set. Events for $b$-quark associated production of \A\ bosons (\bbA) were simulated with \MGNLO[2.2.3]~\cite{Alwall:2014hca} using the four-flavour scheme at
next-to-leading order (\NLO) with massive $b$-quarks and the \CT[10nlo\,nf4] PDF set, and interfaced with \Pythia[8.186], which used the A14 set of tuned parameters.
Events were generated for a range of \A\ boson masses from 220\,\GeV\ to 2\,\TeV, assuming a narrow width. In the 2HDM, the Higgs boson branching fractions
vary strongly as a function of the relevant model parameters, and therefore the \A\ boson signals were generated only for decays of the Higgs boson $h$ into
a $b\bar{b}$ pair. The $Z$ boson was required to decay either via $Z\rightarrow \nu\bar{\nu}$ or via $Z\rightarrow \ell^+\ell^-$. \par
 
The production of \W\ and \Z\ bosons in association with jets was modelled by \Sherpa[2.2.1]~\cite{Bothmann:2019yzt}
with the \NNPDF[3.0nlo] set of PDFs~\cite{Ball:2014uwa}. Diagrams with up to two additional parton emissions were
simulated at NLO precision, and those with three or four additional parton emissions at LO accuracy,
using the Comix~\cite{Gleisberg:2008fv} and \OPENLOOPS~\cite{Buccioni:2019sur,Cascioli:2011va,Denner:2016kdgabc} libraries.
Matrix elements (ME) were merged with the \SHERPA parton shower~\cite{Schumann:2007mgabc} (PS) using the \MEPSatNLO prescription~\cite{Hoeche:2011fdabc,Hoeche:2012yf,Catani:2001ccabc,Hoeche:2009rjabc}
and the set of tuned parameters developed by the \SHERPA authors. The inclusive production cross section for these samples was normalised to match a next-to-next-to-leading-order (NNLO)
prediction~\cite{Anastasiou:2003dsabc}.\par
 
The production of top-quark pair (\ttbar) events was modelled using the \POWHEGBOX[v2]~\cite{Frixione:2007nwabc,Nason:2004rx,Frixione:2007vwabc,Alioli:2010xdabc} generator.
The matrix elements were calculated at NLO precision in QCD using the five-flavour scheme and the \NNPDF[3.0nlo] PDF set, assuming
a top-quark mass of $172.5\,\GeV$. The PS, hadronisation, and underlying event were modelled with the \Pythia[8.230]~\cite{Sjostrand:2014zea} generator using the A14 set of
tuned parameters and the \NNPDF[2.3lo] set of PDFs. The \hdamp\ parameter\footnote{The \hdamp\ parameter is a resummation damping factor and one of the
parameters that control the matching of \POWHEG matrix elements to the PS, effectively regulating the high-\pt\ radiation against which the \ttbar\ system
recoils.} was set to 1.5\,\mtop~\cite{ATL-PHYS-PUB-2016-020}. The cross section was calculated at \NNLO precision, including the resummation of next-to-next-to-leading logarithmic (NNLL) soft gluon terms with
\TOPpp[2.0]~\cite{Beneke:2011mq,Cacciari:2011hy,Baernreuther:2012ws,Czakon:2012zr,Czakon:2012pz,Czakon:2013goa,Czakon:2011xx}.
 
The $\ttbar+h$ samples were generated at NLO accuracy using the \MGNLO[2.3.2] generator with the \NNPDF[3.0nlo] PDF set and interfaced to \Pythia[8.186], which used the
A14 set of tuned parameters and the \NNPDF[2.3lo] PDF set.
 
The production of $\ttbar+V$ events (where $V = W, Z$) was modelled using the \MGNLO[2.3.3] generator at NLO with the
\NNPDF[3.0nlo] PDF set. The events were interfaced to \Pythia[8.230], which used the A14 set of tuned parameters and the \NNPDF[2.3lo] PDF set. \par
 
The associated production of top quarks with \W\ bosons ($Wt$) and the $s$- and $t$-channel production of single top quarks were
modelled by the \POWHEGBOX[v2] generator at NLO in QCD using the five-flavour scheme and the \NNPDF[3.0nlo] set of PDFs.
The diagram removal (DR) scheme~\cite{Frixione:2008yi} was used to remove interference and overlap with \ttbar\ production.
The events were interfaced to  \Pythia[8.230], which used the A14 set of tuned parameters and the \NNPDF[2.3lo] set of PDFs.\par
 
Diboson events ($WW$, $WZ$, $ZZ$) with semileptonic decays were simulated using \Sherpa[2.2.1] with the \NNPDF[3.0nnlo] PDF set, including off-shell effects and
Higgs boson contributions where appropriate. Diagrams with up to one additional emission were calculated at NLO accuracy in QCD, while diagrams with two or three
parton emissions were described at LO accuracy~\cite{ATL-PHYS-PUB-2017-005}. They were merged and matched using the \MEPSatNLO prescription. Loop-induced
diboson processes that are initiated via the $gg$ production mode were simulated at LO in QCD for diagrams with up to one additional parton emission in the
matrix element using \OPENLOOPS in \Sherpa[2.2.2], which used the \NNPDF[3.0nnlo] PDF set. Diboson events with fully leptonic decays were modelled using
\Sherpa[2.2.2] and the \NNPDF[3.0nnlo] PDF set. \par
 
Finally, the production of a SM Higgs boson in association with a vector boson was simulated using \POWHEGBOX[v2], interfaced with \Pythia[8.212] for PS
and non-perturbative effects. The \POWHEG prediction is accurate to next-to-leading order for production of \vh\ plus one jet. The loop-induced $gg\to \Zh$
process was generated separately at LO. The \PDFforLHC[15] PDF set~\cite{Butterworth:2015oua} and the \AZNLO set of tuned parameters~\cite{STDM-2012-23}
of \Pythia[8.212] were used. The $gg\to\Zh$ production cross section was calculated at NLO precision, including the resummation of next-to-leading logarithmic (NLL)
soft gluon terms~\cite{Harlander:2014wda}. For the generation of \vh\ events, the Higgs boson mass was set to $125\,\GeV$. \par
 
A summary of MC generators and programs used to model the underlying event and parton shower (UEPS) for the simulation of background
processes is provided in Table~\ref{tab:generators}.

% The next lines are included from the .//texfiles/generators.tex input file
 
\begin{table}[h!]
\caption{Summary of the Monte Carlo generators used to produce the various background
processes. The perturbative accuracy (in QCD and, if relevant, in EW corrections) of the
total production cross section $\sigma_\mathrm{prod}$ is stated for each process. The
order at which the corresponding matrix elements are calculated in the Monte Carlo simulation is not
necessarily the same as for the cross section. Details on the final state are also provided, where $\ell = e, \mu, \mathrm{and}\,\tau$.}
\centering
\renewcommand{\arraystretch}{1.1}
\resizebox{1.0\textwidth}{!}{
\begin{tabular}{c c c}
\toprule
Process          & Generator  &  Perturbative accuracy of $\sigma_\mathrm{prod}$  \\
\toprule
$q \bar{q} \to Z^{\prime} \to Zh \to \ell\ell/\nu\nu+bb/cc$ & $\MADGRAPH5$\,2.3.3 + \Pythia[8.186] & LO \\
$q \bar{q} \to W^{\prime} \to Wh \to \ell\nu+bb/cc$     & $\MADGRAPH5$\,2.3.3 + \Pythia[8.186] & LO \\
$gg \to A \to Zh \to \ell\ell/\nu\nu+bb$                & $\MADGRAPH5$\,2.3.3 + \Pythia[8.186] & LO \\
$gg \to b\bar{b}A \to Zh \to \ell\ell/\nu\nu+bb$        & \MGNLO[2.2.3] + \Pythia[8.186] & NLO \\
\midrule
$W \to \ell\nu$, $Z\to \ell\ell$, $\PZ \to \Pgn \Pgn$ & \Sherpa[2.2.1]            &  NNLO \\
\midrule
$\ttbar$                                     & \POWHEGBOX[v2] + \Pythia[8.230]   &  NNLO+NNLL  \\
single top quarks ($s$- and $t$-channels)    & \POWHEGBOX[v2] + \Pythia[8.230]   &  NLO  \\
single top quarks ($Wt$-channel)             & \POWHEGBOX[v2] + \Pythia[8.230]   &  approx.~NNLO  \\
$\ttbar+h$                                   & \MGNLO[2.3.2] + \Pythia[8.186]    &  NLO (QCD) and NLO (EW)  \\
$\ttbar+V$                                   & \MGNLO[2.3.3] + \Pythia[8.230]    &  NLO \\
\midrule
$qg/q\bar{q} \to VV \to \ell\ell/\ell\nu/\nu\nu +q\bar{q}$ & \SHERPA~2.2.1 & NLO \\
$gg \to VV \to \ell\ell/\ell\nu/\nu\nu +q\bar{q}$          & \SHERPA~2.2.2 & NLO \\
$qg/q\bar{q} \to \ell\ell \nu\nu$                              & \SHERPA~2.2.2 & NLO \\
\midrule
$qq \to Wh \to \ell\nu + b\bar{b}$       & \POWHEGBOX[v2] + \Pythia[8.212]     & NNLO (QCD) and NLO (EW)  \\
$qq \to Zh \to \ell\ell/\nu\nu + b\bar{b}$  & \POWHEGBOX[v2] + \Pythia[8.212]     & NNLO (QCD) and NLO (EW)  \\
$gg \to Zh \to \ell\ell/\nu\nu + b\bar{b}$  & \POWHEGBOX[v2] + \Pythia[8.212]     & NLO+NLL  \\
\bottomrule
\end{tabular}}
\label{tab:generators}
\end{table}

% End of text imported from the .//texfiles/generators.tex input file

All simulated event samples include the effect of multiple \pp\ interactions in the same and neighbouring bunch crossings (pile-up) by overlaying simulated minimum-bias events on each generated signal or background event. The minimum-bias events were simulated with the
single-, double- and non-diffractive \pp\ processes of \Pythia[8.186] using the A3 set of tuned parameters~\cite{ATL-PHYS-PUB-2016-017} and the \NNPDF[2.3lo] PDF. For all \MADGRAPH and \POWHEGBOX samples, the \EVTGEN[1.6.0] program~\cite{EvtGen} was used for the bottom and charm hadron
decays. The generated samples were processed using the \GEANT-based ATLAS detector simulation~\cite{SOFT-2010-01,GEANT4} and the simulated events were reconstructed using the same algorithms as were used for the data events. \par
Simulated events were corrected to compensate for differences between data and simulations regarding the energy (or momentum) scale and resolution of leptons and jets, the efficiencies for the reconstruction, identification, isolation and triggering of
leptons, and the tagging efficiency for heavy-flavour jets.

% End of text imported from the .//texfiles/DataMCSamples.tex input file

% The next lines are included from the .//texfiles/Reconstruction.tex input file
\section{Event reconstruction}
\label{sec:Reconstruction}

Collision vertices are reconstructed from at least two ID tracks, each with a transverse momentum \linebreak $\pt > 500$~\MeV~\cite{ATL-PHYS-PUB-2015-026}.
Among all vertices, the one with the highest $\pt^2$ sum of all associated tracks is chosen to be the primary vertex (PV) of the event.
 
Electrons are reconstructed from ID tracks that are matched to energy clusters in the electromagnetic calorimeter and which originate from the PV. The latter condition is satisfied by a requirement on the
transverse impact parameter significance, $|d_0|/\sigma(d_0) < 5.0$, and on the longitudinal impact parameter, $|z_0\sin(\theta)| < 0.5$~mm. Electrons must satisfy requirements for the
electromagnetic shower shape, track quality, and track--cluster matching, using a likelihood-based approach. A \textit{Tight} operating point is used for the 1-lepton
channel and a \textit{Loose} operating point is used for the 2-lepton channel as well as the electron veto in the 0-lepton channel~\cite{PERF-2018-01}.
Electrons also have to be isolated from hadronic activity in the event: both the transverse energy sum and the scalar sum of transverse momenta of all ID tracks within
a cone of variable size around the electron have to be smaller than $0.06$ times the electron's transverse energy \ET. The maximum cone size is $\Delta R = 0.2$,
shrinking for larger \ET~\cite{PERF-2018-01}. Electron candidates are required to have a minimum \pt\ of 7~\GeV\ and to lie within $|\eta|< 2.47$.
 
Muons~\cite{MUON-2018-03} are identified by matching tracks found in the ID to either energy deposits in the calorimeter system consistent with a minimum-ionising particle
(\textit{calorimeter-tagged muons}) or full tracks (\textit{combined muons}) or track segments (\textit{segment-tagged muons}) reconstructed in the muon
spectrometer. Muons reconstructed as stand-alone tracks in the muon spectrometer are also considered.
Like electrons, they have to fulfil impact-parameter requirements: $|d_0|/\sigma(d_0) < 3.0$  and $|z_0\sin(\theta)| < 0.5$~mm. Muons are required to pass the \textit{Tight}
identification operating point for the 1-lepton channel or the \textit{Loose} identification operation point for both the 2-lepton channel and the muon veto in the 0-lepton channel. Muon candidates
with $\pt > 300~\GeV$ must satisfy tighter identification requirements in the muon spectrometer in order to improve the \pT\ resolution~\cite{MUON-2018-03}. Muons must also be isolated:  in the ID system, the \pT\ sum
within a variable-size cone around the combined track has to be smaller than $0.06$ times the muon's transverse momentum. The maximum cone size is $\Delta R = 0.3$, shrinking for
larger muon \pT. Muon candidates are required to have a minimum \pt\ of 7~\GeV\ and to lie within $|\eta| < 2.5$.
 
Three jet types are reconstructed, using the anti-\kt\ \cite{Cacciari:2008gp} algorithm implemented in the \textsc{FastJet} package~\cite{Cacciari:2011ma}.
\SmallR\ jets are reconstructed from noise-suppressed topological clusters in the calorimeter~\cite{PERF-2014-07,ATL-LARG-PUB-2008-002} using a radius parameter of $R=0.4$. They are required to
have $\pt>20$~\GeV\ for central jets ($|\eta|<2.5$) and $\pt>30$~\GeV\ for forward jets ($2.5<|\eta|<4.5$).
To suppress central jets from pile-up interactions, they are required to pass the jet vertex tagger~\cite{PERF-2014-03} selection if they have $\pt<120$~\GeV\ and $|\eta|<2.5$.
\LargeR\ jets are used to reconstruct high-momentum Higgs boson candidates, for which the $b$-quarks are emitted with small angular separation.
These jets are built using a radius parameter of $R=1.0$ and a combination of ID tracks and calibrated clusters of energy deposits in calorimeter cells,
referred to as Track-CaloClusters~\cite{ATL-PHYS-PUB-2017-015}. Trimming~\cite{Krohn:2009th} is applied to remove the energy of clusters that originate from initial-state radiation, pile-up interactions or the underlying event. This is done by reclustering the constituents of the initial jet, using the \kt\ algorithm~\cite{Catani:1993hr,Ellis:1993tq}, into smaller $R_{\mathrm{sub}}=0.2$ subjets and then removing any subjet that has a \pt\ less than 5\% of the \pt\ of the parent jet~\cite{ATL-PHYS-PUB-2015-033}.
The \largeR\ jets are required to have $\pt>250\,\GeV$ and $|\eta|<2.0$.
The momenta of both the \largeR\ and \smallR\ jets are corrected for energy losses in passive material and for the non-compensating response of the calorimeter. \SmallR\ jets are also corrected for the average additional energy due to pile-up interactions~\cite{PERF-2016-04,PERF-2011-04}.
The third type of jets are clustered from ID tracks using a variable radius (VR) that shrinks with increasing $p_\mathrm{T}$ of the studied proto-jet \cite{ATL-PHYS-PUB-2017-010}. VR track-jets are used in this analysis to identify $b$-jets from decays of boosted Higgs bosons. They must contain at least two ID tracks compatible with the primary vertex and must have $\pT > 10~\GeV$ as well as $|\eta| < 2.5$.
The VR track-jets are spatially matched to \largeR\ jets using ghost-association~\cite{XbbTagging}.\par
 
\SmallR\ jets  and VR track-jets containing $b$-hadrons are identified using the MV2c10 $b$-tagging
algorithm~\cite{FTAG-2018-01, ATL-PHYS-PUB-2017-013, ATLAS-CONF-2018-001, ATLAS-CONF-2018-006},
with an operating point that corresponds to a selection efficiency for $b$-jets of $70\%$, as measured in simulated \ttbar\ events.
Applying the $b$-tagging algorithm reduces the number of light-flavour and gluon jets, jets containing hadronically decaying $\tau$-leptons,
and $c$-quark jets, by a factor of 300 (304), 36 (55) and 9 (9), respectively, for \smallR\ jets (VR track-jets)~\cite{FTAG-2018-01}.\par
 
Hadronically decaying $\tau$-lepton candidates~\cite{TauEnergyCalibration, TauReconstruction} are used in the \vvbb\ channel to reject backgrounds with real
hadronically decaying $\tau$-leptons. They are reconstructed using calorimeter-based \smallR\ jets. They are required to have either one or three associated tracks,
$\pt > 20~\GeV$ and  $|\eta| < 1.37$ or  $1.52<|\eta|<2.5$. They are identified using a multivariate tagging algorithm, which is $55\%$ ($40\%$) efficient
for one-track (three-track) $\tau$-lepton candidates~\cite{TauReconstruction}. \par
 
The missing transverse momentum, with magnitude \met, is calculated as the negative vector sum of the transverse momenta of calibrated leptons and jets, plus a
track-based soft term, i.e.\ all tracks compatible with the primary vertex and not associated with any lepton or jet used in the \met\ calculation~\cite{PERF-2016-07, ATLAS-CONF-2018-023}.
In addition, a track-based missing transverse momentum estimator \vecmpt\ is built as the negative vector sum of the transverse momenta of all tracks from the
primary vertex. In order to reduce backgrounds from mismeasurements of jet and lepton energies, an object-based \met\ significance $\mathcal{S}$~\cite{ATLAS-CONF-2018-038}
is exploited. This observable takes the resolutions of objects entering the \met\ calculation into account and quantifies how likely it is that there is
significant \met\ from undetectable particles (such as neutrinos).\par
 
Electrons, muons, hadronically decaying $\tau$-leptons and jets are reconstructed and identified independently. This can lead to ambiguous identifications when these objects are spatially close to each other.
An overlap removal procedure is therefore applied to uniquely identify these objects. \SmallR\ jets are discarded if they are within a
cone of size $\Delta R = 0.2$ around an electron or $\tau$-lepton or if they have fewer than three associated tracks and are within a cone of $\Delta R = 0.2$
around a muon candidate. A $\tau$-lepton is removed if it is within a cone of $\Delta R = 0.2$ around an electron or a muon. Finally, electrons and muons are discarded
if they are within $\Delta R = \min(0.4, 0.04 + 10\,\GeV/p^\ell_\mathrm{T})$ of the axis of any surviving jet, where $p^\ell_\mathrm{T}$ is the transverse momentum
of the electron or muon.
 
Two dedicated correction procedures are applied to improve the mass resolution of the reconstructed Higgs boson candidates. If any muons are found within a cone of
\pT-dependent size around the jet axis, the four-momentum of the closest muon in $\Delta R$ with \pT\ larger than $5~\GeV$ is added to the jet four-momentum after
subtracting the energy contribution deposited in the calorimeter by the muon~\cite{EXOT-2016-10,HIGG-2018-51} (refered to as the muon-in-jet correction). For this
correction, muons are not required to pass any isolation requirements. Additional flavour-specific energy corrections are applied, after the
standard jet-energy-scale calibration~\cite{PERF-2016-04}, to both $b$-tagged \smallR\  and \largeR\ jets to account for biases in the response to jets with
leptonic or hadronic decays of heavy-flavour hadrons (refered to as the \pT-reconstruction correction). The \pT-reconstruction corrections are determined from
simulated SM $Vh(h\rightarrow b\bar{b})$ events by calculating the ratio of the \pT\ of the generator-level $b$-jets from the Higgs boson decay to the \pT\ of
the reconstructed $b$-tagged jets after the muon-in-jet correction is applied.\par

% End of text imported from the .//texfiles/Reconstruction.tex input file

% The next lines are included from the .//texfiles/AnalysisStrategy.tex input file
\section{Analysis strategy and event selection}
\label{sec:AnalysisStrategy}
 
The search for the \Zprime\ and \A\ bosons in the $Z\h \rightarrow \vvbb$ and $Z\h \rightarrow \llbb$ decay modes uses event selections
wherein the number of reconstructed charged leptons is exactly zero or two (0-lepton and 2-lepton channels). In the 0-lepton channel,
large $\met$ signals the presence of two neutrinos. For the \Wprime\ search in the $W\h\rightarrow\lpmnu\bbbar$ channel, events with
exactly zero or one charged lepton are used (0-lepton or 1-lepton channels).
 
For the 0-lepton channel and the 1-muon channel, \met\ triggers \cite{TRIG-2019-01} with thresholds of 70--110~\GeV\ are used for the
various data-taking periods, corresponding to the increasing instantaneous luminosity. The \met\ triggers are more efficient than single-muon triggers for the
relatively large momentum of the $W$ boson required in the analysis\footnote{Since muons are not included in the online \met\ calculation,
\met\ and $p_{\mathrm{T}, \textit{W}}$ are equivalent at the trigger level.} ($p_{\mathrm{T}, \textit{W}} > 150~\GeV$)~\cite{Aaboud:2018zhk}. The trigger efficiency
is above 80\% for events with $\met > 150~\GeV$ and essentially $100\%$ above $200~\GeV$.
 
In the 2-lepton channel and the 1-electron channel, events were recorded using a combination of single-lepton triggers~\cite{TRIG-2018-05,TRIG-2018-01} with
isolation requirements. The lowest \pT thresholds range from $24~\GeV$ to $26~\GeV$ in the electron channel and from $20~\GeV$ to $26~\GeV$ in the muon channel.
Additional triggers without an isolation requirement are used to recover efficiency for leptons with $\pt>60~(50)$~\GeV\ in the electron (muon) channel.
The trigger-level electron (muon) is required to match a reconstructed electron (muon) with $\pt > 27$~\GeV. The trigger efficiencies for the
combined single-electron and combined single-muon triggers are larger than $95\%$.
 
The Higgs boson candidate is reconstructed from the four-vectors of its decay products, the $b$-quarks. When the Higgs boson has
relatively low \pT, the $b$-quarks can be reconstructed as two \smallR\ jets (\textit{resolved} category). As the momentum of the
Higgs boson increases, the two $b$-quarks become more collimated and a selection using a single \largeR\ jet becomes more
efficient (\textit{merged} category).
 
For the resolved signal region, two central \smallR\ jets are required to have an invariant mass \mjj\ in the range
110--140~\GeV\ for the 0-lepton and 1-lepton channels and in the range 100--145~\GeV\ for the 2-lepton channel.
The latter selection is looser in order to take advantage of the lower background contribution in this channel.
When two $b$-tagged jets are present in the event, the dijet system representing the Higgs boson candidate
is defined by the two $b$-tagged \smallR\ jets. In events with more than two $b$-tagged jets, the dijet system representing
the Higgs boson candidate is defined by
the two leading $b$-tagged jets. In the case where only one $b$-tagged jet is present, the dijet pair is defined by the $b$-tagged
jet and the leading \smallR\ jet in the remaining set. The leading jet in the pair must have $\pT>45$~\GeV.
For the merged signal region, a \largeR\ jet is required with mass \mjet\ in the range 75--145~\GeV\ and at least one
associated $b$-tagged VR track-jet. Events which satisfy the selection requirements of both the resolved and merged categories are assigned
to the resolved category because of its better dijet mass resolution. In resolved event topologies, the four-momentum of the dijet system is
scaled by 125~\gev\,/\,\mjj.
 
Higgs boson candidates with one or two $b$-tagged jets define the \textit{1~$b$-tag} and \textit{2~$b$-tags} categories, respectively. For the merged selection,
only one or two leading VR track-jets associated with the \largeR\ jet are considered in this counting.
In the searches for \Zprime, \Wprime\ and $A$ boson production via the \ggA\ process, events with exactly one or exactly two $b$-tagged jets are considered,
while signal production via the \bbA\ process is addressed by introducing further categories for events with additional $b$-tagged jets. In the resolved region a category with at least
three $b$-tagged jets (\textit{3+~$b$-tags}) is used, while in the merged regions, events are assigned to additional categories based on the number of $b$-tagged VR
track-jets outside of the \largeR\ jet. The 0-lepton channel uses a category for events in which two $b$-tagged VR track-jets are associated with the \largeR\ jet
and at least one $b$-tagged VR track-jet is outside of the \largeR\ jet (\textit{2~$b$-tags \& 1+~add}).\footnote{The category for events with exactly one $b$-tagged VR track-jet associated with the \largeR\ jet
and at least one $b$-tagged VR track-jet outside of the \largeR\ jet (i.e.\ \textit{1~$b$-tag \& 1+~add}) does not add sensitivity to the 0-lepton channel and is therefore not used for the $A$ boson
search.} Furthermore, the 2-lepton channel uses a category in which at least one $b$-tagged VR track-jet is associated with the \largeR\ jet and at least one $b$-tagged VR track-jet is outside of
the \largeR\ jet (\textit{1+~$b$-tags \& 1+~add}).
 
The calculation of the reconstructed mass \mvh\ of the \Vh\ resonance depends on the decay channel.
In the 0-lepton channel, where it is not possible to fully reconstruct the \Vh\ system, the transverse mass is used as the final discriminant, defined
as: $m_{\mathrm{T},Vh} = \sqrt{\left(E_{\mathrm{T},h}+E^\mathrm{miss}_\mathrm{T}\right)^2-\left(\vec{p}_{\mathrm{T},h}+\vec{E}^\mathrm{miss}_\mathrm{T}\right)^2}$, where $E_{\mathrm{T},h}$ is the transverse energy of the Higgs boson candidate, while $\vec{E}^\mathrm{miss}_\mathrm{T}$ and $\vec{p}_{\mathrm{T},h}$ are vectors of the missing transverse momentum and Higgs boson transverse momentum, respectively.
In the 1-lepton and 2-lepton channels, \mvh\ is defined as the invariant mass calculated from the sum of four-momenta of the vector boson and Higgs boson
candidates. In the 1-lepton channel, the vector boson is reconstructed from the sum of the four-momenta of the charged lepton and the neutrino. The
momentum of the neutrino in the $z$-direction, $p_{z, \nu}$, is obtained by imposing a \W\ boson mass constraint on the lepton--neutrino system, which results in a
quadratic equation. Here $p_{z, \nu}$ is taken as either the real component of the complex solutions or the real solution with the smaller absolute magnitude.
The transverse mass of the \W\ boson, $m_{\mathrm{T},W}$, is calculated from the transverse component of the lepton momentum and the $\vec{E}^\mathrm{miss}_\mathrm{T}$.
In the 2-lepton channel, the vector boson is reconstructed from the sum of the four-momenta of the two charged leptons. In events with two muons, the mass resolution
in the signal region is improved by scaling the four-momentum of the dimuon system by 91.2~\gev\,/\,$m_{\mu\mu}$ to account for poorer dimuon mass resolution at
high $p_\mathrm{T}$.
 
An event-cleaning procedure is applied to all lepton channels. Events are removed if they contain overlaps between one of the VR track-jets used for $b$-tagging and
at least one VR track-jet with a $\pt$ above $5~\GeV$ and with at least two associated tracks, in order to prevent the selection of jets with ambiguously matched
tracks from heavy-flavour hadron decays. Additional selections are applied for each lepton channel, as outlined below, to reduce the main backgrounds and enhance the
signal sensitivity.
 
The 0-lepton channel vetoes electrons and muons and, for the \Zprime, \ggA\ and \bbA\ searches, also hadronically decaying $\tau$-lepton candidates ($\tauhad$).
In order to recover $W'\rightarrow Wh\rightarrow\tauhad\nu bb$ events via the 0-lepton channel, the veto on $\tauhad$ is relaxed for the \Wprime\ search such
that events with at most one $\tauhad$ are selected.
The multijet background is reduced by requiring a missing transverse momentum of $\met > 150~(200)~\GeV$ in the resolved (merged) category, an $\met$ significance $\mathcal{S}$
of 9.0--13.6, increasing linearly as a function of $m_{\mathrm{T},Vh}$, a track-based missing transverse momentum of $\mpt > 60~\GeV$, an angular separation between $\met$ and
$\mpt$ of $\Delta\phi(\vec{E}^\mathrm{miss}_\mathrm{T},\vec{p}^\mathrm{miss}_\mathrm{T}) < \pi/ 2$, an angular separation between $\met$ and the Higgs boson candidate of
$\Delta\phi(\vec{E}^\mathrm{miss}_\mathrm{T},h) > 2\pi/3$, and the minimum angular separation between $\met$ and the jets that are used in the event to satisfy
$\min\left[\Delta\phi(\vec{E}^\mathrm{miss}_\mathrm{T}, {\textrm{small-}R~\mathrm{jet}})\right] > \pi/9~(\pi/6)$ for events with up to three (more than three) jets.
In the resolved channel, the angular separation between the two Higgs boson candidate jets is required to be $\Delta\phi_{jj} < 7\pi/9$, and the scalar sum of all
jet $\pt$ values must satisfy $\sum |\pt| = S_\mathrm{T} > 120~(150)~\GeV$ if the event contains two (more than two) central jets. Finally, the resolved channel requires at
least two central \smallR\ jets, and the merged channel at least one \largeR\ jet.
 
For the 1-lepton channel, selected events are required to have exactly one electron or one muon with $\pT > 27~\GeV$. The multijet background is reduced by requiring $\met > 80\,(40)~\GeV$ for events with an
electron (muon) in the resolved category, and in the merged category the requirement is increased to $\met > 100~\GeV$ for both electrons and muons.
A $m_{Wh}$-dependent requirement is placed on $p_{\mathrm{T},W}$: it has to be larger than both $150\,\GeV$ and $\left[710 - (3.3\cdot 10^5)/m_{Wh}\right]\,\GeV$
for the resolved channel, and larger than both $150\,\GeV$ and $\left[394\cdot\ln(m_{Wh}) - 2350\right]\,\GeV$ for the merged channel,
where $m_{Wh}$ is expressed in $\GeV$. The selection requirements on $p_{\mathrm{T},W}$ exploit the fact that the $m_{Wh}$ and $p_{\mathrm{T},W}$ observables are more
strongly correlated for the signal process than for the various backgrounds.
Finally, the transverse mass of the \W\ boson candidate is required to be less than $300~\GeV$ in order to remove events in which the $E^\mathrm{miss}_\mathrm{T}$
or lepton momentum is not well reconstructed. In addition, the angular separation between the charged lepton and the Higgs boson candidate must
fulfil $\Delta R(\ell,h) > 2.0$.
 
In the 2-lepton channel, same-flavour leptons ($ee$ or $\mu\mu$) are selected with $\pt > 27~\GeV$ for the leading lepton and $\pt > 20~(25)~\GeV$ for the subleading
lepton in the resolved (merged) category. Three kinematic selections are optimised as a function of the resonance mass \mvh\ to reduce the \ttbar\ and \Zjets\ backgrounds.
The selection requirement on the mass of the dilepton system, max[$40~\GeV,\ 87~\GeV - 0.030\cdot \mvh$] $< \mll < 97~\GeV + 0.013\cdot\mvh$, and the requirement $\met/\sqrt{H_{{\mathrm T}}} < 1.15 + (8\times 10^{-3})\cdot \mvh\,/\,(1\ \GeV)$
vary as a function of the \mvh, and these requirements become looser for larger resonance masses to account for smaller backgrounds and a degradation of the dilepton invariant mass and
energy resolution with increasing boost of the final-state particles. The $H_{{\mathrm T}}$
is calculated as the scalar sum of the \pt\ of the leptons and \smallR\ jets in the event. For the third kinematic selection, the momentum of the dilepton system (\ptll) is required to be greater than
$\left[20 + 9 \cdot \sqrt{\mvh/(1~\GeV) - 320}\right]\,\GeV$ for \mvh\ greater than 320~\GeV. In the resolved dimuon category, an opposite-charge
requirement is applied to further reduce diboson backgrounds and because the probability of mismeasuring the charge of individual muons is very low.
 
All signal-region selections are summarised in Table~\ref{tab:selKinEvt}. The products of kinematic acceptance and reconstruction efficiency for
$\Zprime\ \rightarrow \Zh (h\rightarrow b\bar{b}/c\bar{c})$, $\bbA \rightarrow \Zh (h \rightarrow b\bar{b})$ and $\Wprime\ \rightarrow \Wh (h\rightarrow b\bar{b}/c\bar{c})$
are shown in Figure~\ref{fig:acceffggA} as a function of the resonance mass. In this context, the acceptance is defined as the fraction of simulated
signal events for which the expected final state particles pass all relevant object definition requirements, and the reconstruction efficiency is
defined as the fraction of simulated signal events passing the various selection criteria of a given signal region. The acceptance times efficiency
curves are complex functions of the resonance mass. Contributions from the resolved categories are dominant for relatively low masses, but
become insignificant for resonances with masses of multiple \TeV, while the contributions from the merged categories become dominant
for masses above 1.4\,\TeV. The overall acceptance times efficiency is relative stable over a large range of considered resonance masses
in the 0- and 1-lepton channels, while the 2-lepton channel suffers from a decline of lepton reconstruction, identification, and isolation
efficiencies for highly boosted $Z\rightarrow \ell^+\ell^-$ decays~\cite{ATL-PHYS-PUB-2022-056}, which leads to a degradation of the
acceptance times efficiency for resonance masses above 2\,\TeV.
 
The \mvh\ resolution is expected to be in the range 2\%--5\% (5\%--9\%) for the resolved (merged) \llbb\ channel and in the range
7\%--12\% (9\%--12\%) for the resolved (merged) \lvbb\ channel, while the $m_{{\mathrm T}, V\h}$ resolution is expected to range from
$8\%$ to about $28\%$ (from $12\%$ to $17\%$) for the resolved (merged) \vvbb\ channel. The resolution of the $m_{{\mathrm T}, V\h}$ and
\mvh\ deteriorates in the resolved (merged) category with increasing (decreasing) resonance mass.
 
In addition to the signal regions, dedicated control regions are used to improve the modelling of major backgrounds.
These control regions are constructed by inverting one of the selection requirements of the signal region:
in the 2-lepton resolved category, \ttbar\ control regions are created by selecting events with different-flavour oppositely charged leptons $e\mu$, and without
the $\met/\sqrt{H_{{\mathrm T}}}$ requirement. The \ttbar\ purity of this selection is greater than 90\%.
These events are used to constrain the \ttbar\ normalisation in the final fit. Dedicated \mjj\ (\mjet) sideband regions are defined in the resolved (merged)
$0$- and $1$-lepton regions by requiring \mjj (\mjet) to be between $50~\GeV$ and $200~\GeV$ but outside the respective signal-region windows defined above.
They are used to improve the modelling  of the \WZjets\ kinematics and to constrain the normalisation of the \ttbar\ and \WZjets\ backgrounds.
As the \Zjets\ backgrounds are by far the largest in the 2-lepton signal region, yielding up to $90\%$ of the total background,
and due to the significant shape difference between the \mvh\ templates for the hypothetical signal processes and the \Zjets\ backgrounds,
no additional control regions are required to constrain the \Zjets\ contributions. Thus, the 2-lepton \mjj and \mjet\ sideband regions are not included
into the likelihood fit. Instead, these sidebands are used to correct the kinematics of the \Zjets\ background (see Section~\ref{sec:Background}).
The channels and regions used for the different fits are detailed in Table~\ref{tab:regionsFit}.

% The next lines are included from the .//texfiles/eventSelectionTable.tex input file
{
\def\arraystretch{1.35}
\begin{table}[h!]
\caption{
Topological and kinematic selections for each channel and category as described in the text.
$^{(\ast)}$ Applies in the case of only two central jets.
$^{(\ast\ast)}$ Applied only for the \Wprime\ search.
$^{(\dagger)}$ A higher threshold (80~\GeV) is used for the single-electron channel.
$^{(\dagger\dagger)}$ Applied only for $m_{Vh} > 320$~\GeV.
$^{(\ddagger\ddagger)}$ Only the two leading VR track-jets matched by ghost-association to the large-$R$ jet are considered when classifying events into $b$-tag categories.
Events are further classified according to the number of $b$-tagged jets in the events.
}
\label{tab:selKinEvt}
\begin{center}
\resizebox{1.0\textwidth}{!}{
\begin{tabular}{l| c| c}
\hline\hline
Variable                                                                      & Resolved                                                                                                             & Merged                                                         \\
\hline
\multicolumn{1}{c}{}                                                          & \multicolumn{2}{c}{Common selection}                                                                                                                                                                                                                                  \\
\hline
\multirow{2}{*}{Number of jets}                                               & $\geq$2 ${\textrm{small-}R~\mathrm{jets}}$ (0, 2-lep.)                                                               & $\geq$1 large-$R$ jet                         \\
& 2 or 3 ${\textrm{small-}R~\mathrm{jets}}$ (1-lep.)                                                                                        &   $\geq 1$ VR track-jets (matched to leading large-$R$ jet)$^{\ddagger\ddagger}$                                                              \\
Leading jet \pt[\GeV]                                                       & $> 45$                                                                                                               & $> 250$                                                        \\
$m_{h}$  [\GeV]                                                               & 110--140 (0,1-lep.), 100--145 (2-lep.)                                                                               & 75--145                                                        \\
\hline
\multicolumn{1}{c}{}                                                          & \multicolumn{2}{c}{0-lepton selection}                                                                                                                                                                                                                                \\
\hline
\met\ [\GeV]                                                                  & $>150$                                                                                                               & $>200$                                                         \\
$S_\mathrm{T}$ [\GeV]                                                         & $>150$ $(120{}^{\ast})$                                                                                              & --                                                             \\
$\Delta\phi_{jj}$                                                             & $<7\pi/9$                                                                                                            & --                                                             \\\cline{2-3}
$p^\mathrm{miss}_\mathrm{T}$\ [\GeV]                                            & \multicolumn{2}{c}{$>60$}                                                                                                                                                  \\
$\Delta\phi(\vec{E}^\mathrm{miss}_\mathrm{T},\vec{p}^\mathrm{miss}_\mathrm{T})$ & \multicolumn{2}{c}{$<\pi/2$}                                                                                                                                                          \\
$\Delta\phi(\vec{E}^\mathrm{miss}_\mathrm{T},h)$                                & \multicolumn{2}{c}{$>2\pi/3$}                                                                                                                                                         \\
$\min\left[\Delta\phi(\vec{E}^\mathrm{miss}_\mathrm{T}, {\textrm{small-}R~\mathrm{jet}})\right]$            & \multicolumn{2}{c}{$>\pi/9$ (2 or 3 jets), $>\pi/6$ ($\geq$ 4 jets) }                                                                                                                 \\
$N_{\tau_{\mathrm{had}}}$                                                       & \multicolumn{2}{c}{0 ($\leq$ 1$^{\ast\ast}$)}                                                                                                                                                   \\
\met\ significance $\mathcal{S}$                                                  & \multicolumn{2}{c}{
$\begin{cases}
> 9     & \text{if } m_{Vh} < 240\,\GeV, \\
> 6.6 + 0.01 \cdot m_{Vh}  & \text{if } 240\,\GeV \leq m_{Vh} < 700\,\GeV, \\
> 13.6  & \text{if } m_{Vh} > 700\,\GeV,
\end{cases}$ }\\
\hline
\multicolumn{1}{c}{}                                                          & \multicolumn{2}{c}{1-lepton selection}                                                                                                                                                                                                                                \\
\hline
Leading lepton \pt\ [\GeV]                                                    & $>27$                                                                                                                & $>27$                                                          \\
\met\ [\GeV]                                                                  & $>40\ (80^{\dagger})$                                                                                                & $>100$                                                         \\
$p_{\mathrm{T},W}$\ [\GeV]                                                    & $ > \max\left[150,\ 710 - (3.3 \cdot 10^5\ \GeV) / m_{Vh}\right]$                                                    & $>\max\left[150,\ 394\cdot\log(m_{Vh}/(1\ \GeV)) - 2350\right]$ \\\cline{2-3}
$m_{\mathrm{T},W}$\ [\GeV]                                                    & \multicolumn{2}{c}{ < 300 }                                                                                                                                                           \\
$\Delta R(\ell,h)$                                                            & \multicolumn{2}{c}{> 2.0} \\
\hline
\multicolumn{1}{c}{}                                                          & \multicolumn{2}{c}{2-lepton selection}                                                                                                                                                                                                                                \\
\hline
Leading lepton \pt\ [\GeV]                                                    & $>27$                                                                                                                & $>27$                                                          \\
Subleading lepton \pt\ [\GeV]                                                & $>20$                                                                                                                 & $>25$                                                          \\ \cline{2-3}
$\met/\sqrt{H_{{\mathrm T}}}$ [$\sqrt{\mathrm{\GeV}}$]                        & \multicolumn{2}{c}{$< 1.15\ + 8\times 10^{-3}\cdot m_{Vh} / (1\ \GeV)$}                                                                                                               \\
$p_{\mathrm{T},\ell\ell}$\ [\GeV]                                             & \multicolumn{2}{c}{$> 20 + 9\cdot \sqrt{ m_{Vh}/(1\ \GeV) - 320}^{\dagger\dagger}$}                                                                                                   \\
$m_{\ell\ell}$\ [\GeV]                                                        & \multicolumn{2}{c}{ $\in\Bigl[\max\left[40\ ,\ 87 - 0.030\cdot m_{Vh} / (1\ \GeV)\right], 97 + 0.013\cdot m_{Vh} / (1\ \GeV) \Bigr]$ }                                                                  \\
\hline\hline
\end{tabular}}
\end{center}
\end{table}
}

% End of text imported from the .//texfiles/eventSelectionTable.tex input file

% The next lines are included from the .//texfiles/selectionFiguresAcceff.tex input file
\begin{figure}[h!]
\centering
\subfloat[]{
\includegraphics[width=0.45\textwidth]{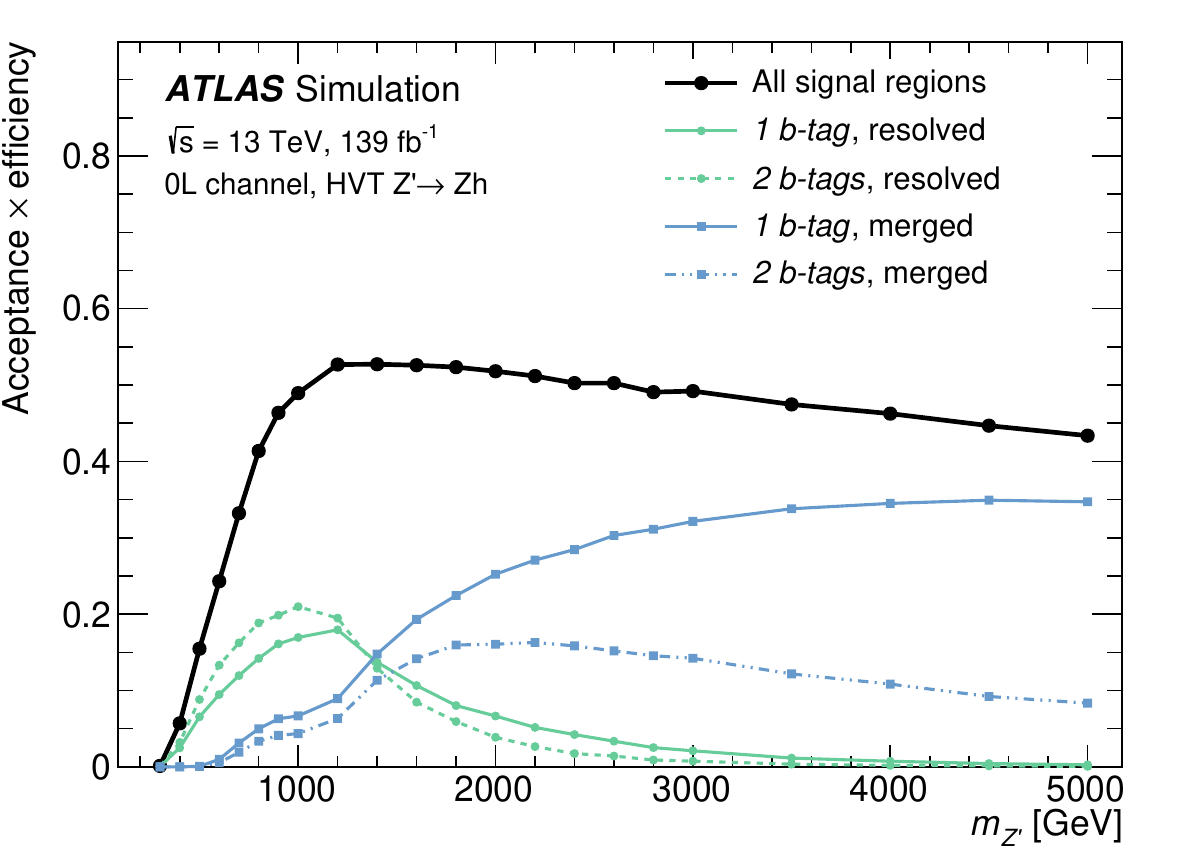}
}
\subfloat[]{
\includegraphics[width=0.45\textwidth]{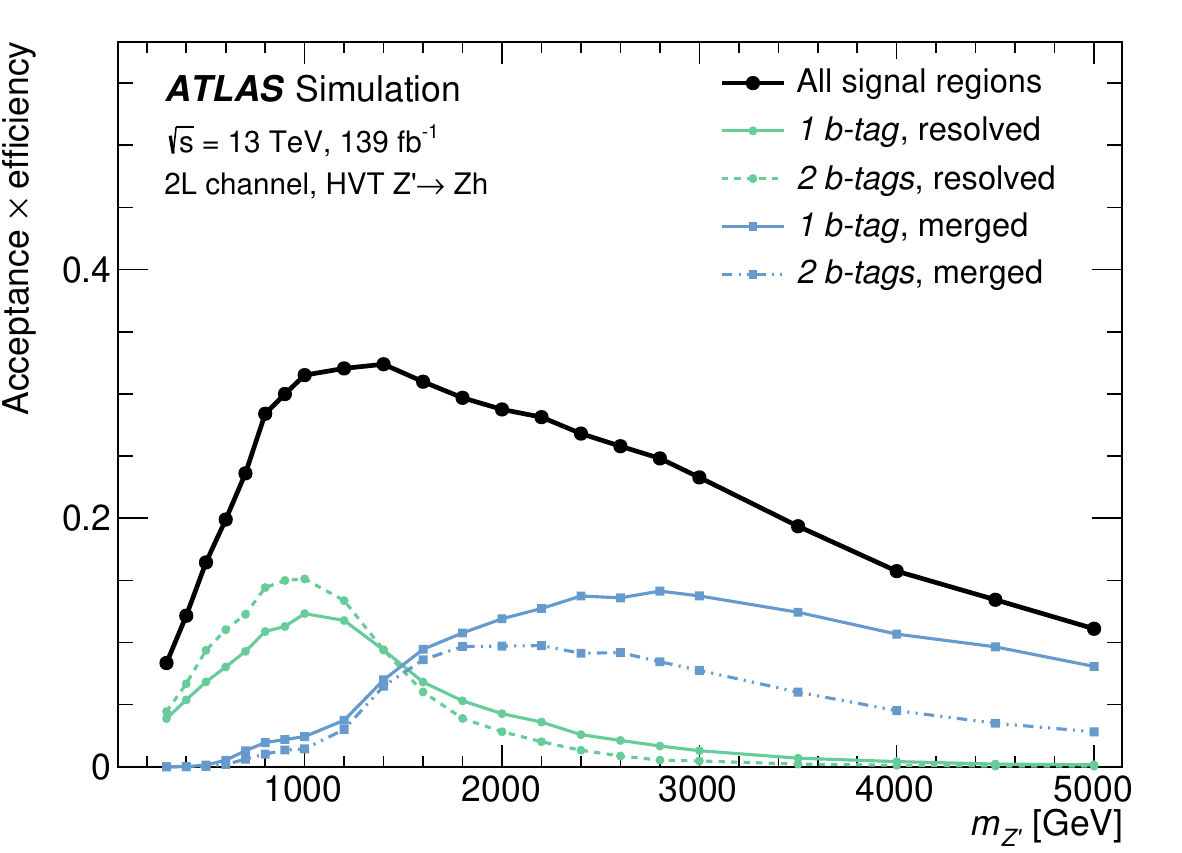}
}\\
\subfloat[]{
\includegraphics[width=0.45\textwidth]{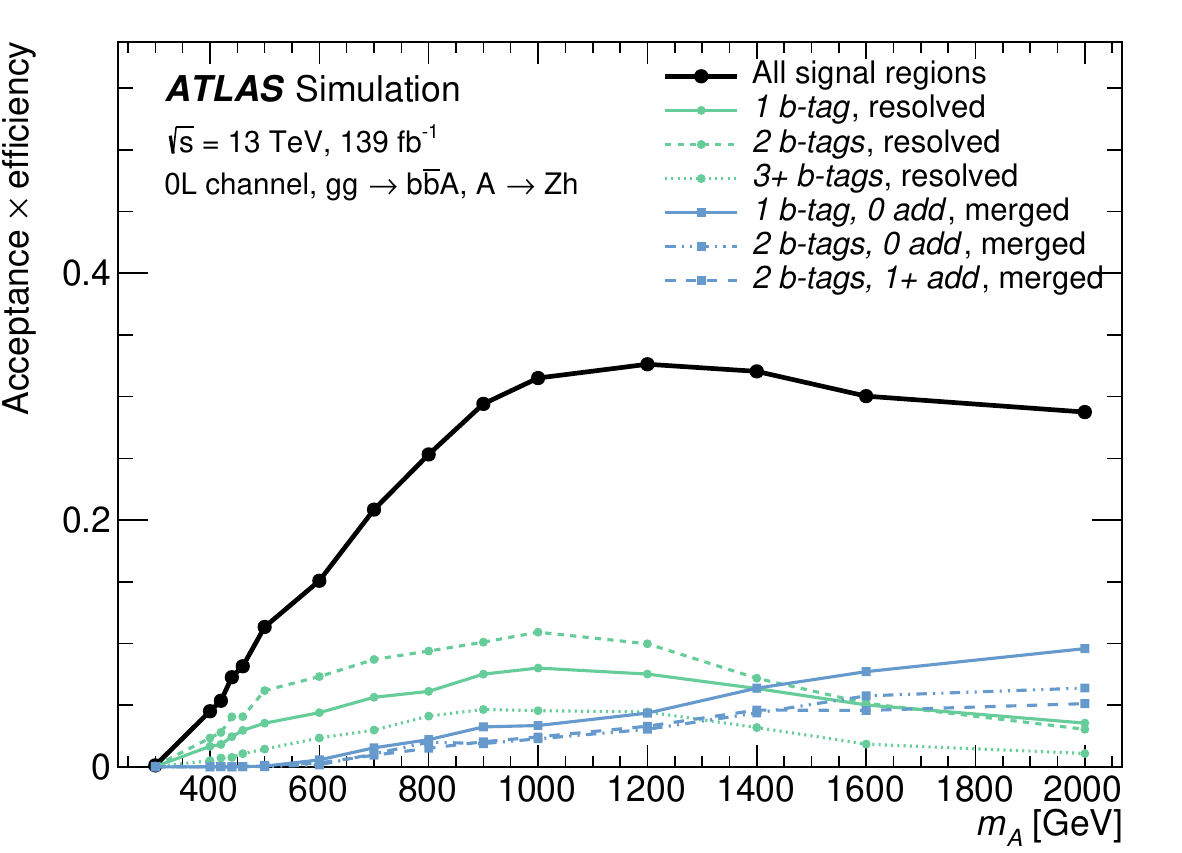}
}
\subfloat[]{
\includegraphics[width=0.45\textwidth]{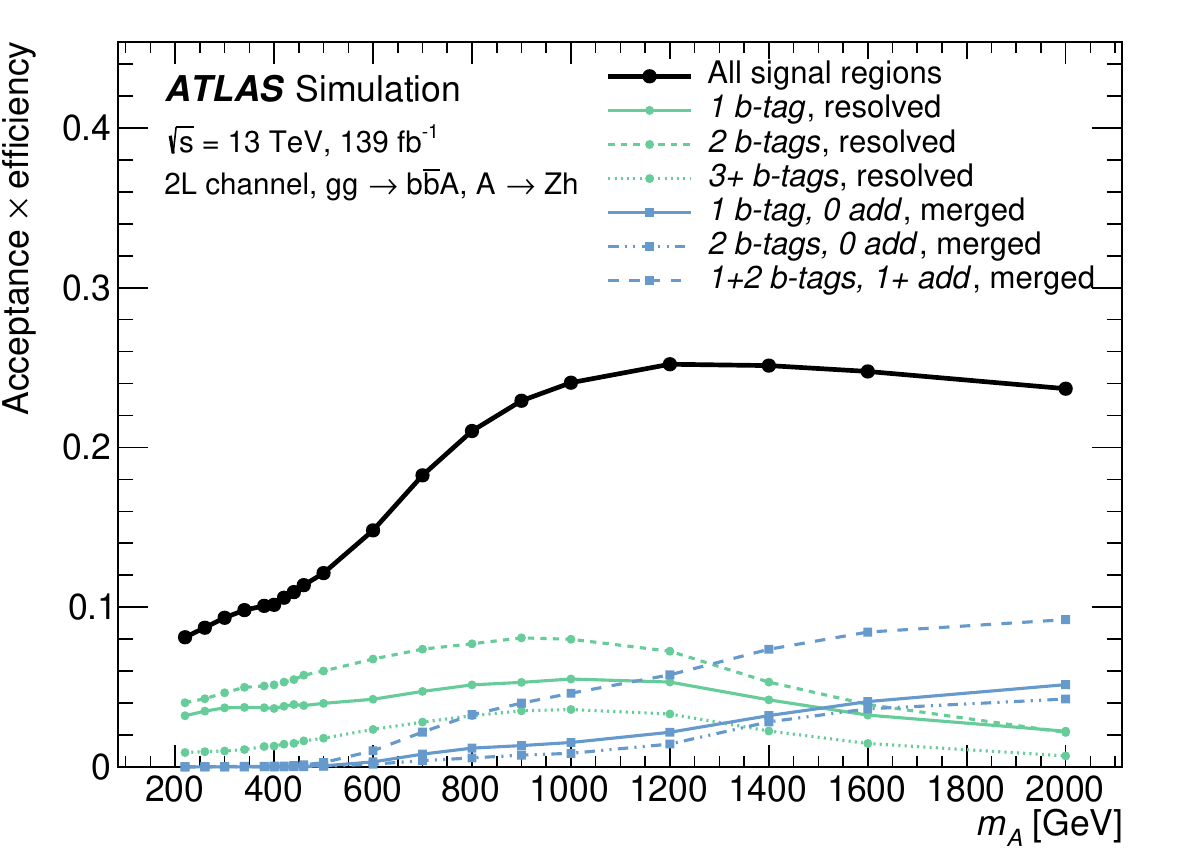}
}\\
\subfloat[]{
\includegraphics[width=0.45\textwidth]{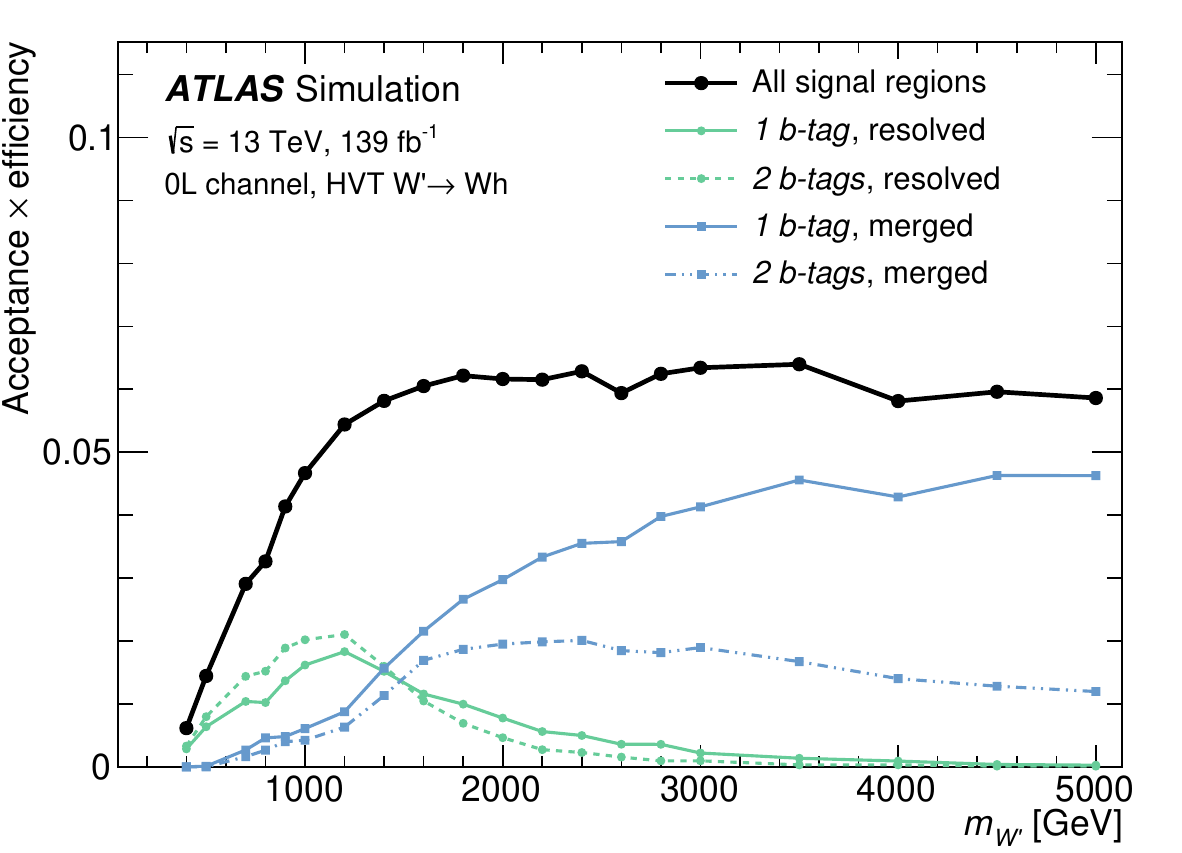}
}
\subfloat[]{
\includegraphics[width=0.45\textwidth]{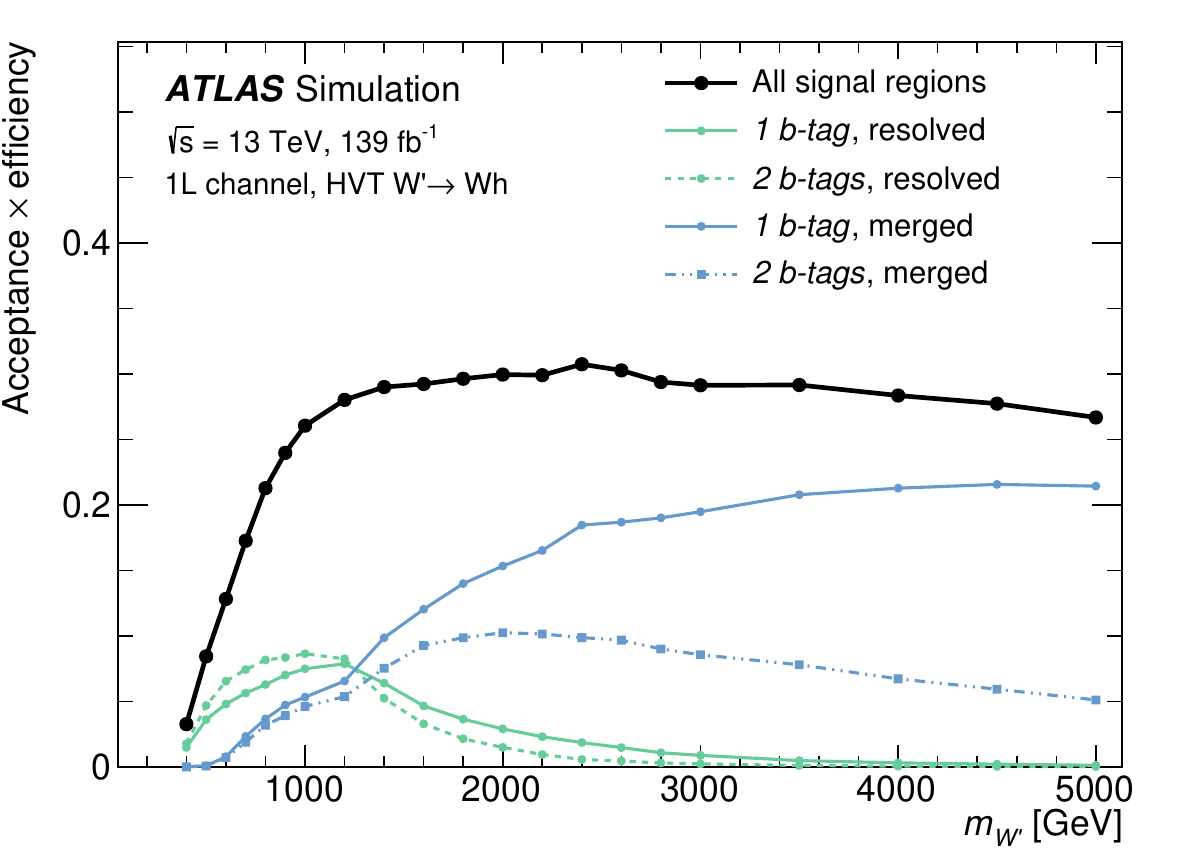}
}
\caption{
Product of acceptance and efficiency for (a) $\Zprime \rightarrow \Zh \rightarrow \vvbbcc$, (b) $\Zprime \rightarrow \Zh \rightarrow \llbbcc$,
(c) $\bbA\left(\rightarrow\Zh\rightarrow\vvbb\right)$, (d) $\bbA\left(\rightarrow\Zh\rightarrow\llbb\right)$ and (e, f) $\Wprime \rightarrow \Wh \rightarrow \lvbbcc$ as a function of the resonance mass for the 0-lepton signal regions (a, c, e), the 2-lepton signal regions (b, d), and the 1-lepton signal regions (f).}
\label{fig:acceffggA}
\end{figure}
\clearpage

% End of text imported from the .//texfiles/selectionFiguresAcceff.tex input file

% The next lines are included from the .//texfiles/fitRegionsTable.tex input file
{\def\arraystretch{1.25}
\begin{table}[h!]
\caption{\label{tab:regionsFit}
A list of the signal and control regions included in the statistical analysis of (a) the \ggA, \Zprime\ and \Wprime\
signal model hypotheses and (b) the \bbA\ signal hypothesis. The searches for the pseudoscalar boson and the \Zprime\ boson
are based on the 0- and 2-lepton channels, while the search for the \Wprime\ boson is based on the 0- and 1-lepton channels.
The notation 1+2 $b$-tag indicates that the 1 and 2 $b$-tag regions are combined.}
\begin{center}
\begin{footnotesize}
\subfloat[Signal and control regions for the \ggA, \Zprime and \Wprime searches]{
\resizebox{1.0\textwidth}{!}{
\begin{tabular}{c | c | c | c | c}
\hline\hline
Channel & Region    & Signal regions  & Control regions & Signal hypotheses \\
\hline
\multicolumn{5}{c}{Resolved}   \\
\hline{\multirow{2}{*}{0-lepton}}
& $b$-tags & \textit{1, 2 $b$-tags}  & \textit{1, 2 $b$-tags} &  {\multirow{2}{*}{\ggA, \Zprime, \Wprime}}\\
& Mass window & $ 110 < m_{jj} ~(\GeV) < 140 $ &   $50 < m_{jj} ~(\GeV) < 110~||~140 < m_{jj} ~(\GeV) < 200$ & \\
\hline{\multirow{2}{*}{1-lepton}}
& $b$-tags & \textit{1, 2 $b$-tags}  & \textit{1, 2 $b$-tags} & {\multirow{2}{*}{\Wprime}}\\
& Mass window & $ 110 < m_{jj} ~(\GeV) < 140 $ &   $50 < m_{jj} ~(\GeV) < 110~||~140 < m_{jj} ~(\GeV) < 200$ & \\
\hline{\multirow{3}{*}{2-lepton}}
& $b$-tags        & \textit{1, 2 $b$-tags}          &   \textit{1 + 2 $b$-tags} & {\multirow{3}{*}{\ggA, \Zprime}} \\
& Mass window     &  $ 100 < m_{jj} ~(\GeV) < 145 $   &   $ 100 < m_{jj} ~(\GeV) < 145 $ & \\
& Lepton flavour  & $ee$, $\mu\mu$                 &   $e\mu$ & \\
\hline
\multicolumn{5}{c}{Merged}   \\
\hline{\multirow{2}{*}{0-lepton}}
& $b$-tags & \textit{1, 2 $b$-tags}  & \textit{1, 2 $b$-tags}  &  {\multirow{2}{*}{\ggA, \Zprime, \Wprime}} \\
& Mass window & $75 < \mjet ~(\GeV) < 145 $ &  $50 < \mjet ~(\GeV) < 75~|| ~145 < \mjet ~(\GeV) < 200$ & \\
\hline{\multirow{2}{*}{1-lepton}}
& $b$-tags & \textit{1, 2 $b$-tags}  & \textit{1, 2 $b$-tags} & {\multirow{2}{*}{\Wprime}} \\
& Mass window & $75 < \mjet ~(\GeV) < 145 $ &  $50 < \mjet ~(\GeV) < 75~|| ~145 < \mjet ~(\GeV) < 200$ &  \\
\hline{\multirow{3}{*}{2-lepton}}
& $b$-tags & \textit{1, 2 $b$-tags}            &  {\multirow{3}{*}{--}} & {\multirow{3}{*}{\ggA, \Zprime}} \\
& Mass window & $75 < \mjet ~(\GeV) < 145 $ &  & \\
& Lepton flavour  & $ee$, $\mu\mu$          &  & \\
\hline\hline
\end{tabular}}}
 
\subfloat[Signal and control regions for the \bbA\ search]{
\begin{tabular}{c | c | c | c | c}
\hline\hline
Channel & Region    & Signal regions  & Control regions & Signal hypothesis \\
\hline
\multicolumn{5}{c}{Resolved}   \\
\hline{\multirow{2}{*}{0-lepton}}
& $b$-tags    & \textit{1, 2, 3+ $b$-tags}  & {\multirow{2}{*}{--}} & {\multirow{2}{*}{\bbA}} \\
& Mass window & $ 110 < m_{jj} ~(\GeV) < 140 $ &  \\
\hline{\multirow{3}{*}{2-lepton}}
& $b$-tags        & \textit{1, 2, 3+ $b$-tags}    &   \textit{1+2, 3+ $b$-tags}  & {\multirow{3}{*}{\bbA}} \\
& Mass window     & $ 100 < m_{jj} ~(\GeV) < 145 $   &   $ 100 < m_{jj} ~(\GeV) < 145 $ & \\
& Lepton flavour  & $ee$, $\mu\mu$                &   $e\mu$ & \\
\hline
\multicolumn{5}{c}{Merged}   \\
\hline{\multirow{2}{*}{0-lepton}}
& $b$-tags    & \textit{1, 2 $b$-tag, 2 $b$-tags \& 1+ add}       &  {\multirow{2}{*}{--}} & {\multirow{2}{*}{\bbA}} \\
& Mass window & $75 < \mjet ~(\GeV) < 145 $ &                     &    \\
\hline{\multirow{3}{*}{2-lepton}}
& $b$-tags    & \textit{1, 2 $b$-tag, 1+ $b$-tags \& 1+ add}   &  {\multirow{3}{*}{--}} & {\multirow{3}{*}{\bbA}} \\
& Mass window & $75 < \mjet ~(\GeV) < 145 $          &  &  \\
& Lepton flavour  & $ee$, $\mu\mu$                &  & \\
\hline\hline
\end{tabular}}
\end{footnotesize}
\end{center}
\end{table}}
\clearpage

% End of text imported from the .//texfiles/fitRegionsTable.tex input file

% End of text imported from the .//texfiles/AnalysisStrategy.tex input file

% The next lines are included from the .//texfiles/Background.tex input file
\section{Background estimation}
\label{sec:Background}
 
The background composition in the signal region depends on the charged-lepton and $b$-jet multiplicities. In the 0-lepton channel, the dominant background sources are \Zjets\ and \ttbar\ events,
with a significant contribution from \Wjets. In the 2-lepton channel, \Zjets\ production is the dominant background followed by  \ttbar\ production. The dominant background sources for
the 1-lepton channel are  \Wjets\ and \ttbar. Contributions from diboson, SM \Vh, $\ttbar+h$ and $\ttbar+V$ production are small in all channels.
 
Background contributions from multijet events were studied and found to be negligible for the 0-lepton channels, the 2-lepton channels and the merged 1-lepton
channel, while a small multijet background contribution (a few per mille of the total background) is found in the resolved 1~$b$-tag category for the 1-lepton channel.
This contribution is estimated using a data-driven method: a template is produced in a dedicated control region enriched with multijet events, constructed by inverting
the lepton isolation requirements decribed in Section~\ref{sec:Reconstruction}. The shape of the \mvh template is obtained by subtracting the other
backgrounds from data, assuming that the other background processes are well modelled in Monte Carlo simulation. The normalisation of the template is derived from a
two-component fit (multijet and `rest') to data in the $m_{jj}$ sideband control region. Further details of the template method can be found in Ref.~\cite{Aaboud:2018zhk}.
 
The other background distributions are estimated from the samples of simulated events, with normalisations of the main backgrounds estimated from the data. The
simulated \WZjets\ samples are reweighted as a function of the transverse momentum of the dijet system, \ptbb, based on fits to the data/simulation distribution in
the \mjj\ sideband regions, defined in Section~\ref{sec:AnalysisStrategy}. In the 2-lepton channel, different \ptbb\ reweighting are derived for the 1~$b$-tag and
2~$b$-tag regions. The \ptbb\ reweighting functions are derived after accounting
for differences in the overall normalisation between data and the simulations.\par
 
The \ttbar, single-top-quark, $\ttbar+V$ and $\ttbar+h$ processes are combined into one single component, which is referred to as the top-quark backgrounds.
In the search for a pseudoscalar produced via the \bbA\ mode, the top-quark backgrounds are split into two components. The categorisation is based on the flavour
of the additional jets\footnote{Generator-level particle jets are reconstructed from stable particles (i.e.\ particles with a mean lifetime
of $\tau > 3\cdot 10^{-11}$~seconds) using the anti-$k_{t}$ algorithm with a radius parameter of $R = 0.4$. These jets are required to have $p_\mathrm{T} > 15\,\GeV$ and $|\eta| < 4.5$.}
in the event, following the same strategy as described in Ref.~\cite{HIGG-2017-03}. Events containing at least one $b$- or $c$-quark jet not originating from a top
quark, $W$, $Z$ or Higgs boson decay are classified as \TopHF\ (with $\mathrm{hf}$ denoting `heavy flavour'), while all other events are classified as \Toplight\ (with $\mathrm{lf}$ denoting `light flavour'). \par
 
Simulated samples of \WZjets\ events are split into different jet-flavour components to help improve the grouping of the background components in the final fit because their contributions vary between the 1~$b$-tag and 2~$b$-tag categories.
In the resolved category, the samples are split according to the generated flavour of the two \smallR\ jets forming the Higgs boson candidate. In the merged category, they are split according to the generated flavour of the one or two leading track-jets associated with the \largeR\ jet.
The generated jet flavour is determined by counting generated heavy-flavour hadrons with $\pt > 5$~\GeV\ that are matched by ghost-association to the reconstructed jet.
If a $b$-hadron is found, the jet is labelled as a $b$-jet, otherwise if a $c$-hadron is found the jet is labelled as a $c$-jet. If neither a $b$-hadron nor a $c$-hadron is associated with the reconstructed jet, it is labelled as a light-flavour jet.
Based on this association scheme, the \WZjets\ simulated event samples are split into three components: \WZhf\ (\WZbb\ , \WZbc\ and \WZcc), \WZhl\ (\WZbl\ and \WZcl) and \WZl; in this notation `$\mathrm{l}$' refers to a light-flavour jet.
In the statistical analysis described in Section~\ref{sec:Results}, the global normalisations of the \Zhf, \Zhl, \Whf\ and \Whl\  components
are determined via a fit to data, while the normalisation of the \WZl\ component is taken from simulation. In the \A\ boson interpretation,
the \Zhf\ background normalisation in the 3+~$b$-tag region is determined from this region independently to account for  decreased
perturbative accuracy of the background predictions for events with additional heavy-flavour jets in the final state.
 
In the statistical analysis described in Section~\ref{sec:Results} the global normalisations of the dominant background contributions, i.e.\ the top-quark,
\WZhf\ , and \WZhl\ backgrounds, are determined separately for the 0-, 1- and 2-lepton channels via a fit to data. In the 0- and 1-lepton
channels, the \mjj\ (\mjet) sideband regions are used to constrain the \WZjets\ and top-quark backgrounds in the resolved (merged) event
categories. In the 2-lepton channel, an \emu\ control region is used for resolved event topologies to constrain the normalisation of the
combined top-quark backgrounds.

% End of text imported from the .//texfiles/Background.tex input file

% The next lines are included from the .//texfiles/Systematics.tex input file
\section{Systematic uncertainties}
\label{sec:Systematics}

The distributions of \mvh\ and \mtvh\ are affected by both experimental and modelling uncertainties, which enter the final fits as nuisance parameters. 
Uncertainties in the modelling of physics objects are correlated over signal and background processes, channels, kinematic regions  and distributions of observables.\par
 
The largest experimental systematic uncertainties are associated with the calibration and resolution of the \smallR\ and \largeR\ jet energy and of the \largeR\ jet mass. Further dominant uncertainties are related to the determination of the $b$-tagging efficiency and misidentification rate.
The uncertainties in the \smallR\ jet energy scale have contributions from in situ calibration studies, from the dependency on the pile-up activity, and from the flavour composition of jets~\cite{PERF-2016-04,PERF-2011-04}. An additional uncertainty in the energy calibration of $b$- and $c$-jets is also used. The uncertainty in the scale and resolution of the \largeR\ jet energy and mass is estimated by comparing the ratio of calorimeter-based to track-based measurements in dijet data and simulation~\cite{ATL-PHYS-PUB-2015-033,ATLAS-CONF-2016-035}.
Differences in the $b$-tagging efficiency measured in data and simulation result in correction factors for $b$-jets, $c$-jets and light-flavour jets.
Uncertainties in these correction factors are decomposed into uncorrelated components ~\cite{FTAG-2018-01, ATLAS-CONF-2018-001, ATLAS-CONF-2018-006}.
An additional term is included to extrapolate the measured uncertainties to the high-$p_\mathrm{T}$ region of interest. This term is calculated from simulated events
by considering variations of the quantities affecting the $b$-tagging performance, such as the impact parameter resolution, percentage of
tracks from random combinations of measurements in the ID, description of the detector material, and track multiplicity per jet. \par
Uncertainties in the reconstruction, identification, isolation and trigger efficiencies of muons~\cite{PERF-2015-10}, electrons~\cite{PERF-2017-01}
and $\tau$-leptons~\cite{ATL-PHYS-PUB-2015-045} and in their energy scale and resolution have only a small impact on the result.
Uncertainties in the \met trigger efficiency are taken into account by following the approach described in Ref.~\cite{HIGG-2018-51}.\par
The uncertainties in the energy scale and resolution of the \smallR\ jets and leptons are propagated to the calculation of \met, which also has additional uncertainties from the modelling of the underlying event and the momentum scale, momentum resolution and reconstruction efficiency of the tracks used to compute the soft-term~\cite{PERF-2016-07, ATLAS-CONF-2018-023}.
Finally, a global luminosity uncertainty of $1.7\%$ is applied to the normalisation of all simulated events~\cite{ATLAS-CONF-2019-021,Avoni:2018iuv}.\par
 
Theoretical uncertainties are derived for all simulated signal and background processes and lead to variations in the normalisation and shape of the templates used
for the statistical analysis. These uncertainties are assessed by comparing nominal and alternative event generators and UEPS
models. In general, the perturbative precision and the PDF sets used in these alternative configurations match those of the nominal generators (unless explicitly
stated). Uncertainties due to the PDF set parameterisation are evaluated using replica sets, and uncertainties due to missing higher orders are evaluated by varying
appropriate scale parameters, as described below. Uncertainties due to the modelling of PDFs are evaluated for the signal processes and all relevant
backgrounds by comparing the predictions of their nominal PDF set with those of alternative PDF sets and then comparing the maximum difference with the difference
from the root-mean-square spread of the nominal replica sets. The larger of the two is taken as the uncertainty. \par
 
For the signal processes, uncertainties in the acceptance were derived by replacing the nominal PDF set by the \MSTW[lo]~\cite{Martin:2009iq} and \CT[10]~\cite{Gao:2013xoa}
PDF sets, by using the eigenvariations of the \Pythia[8.186] A14 set of tuned parameter values~\cite{ATL-PHYS-PUB-2014-021}, and by replacing \Pythia[8.186] by
\Herwig[7.0.4]~\cite{HERWIG7}. Renormalisation and factorisation scale variations are also considered for the signal processes.
The effects of these variations are sizeable for both of the two $A$ boson production modes and negligible for the \Zprime\ and \Wprime\ signals. For all the $A$, \Zprime\ and \Wprime\ signals,
the total acceptance uncertainties are between 2\% and 7\%.
 
The modelling uncertainties in \ttbar\ events are derived as follows. To assess potential uncertainties in the matching of the matrix element to the parton shower,
the nominal sample is compared with a sample where \POWHEGBOX is replaced by \MGNLO. The parton shower modelling uncertainty is assessed by replacing \Pythia[8.230] by \Herwig[7.04] together
with the H7UE set of tuned parameters~\cite{HERWIG7}. The uncertainty in the modelling of initial-state radiation (ISR) is addressed by changing the renormalisation and
factorisation scales by a factor of two in combination with the eigenvariations of the A14 set of tuned parameters. The uncertainty in the modelling of
final-state radiation (FSR) is addressed by varying the FSR scale~\cite{ATL-PHYS-PUB-2014-021}.
 
To account for the ambiguities in the interference between top-quark pair and $Wt$ production, an alternative
sample generated with \POWHEGBOX and \Pythia[8.230], based on the diagram subtraction (DS) scheme~\cite{Frixione:2008yi}, is used. The difference between the
DS and DR schemes for $Wt$ production is considered as an additional systematic uncertainty.
 
The relative contributions of the single-top-quark, $\ttbar+V$ and $\ttbar+h$ processes to the combined top-quark background are varied
by $19\%$ for single top quarks and by $50\%$ for $\ttbar+V$ and $\ttbar+h$ production.
 
For \WZjets\ backgrounds, the effects of scale uncertainties are estimated by varying the renormalisation and factorisation scales by a factor of two. In order to account for variations of
both the matrix element and the parton shower, the nominal samples were compared with samples with LO-accurate matrix elements that contain up to four final-state partons
generated with \MGNLO[2.2.2] using the \NNPDF[3.0nlo] PDF set and interfaced to \Pythia[8.186]~\cite{Sjostrand:2007gsChanged} using the CKKW-L merging
procedure~\cite{Lonnblad:2001iq,Lonnblad:2011xxabc}. The A14 set of tuned parameters of \PYTHIA is used with the \NNPDF[2.3lo] PDF set. Decays of bottom and charm
hadrons are performed by \EVTGEN[1.2.0].
 
In the 1-lepton channel, additional \textit{JetBinMigration} uncertainties in the \Wjets\ backgrounds, based on Ref.~\cite{Stewart_2012}, are derived to account for
the modelling of the veto of additional jets in the selection. These uncertainties are evaluated by individually varying the contributions
from \Wjets\ events with either less than two jets or more than three jets at the generator level. In both cases, the contributions to the signal and control regions are
varied upwards/downwards by $50\%$. Only the shape uncertainties are retained because the normalisations of \Wjets\ backgrounds are allowed to float in the final fit. \par
 
For both the \ttbar\ and \WZjets\ backgrounds, uncertainties in the  PDF were determined by the standard deviation of the 100  \NNPDF[3.0nlo] replicas, by using \NNPDF[3.0nlo] PDFs with  varied \alphas~\cite{Ball:2014uwa}, and by comparisons with the nominal \MMHT[nnlo] and the \CT[14nnlo] PDF sets~\cite{Butterworth:2015oua}.\par
 
For the major backgrounds, the theory uncertainties listed above translate into pre-fit probability distributions (priors) for nuisance parameters affecting the global cross section
and/or acceptance, the acceptance ratios between different fit regions,\footnote{Extrapolation effects between the signal and control
regions, the resolved and merged categories, and the different lepton channels are taken into account.} and the shape of the distribution of the final mass
discriminant. For the diboson and SM \vh\  backgrounds, only a normalisation uncertainty is assigned, as these processes give only a
relatively small contribution to the studied phase-space regions. This normalisation uncertainty covers effects on both the cross section
and acceptance~\cite{DibosonUnc}. The largest pre-fit impact on the \mvh\ distribution's slope for \WZjets\ comes from ME+PS variations (10\%--20\% over the
mass range studied), from scale uncertainties (5\%--10\%) and from the comparison with alternative PDF replicas (5\%). The largest pre-fit
impacts on the \mvh\ distribution's slope in \ttbar\ simulation come from the parton shower modelling and the matching of the matrix element to the parton
shower (5\%--20\%), the ISR and FSR (2\%--10\%) and the PDF uncertainties (up to 5\%).
 
The multijet background gives only minor contributions (of a few per mille of the total background) to the 1~$b$-tag signal and control regions of the
1-lepton channel. Various conservative uncertainties are assigned to the data-driven multijet background estimate (see Section~\ref{sec:Background}):
an uncertainty in the normalisation is set to $50\%$ and an uncertainty in the shape is derived from the observed differences when varying the \met selection
requirement by $\pm25\%$. This variation yields a shape uncertainty of up to $40\%$ for the \mvh\ template. However, due to the small size of the
multijet background, its uncertainties do not have a significant impact on the final results.
 
Additional uncertainties in the \ptbb\ reweighting of the \WZjets\ backgrounds in the 0-lepton and 2-lepton channels (see Section~\ref{sec:Background}) are
estimated from the full difference between the original and reweighted distributions. Residual mismodellings (i.e.\ non-closures in comparisons between data and simulation)
of the $p^\mathrm{miss}_\mathrm{T}$ distribution in the 0-lepton channel, the $p_{\mathrm{T}, W}$ distribution in the 1-lepton channel, and the transverse
momentum of the leading \largeR\ jet in all three channels, were observed in the respective $m_{jj}$\ and $m_{J}$ sideband regions. The magnitude of these mismodellings
depends on $m_{Vh}$ and ranges from $1\%$ to $10\%$. These are not corrected, but the data-to-simulation ratios are used to define systematic uncertainties in the
relevant shapes and those are applied to all backgrounds as uncorrelated nuisance parameters.
 
The relative MC generator modelling systematic uncertainties (at the pre-fit stage), connected to the normalisation,
cross-region extrapolation, and shape of the signal and background processes are detailed in Tables \ref{tab:syst2} and \ref{tab:syst1}.

% The next lines are included from the .//texfiles/systematicsTable2.tex input file
 
{
\def\arraystretch{1.2}
\begin{table}[h!]
\caption{\label{tab:syst2}
Relative MC generator modelling systematic uncertainties (at the pre-fit stage), connected to the normalisation, cross-region extrapolation, and shape of the \WZjets\ background
processes included in the fits described in the text. An `S' indicates that a shape variation is included for the sources listed,
`$A \leftrightarrow B$' indicates an extrapolation uncertainty between the regions $A$ and $B$, and `norm.' is the sum of cross-section and acceptance variations.
A value of `float' indicates that the parameter is not constrained in the fit. The uncertainty in the normalisation ratios of two regions, determines to what
degree the normalisations of two regions can diverge in the combined fit.  A range of values means the value depends on the lepton channel. Parentheses indicate
when the uncertainty applies only to a given fit or a given region.
}

\vspace{0.5cm}
\begin{footnotesize}
\begin{tabular}{cc}
 
\begin{minipage}{0.5\linewidth}
\begin{center}
\begin{tabular}{l | c | c }
\hline\hline
Process    & Quantity/source & Value \\
\hline

\multirow{2}{*}{\Zhf}
& 0/2 lep. norm. & float\\
& 1-lep. norm. & 50\% \\
& 0/2-lep. resolved $\leftrightarrow$ merged  & 10\%--19\% \\
& 1-lep. resolved $\leftrightarrow$ merged    & ~~9\%--18\% \\
& 0-lep. SR $\leftrightarrow$ CR  & ~~5\%--12\% \\
& 1-lep. SR $\leftrightarrow$ CR  & ~~6\%--20\% \\
& 0-lep. $\leftrightarrow$ 1-lep.  & ~~7\%--27\% \\
& 0-lep. $\leftrightarrow$ 2-lep.  & ~~4\%--16\% \\
& generator, PDF, scale & S \\
& \ptbb\ reweighting & S \\
& $p^\mathrm{miss}_\mathrm{T}$ non-closure (0-lep.) & S \\
& $p_{\mathrm{T},W}$ non-closure (1-lep.) & S \\
& $\LargeR$ jet $ p_\mathrm{T}$ non-closure & S \\
\hline
 
\multirow{2}{*}{\Zhl}
& 0/2 lep. norm.  & float\\
& 1-lep. norm. & 50\% \\
& 0/2 lep. resolved $\leftrightarrow$ merged  & 15\%--28\% \\
& 1-lep. resolved $\leftrightarrow$ merged    & 12\%--13\% \\
& 0-lep. SR $\leftrightarrow$ CR  & ~~3\%--20\% \\
& 1-lep. SR $\leftrightarrow$ CR  & 5\%--7\% \\
& 0-lep. $\leftrightarrow$ 1-lep.  & ~~7\%--27\% \\
& 0-lep. $\leftrightarrow$ 2-lep.  & ~~6\%--17\% \\
& generator, PDF, scale & S \\
& \ptbb\ reweighting & S \\
& $p^\mathrm{miss}_\mathrm{T}$ non-closure (0-lep.) & S \\
& $p_{\mathrm{T},W}$ non-closure (1-lep.) & S \\
& $\LargeR$ jet $ p_\mathrm{T}$ non-closure & S \\
\hline
 
\multirow{2}{*}{\Zl}
& norm. & 19\%\\
& 0/2 lep. resolved $\leftrightarrow$ merged  & ~~8\%--50\% \\
& 1-lep. resolved $\leftrightarrow$ merged    & 10\%--29\% \\
& 0-lep. SR $\leftrightarrow$ CR  & ~~5\%--20\% \\
& 1-lep. SR $\leftrightarrow$ CR  & 29\%--99\% \\
& 0-lep. $\leftrightarrow$ 1-lep. & ~~8\%--39\% \\
& 0-lep. $\leftrightarrow$ 2-lep. & ~~4\%--17\% \\
& generator, PDF, scale & S \\
& \ptbb\ reweighting & S \\
& $p^\mathrm{miss}_\mathrm{T}$ non-closure (0-lep.) & S \\
& $p_{\mathrm{T},W}$ non-closure (1-lep.) & S \\
& $\LargeR$ jet $ p_\mathrm{T}$ non-closure & S \\
\hline

\hline
\end{tabular}
\end{center}
\end{minipage}

\hfill
 
\begin{minipage}{.5\linewidth}
\begin{center}
\begin{tabular}{l | c | c }
\hline\hline
Process    & Quantity/source & Value \\
\hline

\multirow{2}{*}{\Whf}
& 0/2 lep. norm.  & 30\% \\
& 1-lep. norm.  & float \\
& 0/2 lep. resolved $\leftrightarrow$ merged  & 13\%--28\% \\
& 1-lep. resolved $\leftrightarrow$ merged    & 15\%--22\% \\
& 0-lep. SR $\leftrightarrow$ CR  & ~~5\%--28\% \\
& 1-lep. SR $\leftrightarrow$ CR  & 3\% \\
& 0-lep. $\leftrightarrow$ 1-lep. & 4\% \\
& generator, PDF, scale & S \\
& \ptbb\ reweighting (0-lep.) & S \\
& $p^\mathrm{miss}_\mathrm{T}$ non-closure (0-lep.) & S \\
& $p_{\mathrm{T},W}$ non-closure (1-lep.) & S \\
& $\LargeR$ jet $ p_\mathrm{T}$ non-closure & S \\
& JetBinMigration (1-lep.) & S \\
\hline
 
\multirow{2}{*}{\Whl}
& 0/2 lep. norm. & 30\%\\
& 1-lep. norm. & float \\
& 0/2 lep. resolved $\leftrightarrow$ merged  &  ~~2\%--43\% \\
& 1-lep. resolved $\leftrightarrow$ merged    & 12\%--13\% \\
& 0-lep. SR $\leftrightarrow$ CR  & ~~2\%--20\% \\
& 1-lep. SR $\leftrightarrow$ CR  & 1\%--2\% \\
& 0-lep. $\leftrightarrow$ 1-lep. & 4\% \\
& generator, PDF, scale & S \\
& \ptbb\ reweighting (0-lep.) & S \\
& $p^\mathrm{miss}_\mathrm{T}$ non-closure (0-lep.) & S \\
& $p_{\mathrm{T},W}$ non-closure (1-lep.) & S \\
& $\LargeR$ jet $ p_\mathrm{T}$ non-closure & S \\
& JetBinMigration (1-lep.) & S \\
\hline
 
\multirow{2}{*}{\Wl}
& 0/2 lep. norm. & 20\% \\
& 1-lep. norm. & float \\
& 0/2 lep. resolved $\leftrightarrow$ merged  & 14\%--18\% \\
& 1-lep. resolved $\leftrightarrow$ merged    & 20\%--21\% \\
& 0-lep. SR $\leftrightarrow$ CR  & ~~4\%--20\% \\
& 1-lep. SR $\leftrightarrow$ CR  & 2\%--4\% \\
& generator, PDF, scale & S \\
& \ptbb\ reweighting (0-lep.) & S \\
& $p^\mathrm{miss}_\mathrm{T}$ non-closure (0-lep.) & S \\
& $p_{\mathrm{T},W}$ non-closure (1-lep.) & S \\
& $\LargeR$ jet $ p_\mathrm{T}$ non-closure & S \\
& JetBinMigration (1-lep.) & S \\
\hline

\hline
\end{tabular}
\end{center}
\end{minipage}\\
\end{tabular}
\end{footnotesize}
\end{table}}
\clearpage

% End of text imported from the .//texfiles/systematicsTable2.tex input file

% The next lines are included from the .//texfiles/systematicsTable1.tex input file
{
\def\arraystretch{1.2}
\begin{table}[h!]
\caption{\label{tab:syst1}
Relative MC generator modelling systematic uncertainties (at the pre-fit stage), connected to the normalisation, cross-region extrapolation, and shape of signal
and the top, SM \Vh\ and diboson background processes included in the fits described in the text. An `S' indicates that a shape
variation is included for the sources listed, `$A \leftrightarrow B$' indicates an extrapolation uncertainty between the regions $A$ and
$B$, and `norm.' is the sum of cross section and acceptance variations. A value of `float' indicates that the parameter is
not constrained in the fit. The uncertainty in the normalisation ratios of two regions, determines to what degree the normalisations of
two regions can diverge in the combined fit. A range of values means the value depends on the lepton channel. Parentheses indicate when the
uncertainty applies only to a given fit or a given region. }
\vspace{0.5cm}
\begin{footnotesize}
\begin{tabular}{cc}
 
\begin{minipage}{0.5\linewidth}
\begin{center}
\begin{tabular}{l | c | c }
\hline\hline
Process    & Quantity/source & Value \\
\hline
 
\multirow{1}{*}{ Signal }
& PS, ISR/FSR, PDF  &  2\%--7\%, S \\
\hline
 
\multirow{1}{*}{SM \V\h }
& 0/1/2-lep norm. & 32\% \\
& $p^\mathrm{miss}_\mathrm{T}$ non-closure (0-lep.) & S \\
& $p_{\mathrm{T},W}$ non-closure (1-lep.) & S \\
& $\LargeR$ jet $ p_\mathrm{T}$ non-closure & S \\
\hline
 
\multirow{1}{*}{Diboson }
& 0/1-lep norm. & 50\% \\
& 2-lep norm. & 20\% \\
& $p^\mathrm{miss}_\mathrm{T}$ non-closure (0-lep.) & S \\
& $p_{\mathrm{T},W}$ non-closure (1-lep.) & S \\
& $\LargeR$ jet $ p_\mathrm{T}$ non-closure & S \\

\hline
\end{tabular}
\end{center}
\end{minipage}

\hfill
 
\begin{minipage}{.5\linewidth}
\begin{center}
\begin{tabular}{l | c | c }
\hline\hline
Process    & Quantity/source & Value \\
\hline
 
\multirow{2}{*}{Top quark} & 0/1/2-lep. norm. & {\multirow{2}{*}{float}} \\
& (\bbA: separate \ttbar+hf norm.) & \\
& single-top-quark contribution    & 19\% \\
& $\ttbar+V$ contribution          & 50\% \\
& $\ttbar+h$ contribution          & 50\% \\
& 0-lep. resolved $\leftrightarrow$ merged  & ~~9\%--20\% \\
& 1-lep. resolved $\leftrightarrow$ merged  & 18\%--20\% \\
& 2-lep. resolved $\leftrightarrow$ merged  & 18\% \\
& 0-lep. SR $\leftrightarrow$ CR  & ~~2\%--12\% \\
& 1-lep. SR $\leftrightarrow$ CR  & 2\%--3\% \\
& 2-lep. SR $\leftrightarrow$ CR  & 1.2\% \\
& PS, ISR/FSR, ME, PDF  & S \\
& DS vs DR scheme ($Wt$) & S \\
& $p^\mathrm{miss}_\mathrm{T}$ non-closure (0-lep.) & S \\
& $p_{\mathrm{T},W}$ non-closure (1-lep.) & S \\
& $\LargeR$ jet $ p_\mathrm{T}$ non-closure & S \\
\hline

\hline
\end{tabular}
\end{center}
\end{minipage}
 
\end{tabular}
 
\end{footnotesize}
\end{table}
}

% End of text imported from the .//texfiles/systematicsTable1.tex input file

% End of text imported from the .//texfiles/Systematics.tex input file

% The next lines are included from the .//texfiles/Results.tex input file
\section{Results}
\label{sec:Results}
 
In order to test for the presence of a massive resonance, the $\mtvh$ and $\mvh$ templates obtained from the signal and background simulated event samples are fitted
to data using a binned maximum-likelihood approach based on the \textsc{RooFit/RooStats} framework~\cite{Moneta:1289965,Verkerke:2003ir,Baak2015}.
Fits are simultaneously performed on the SRs and CRs defined in Table~\ref{tab:regionsFit} in order to constrain the normalisation of backgrounds and the
nuisance parameters (NPs) describing the systematic uncertainties detailed in Section~\ref{sec:Systematics}. Each systematic variation enters
the fit as an individual NP using a log-normal constraint so that the expected event yields remain positive for
all values of the corresponding NPs (even for bins with low event counts). The correlations between systematic uncertainties are maintained across
regions and processes. Fits are performed on the $\mtvh$ distribution in the 0-lepton channel and the $\mvh$ distribution in the 1-lepton and 2-lepton channels
using a binning of the distributions chosen to optimise the search sensitivity while minimising statistical fluctuations. Asimov datasets~\cite{Cowan:2010js} are
used to evaluate the expected performance of each fit.
 
The results are expressed as upper limits at the $95\%$ confidence level (CL) on the production cross section times branching fraction
of \Zprime, \Wprime, and $A$ bosons for a wide range of resonance masses. The limits are evaluated using a modified frequentist method known as CL$_\text{s}$~\cite{Read:2002hq} and the
profile-likelihood-ratio test statistic in an asymptotic approximation~\cite{Cowan:2010js}.
 
Different fit configurations (i.e.\ fit models) are used depending on the targeted signal hypothesis, production mode, or final state:
 
\begin{itemize}
\item Searches for heavy vector bosons produced via quark--antiquark annihilation:
\begin{itemize}
\item The search for a \Zprime\ boson is based on a combined fit in the 0-lepton and 2-lepton channels. The signal cross section is a free parameter, as are the
global normalisations of the top-quark, \Zhf\ and \Zhl\ backgrounds. Common normalisation factors are used for the \Zhf\ and \Zhl\ backgrounds, while the
normalisation of the top-quark backgrounds is determined separately for the 0-lepton and 2-lepton event categories. The results of these searches are presented as
one-dimensional upper limits at $95\%$ CL on the production cross section times
branching ratio, as well as constraints on the parameter space of the HVT model.
\item The search for a \Wprime\ boson is based on a combined fit in the 0-lepton and 1-lepton channels.
The signal cross section is a free parameter, as are the global normalisations of the top-quark, \Zhf\ , \Zhl\ , \Whf\ and \Whl\ backgrounds.
Results are presented as one-dimensional upper limits at $95\%$ confidence level on the production cross section times branching ratio,
as well as constraints on the parameter space of the HVT model.
\end{itemize}
\item Searches for a pseudoscalar $A$ boson based on combined fits in the 0-lepton and 2-lepton channels.
\begin{itemize}
\item The search for the gluon--gluon fusion production mode follows the same strategy as the search for the \Zprime\ boson.
\item In the search for the $b$-associated production mode, the signal cross section is a free parameter, as are the
global normalisations of the \Toplight, \TopHF, \Zhf\ and \Zhl\ backgrounds. An additional normalisation
factor is used for the \Zjets\ background contribution in the event categories with three or more $b$-tagged jets.
\item A simultaneous fit of the \ggA\ and \bbA\ signal hypotheses is performed for a variety of different $\sigma_{\bbA}$ and $\sigma_{\ggA}$ combinations.
A negative log-likelihood (NLL) value is calculated for each $(\sigma_{\bbA}$, $\sigma_{\ggA})$ combination, and the best estimate is obtained at the point
where the NLL curve reaches its minimum. Contours at $68\%$ and $95\%$ CL are determined relative to the minimum. The background model is the same as for
the \bbA\ search.
\item A combined fit of the \ggA\ and \bbA\ production modes with one free parameter is performed to determine the total cross section of the two components.
The fit is performed for different admixtures of \ggA\ and \bbA\ contributions, $\sigma_{\bbA}/(\sigma_{\bbA} + \sigma_{\ggA})$, as well as different $A$ boson widths
$\Gamma_A$. The results of these fits are used to set constraints on the parameter space of the 2HDM. The background model is the same as for
the \bbA\ search.
\end{itemize}
\end{itemize}
 
The search for a \Zprime\ and \Wprime\ boson seeks resonances in the mass range from $300~\GeV$ to $5~\TeV$, while the
search for a pseudoscalar seeks resonances in a mass range from $220~\GeV$ to $2~\TeV$.  In total, 19 simulated signal mass
hypotheses are tested. In addition, parameter morphing~\cite{Baak:2014fta} is applied to the shape and normalisation of the $m_{\mathrm{T},Vh}$ and $m_{Vh}$
distributions for all four signal hypotheses in order to interpolate between the simulated signal mass points. \par
 
The same normalisation factors are used across the various regions entering the likelihood fit. However, dedicated nuisance parameters
exist to account for extrapolation effects between the signal and control regions, between the resolved and merged categories as well as between
the different lepton channels. The corresponding extrapolation uncertainties are detailed in Table~\ref{tab:syst2} and Table~\ref{tab:syst1}.
The normalisation scale factors and their uncertainties obtained from a background-only fit in the various signal and control regions of the
combined 0- and 2-lepton channels, as well as the combined 0- and 1-lepton channels, are summarised in Table~\ref{tab:postFitNFs}.
In the search for the $bbA$ process, separate normalisation scale factors are used for \TopHF\ and \Toplight\ production. Separate
normalisation scale factors for \Toplight\ are used for the 0-lepton and 2-lepton channels, while a common normalisation scale factor
is used for \TopHF\ due to statistical limitations in the 3+ $b$-tags regions of the 2-lepton channel.
The normalisation scale factors are consistent with unity for \Toplight\ and $1.27\pm 0.15$ for \TopHF. Since the \TopHF\ process gives only small
contributions to the 1~$b$-tag and 2~$b$-tag signal regions, no bias was found in the \ggA, \Zprime\ and \Wprime\ fit models, in which
\TopHF\ and \Toplight\ are described by a common normalisation scale factor.

% The next lines are included from the .//texfiles/PostFitNormFactors.tex input file
{\def\arraystretch{1.2}
\begin{table}[!h]
\caption{\label{tab:postFitNFs} Post-fit normalisation scale factors (NF) and their uncertainties obtained from a combined background-only fit to the various signal and control regions of the 0- and 2-lepton channels and from a combined background-only fit of the 0- and 1-lepton channels. Numbers are presented for the background components that are allowed to float in the likelihood fit. The top-quark background is normalised with separate NFs for the 0-, 1- and 2-lepton regions, while the NFs for the \Zhf\ and \Zhl\ backgrounds are correlated between the 0- and 2-lepton channels.}
\begin{center}
\begin{tabular}{c c c c }
\hline\hline
Background & 0-lepton & 2-lepton  & 1-lepton\\
\hline
\multicolumn{4}{c}{Combined \Zprime\ /\ggA\ fit of 0- and 2-lepton channels}\\
\hline
top  & \num{0.92\pm0.05}   &  \num{1.02\pm0.02} & -- \\
\Zhf & \multicolumn{2}{c}{\num{1.18 \pm 0.07}}  & -- \\
\Zhl & \multicolumn{2}{c}{\num{1.16 \pm 0.06}}  & -- \\
\hline
\multicolumn{4}{c}{Combined \Wprime\ fit of 0- and 1-lepton channels}\\
\hline
top  & \num{0.92\pm0.04}    &  --  & \num{0.93\pm0.04} \\
\hline
\Zhf & \num{1.30 \pm 0.16}  &  --  &  -- \\
\Zhl & \num{1.30 \pm 0.16}  &  --  &  -- \\
\hline
\Whf &  -- & -- &  \num{1.35 \pm 0.18}  \\
\Whl &  -- & -- &  \num{1.34 \pm 0.12}  \\
\hline
\multicolumn{4}{c}{Combined \bbA\ fit of 0- and 2-lepton channels}\\
\hline
\TopHF    & \multicolumn{2}{c}{\num{1.27 \pm 0.15}}  & -- \\
\Toplight & \num{0.92\pm0.11}   &  \num{1.03\pm0.08}  & -- \\
\hline
$Z$ + jets ($N^{b\textrm{-tags}}\geq 3$) & \multicolumn{2}{c}{\num{1.19 \pm 0.15}}  &  -- \\
\Zhf & \multicolumn{2}{c}{\num{1.21 \pm 0.07}}  &  -- \\
\Zhl & \multicolumn{2}{c}{\num{1.12 \pm 0.06}}  &  -- \\
\hline\hline
\end{tabular}
 
\end{center}
\end{table}}

% End of text imported from the .//texfiles/PostFitNormFactors.tex input file

The $\mtvh$ and $\mvh$ distributions after a fit of the background templates to data are shown in Figures~\ref{fig:mVH_ZPpostfit_resolved} and~\ref{fig:mVH_ZPpostfit_merged}
for the signal regions used in the search for a \Zprime\ boson. Similar distributions are obtained from the 2HDM \ggA\ fits, with background yields consistent within
the uncertainties. Figure~\ref{fig:mVH_WPpostfit} shows the signal region distributions in the 1-lepton channel after the \Wprime\ fit.
The distributions of the $\mtvh$ and $\mvh$ observables are presented in Figure~\ref{fig:mVH_bbApostfit1} for events with three or more $b$-tags
(i.e.\ regions used in the search for the \bbA\ process) after a fit of the background templates to data.
Control region distributions are presented in Figure~\ref{fig:CR_Summary_Vprime}.
\clearpage
 
The expected and observed event yields after fits of the backgrounds as well as the \bbA\ and \Wprime\ signals to data are shown in Table~\ref{tab:postFitYields}.

% The next lines are included from the .//texfiles/resultFiguresPostfit.tex input file
\begin{figure}[h!]
\centering
\subfloat[]{
\includegraphics[width=0.50\textwidth]{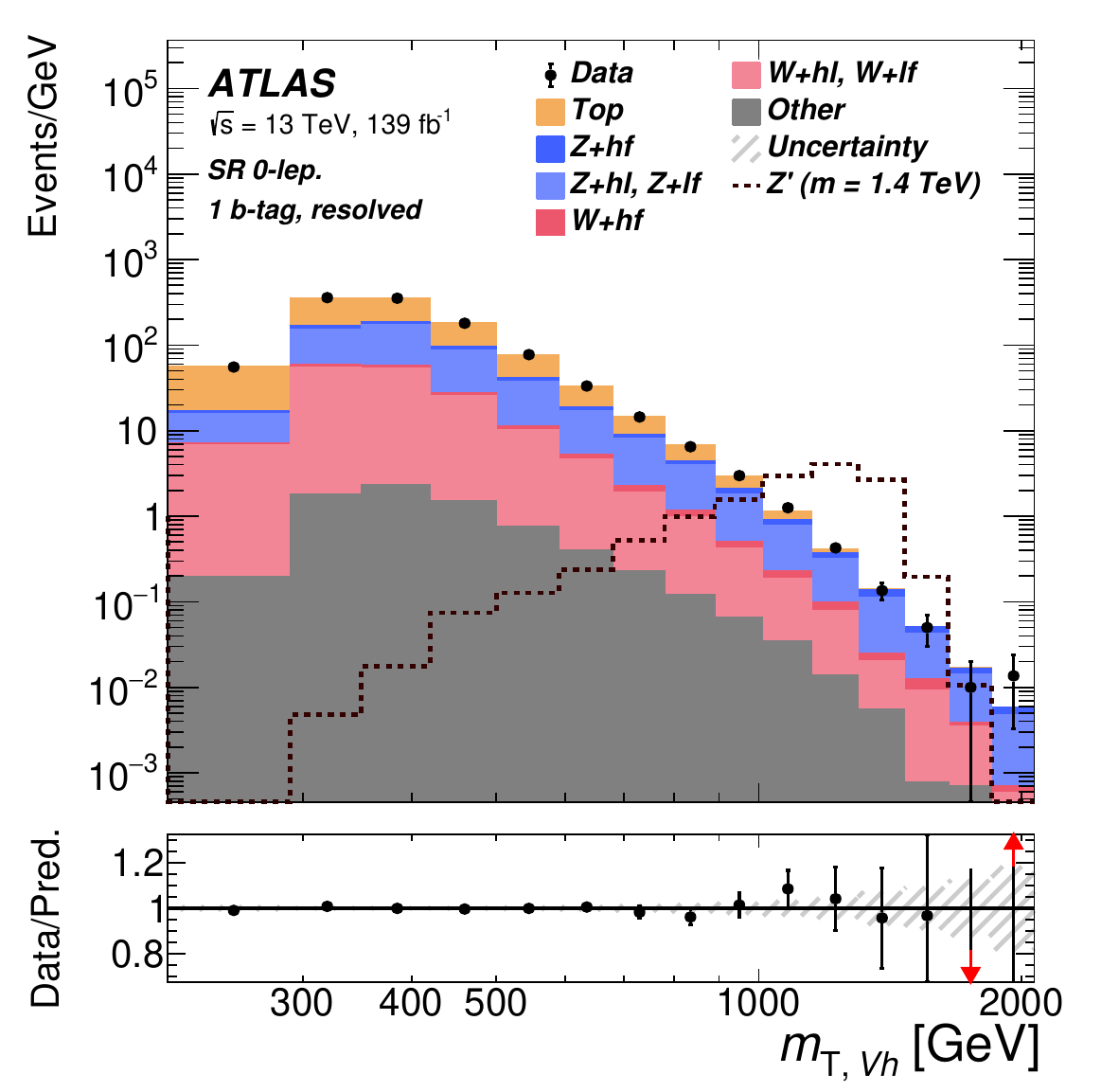}
}
\subfloat[]{
\includegraphics[width=0.50\textwidth]{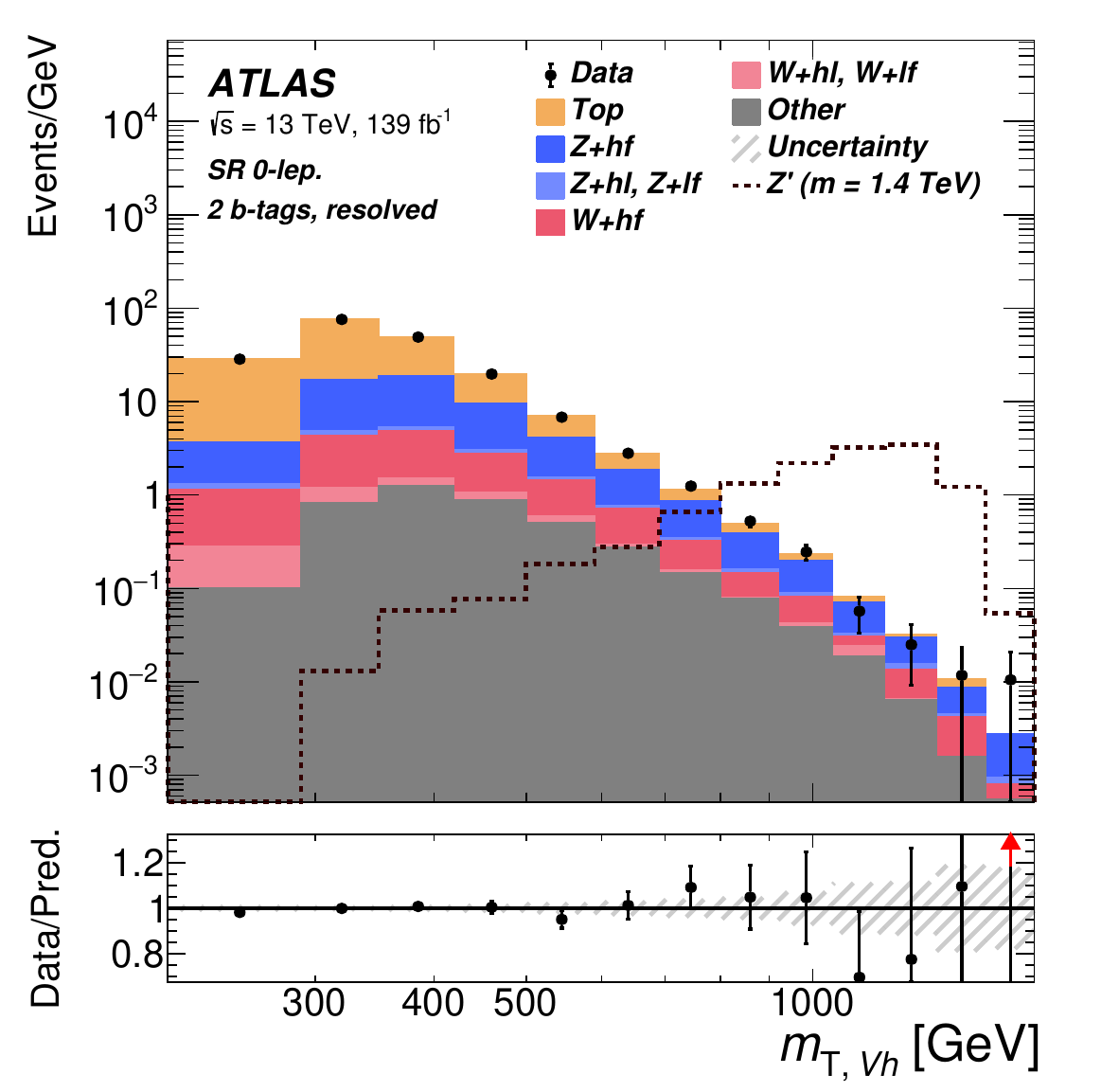}
}\\
\subfloat[]{
\includegraphics[width=0.50\textwidth]{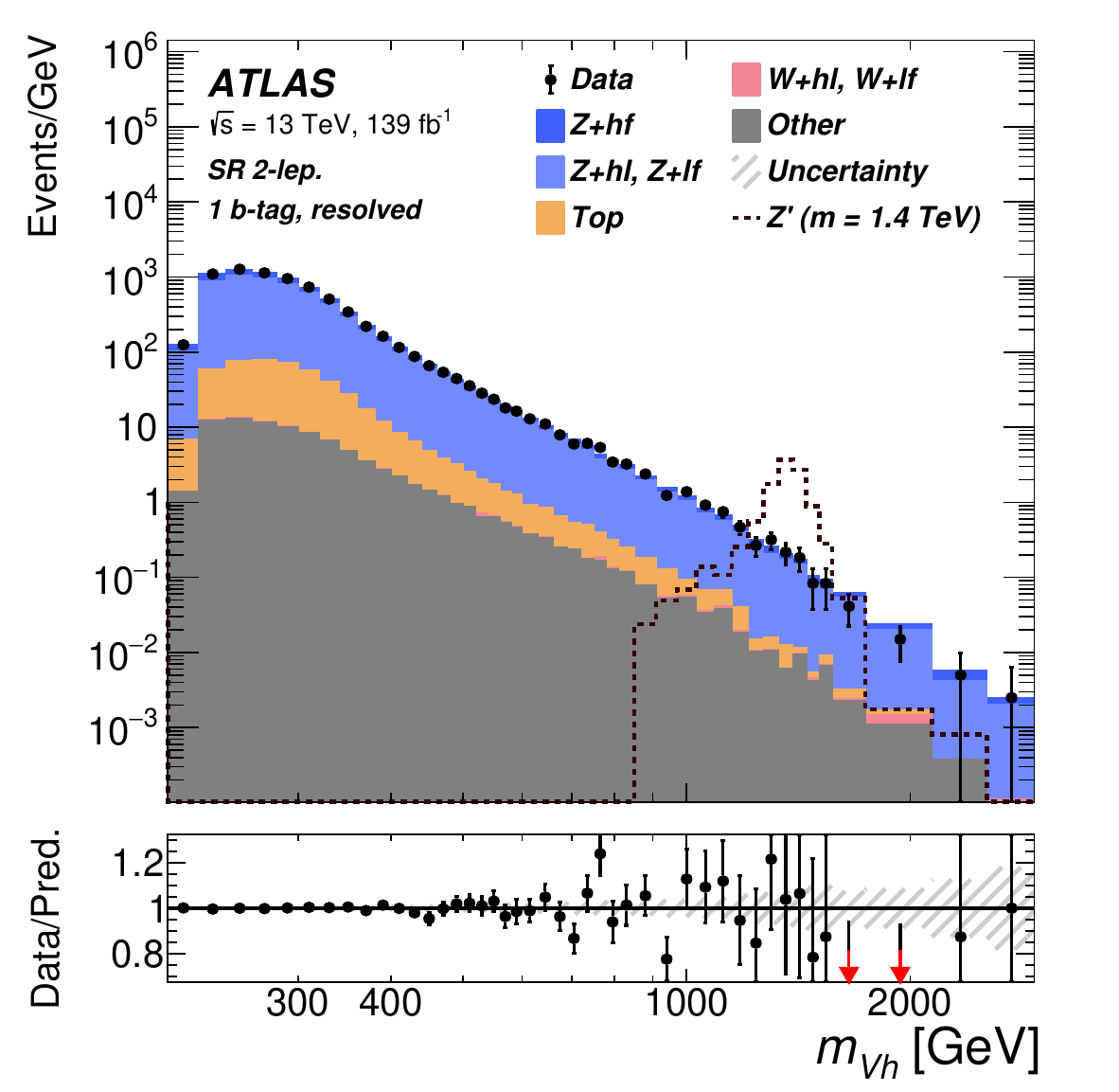}
}
\subfloat[]{
\includegraphics[width=0.50\textwidth]{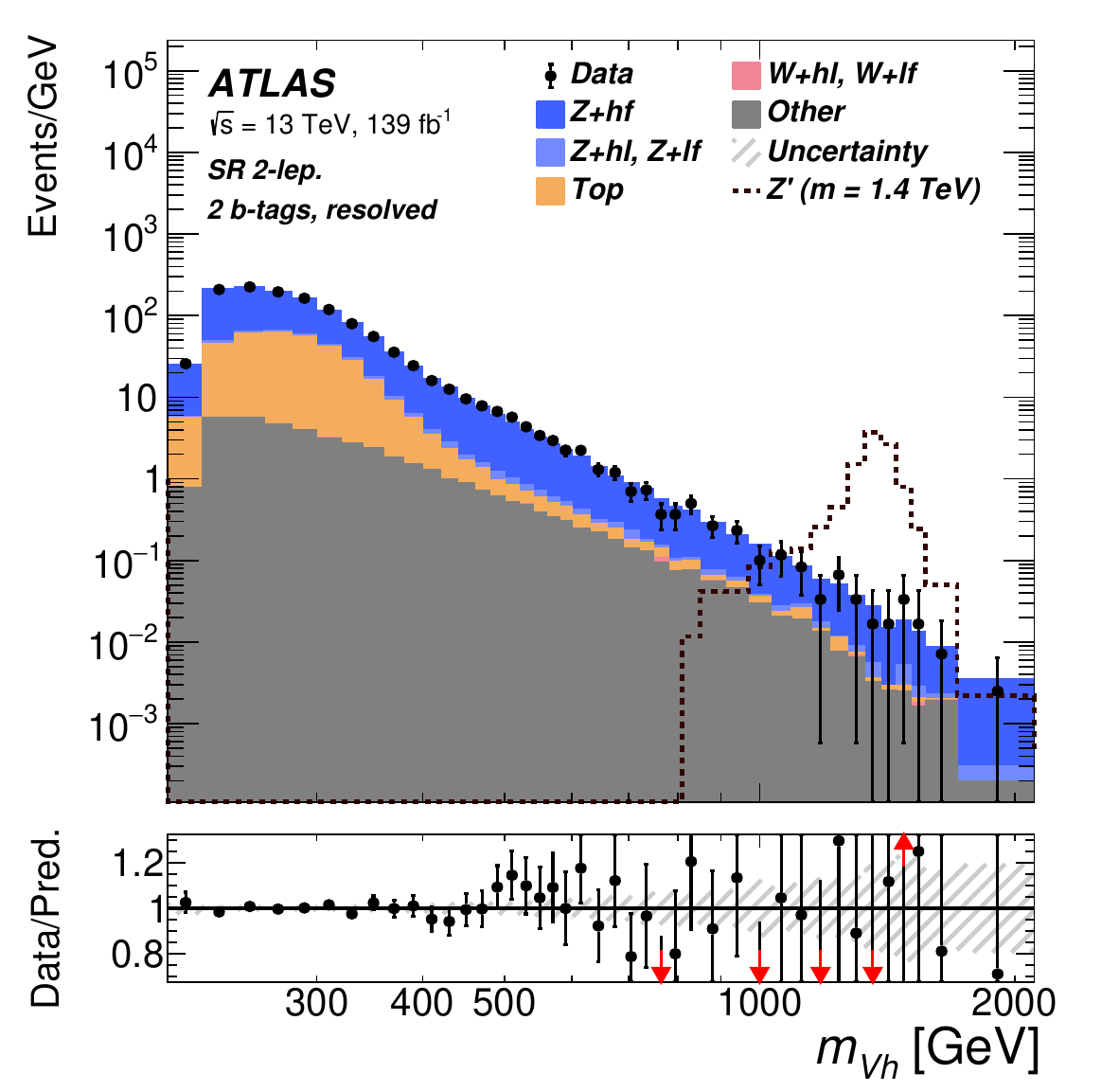}
}
\caption{
Event distributions of \mtvh\ for the 0-lepton SR (a, b) and of \mvh\ for the 2-lepton SR (c, d) in the resolved categories of the \Zprime\ fit.
The quantity on the vertical axis is the number of data events divided by the bin width in \GeV. The term `Top' summarises events from \ttbar, single-top-quark,
$\ttbar + h$  and $\ttbar + V$ contributions. In each plot, the last bin contains the overflow. The background prediction is shown after a background-only maximum-likelihood fit to the data.
The lower panels show the ratio of the observed to the estimated SM background. The background uncertainty band shows the post-fit statistical and systematic components added in quadrature.
The signal for the benchmark HVT \textit{Model A} with $m_{Z'} = 1.4 $~\TeV, normalised to $\sigma \times B(Zh) \times B(h\rightarrow b\bar{b}) = 0.1~\mathrm{pb}$, is shown as a dashed line.
} \label{fig:mVH_ZPpostfit_resolved}
\end{figure}
 
\begin{figure}[h!]
\centering
\subfloat[]{
\includegraphics[width=0.50\textwidth]{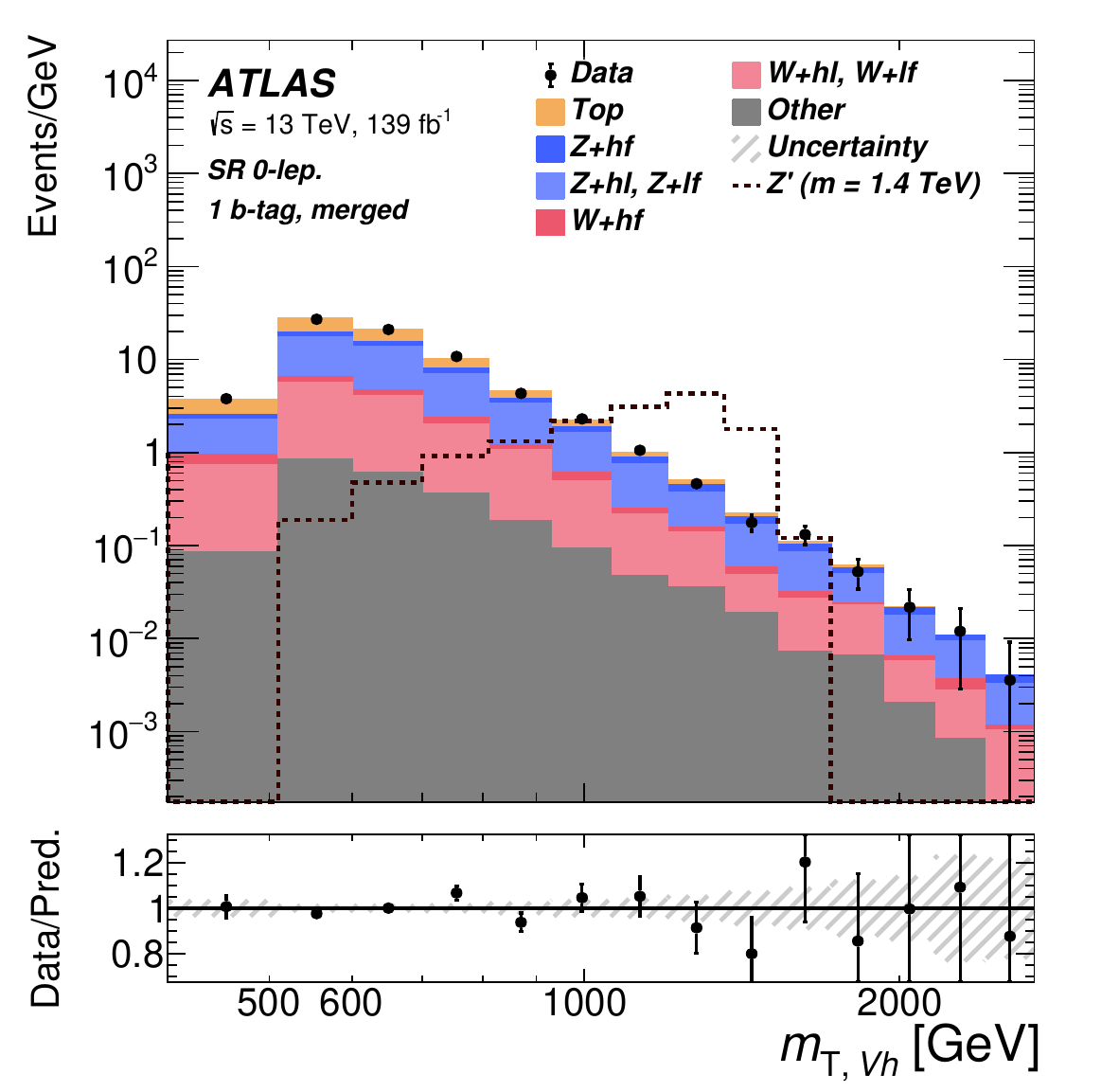}
}
\subfloat[]{
\includegraphics[width=0.50\textwidth]{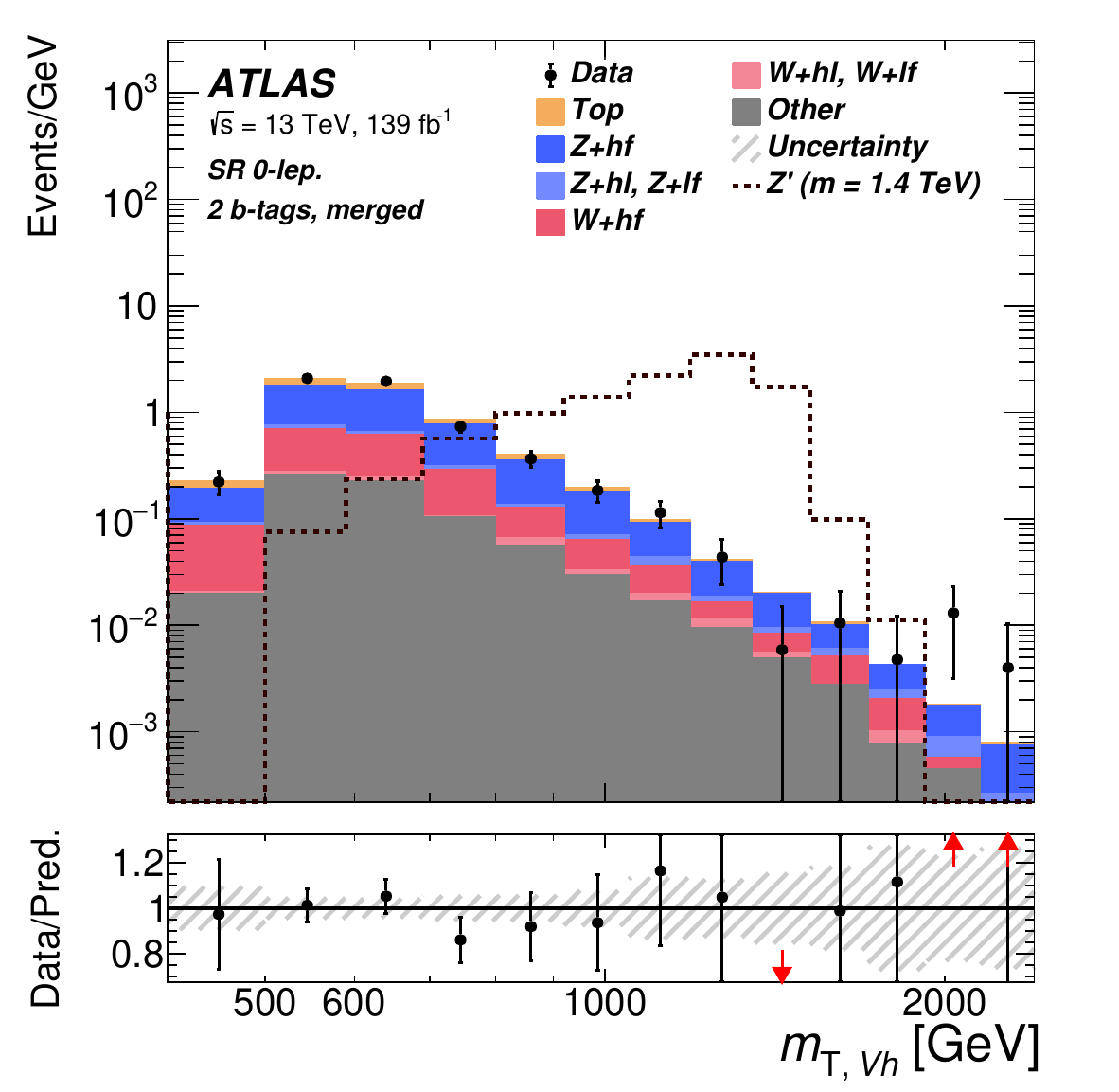}
}\\
\subfloat[]{
\includegraphics[width=0.50\textwidth]{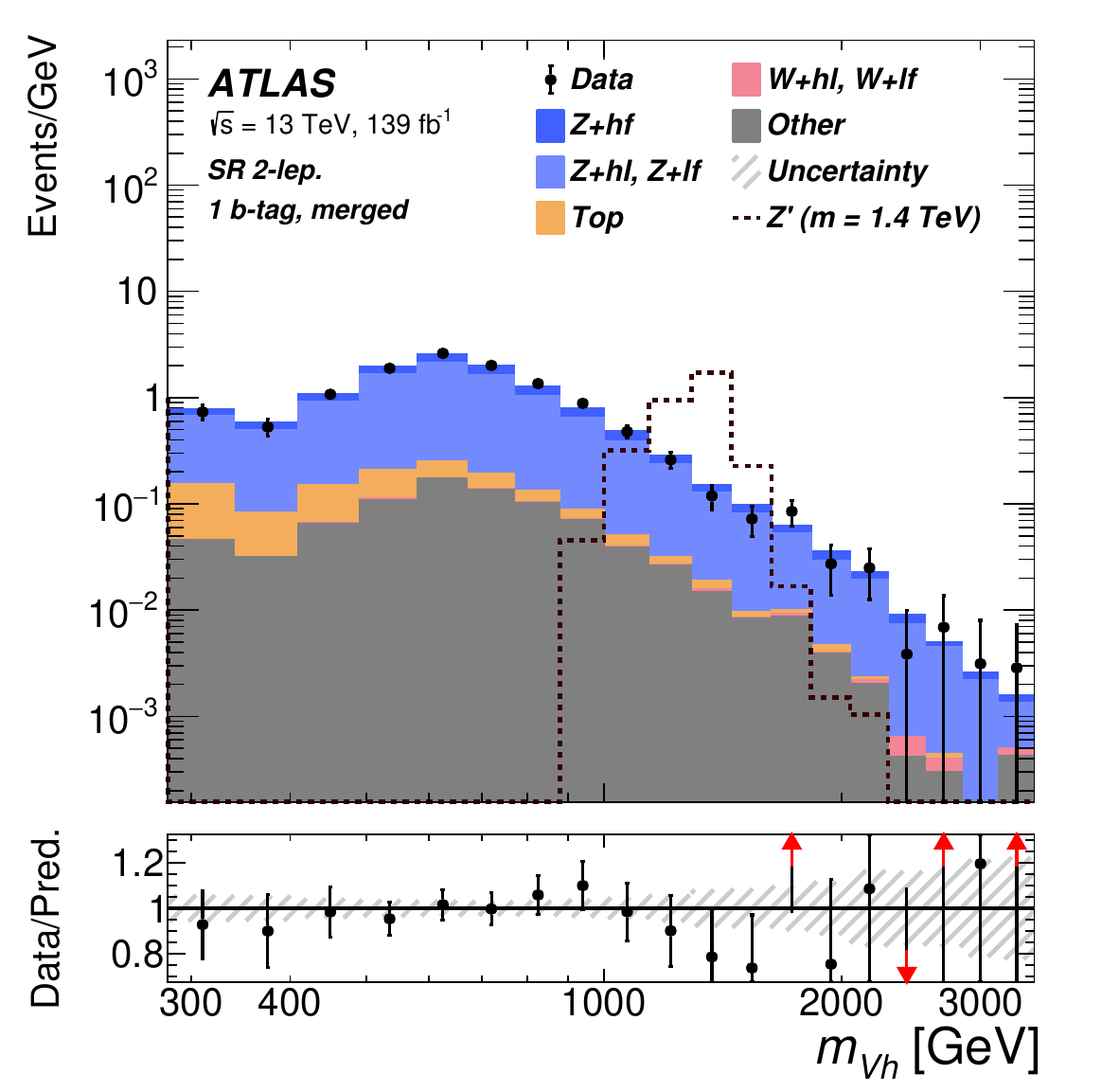}
}
\subfloat[]{
\includegraphics[width=0.50\textwidth]{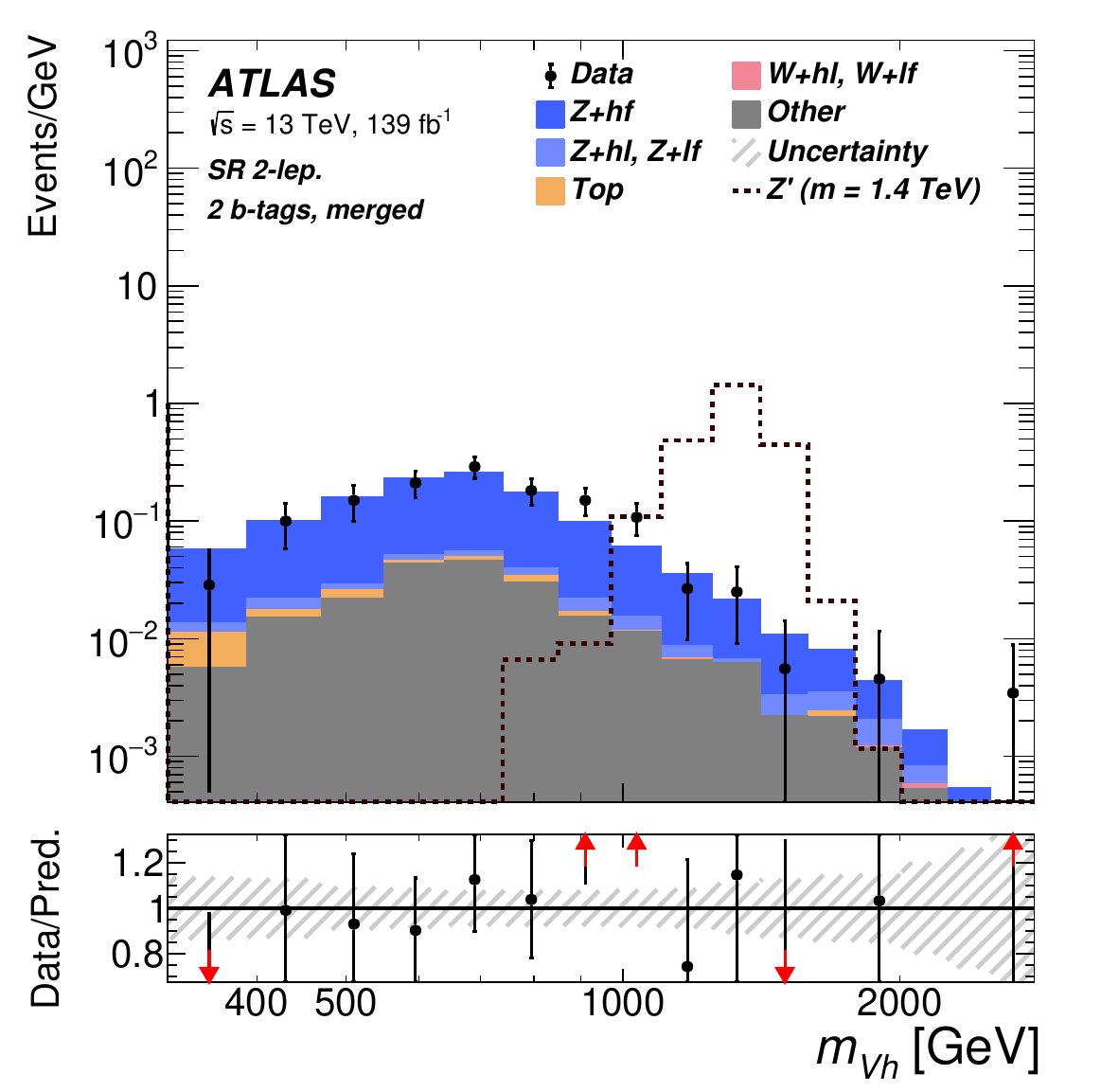}
}
\caption{
Event distributions of \mtvh\  for the 0-lepton channel (a, b) and of \mvh\ for the 2-lepton channels (c, d) in the merged categories of the \Zprime\ fit. The quantity on the vertical axis is the number of data events divided by the bin width in \GeV.
The term `Top' summarises events from \ttbar, single-top-quark, $\ttbar + h$  and $\ttbar + V$ contributions. In each plot, the last bin contains the overflow. The background prediction is shown after a background-only maximum-likelihood fit to the data.
The signal for the benchmark HVT \textit{Model A} with $m_{Z'} = 1.4 $~\TeV, normalised to $\sigma \times B(Zh)\times B(h\rightarrow b\bar{b}) = 0.1~\mathrm{pb}$,  is shown as a dashed line.
The lower panels show the ratio of the observed to the estimated SM background. The background uncertainty band shows the post-fit statistical and systematic components
added in quadrature.} \label{fig:mVH_ZPpostfit_merged}
\end{figure}
 
\begin{figure}[h!]
\centering
\subfloat[]{
\includegraphics[width=0.50\textwidth]{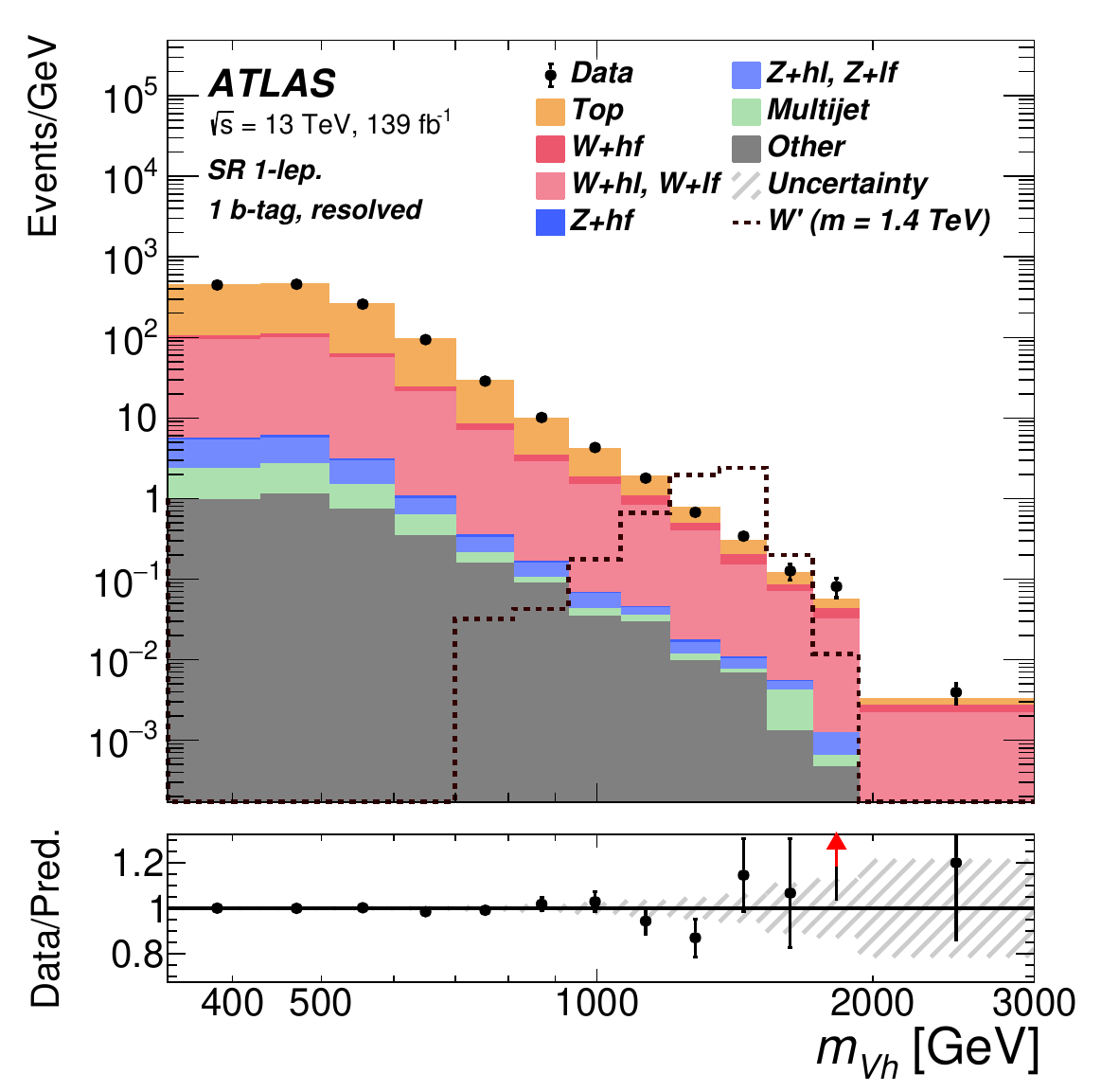}
}
\subfloat[]{
\includegraphics[width=0.50\textwidth]{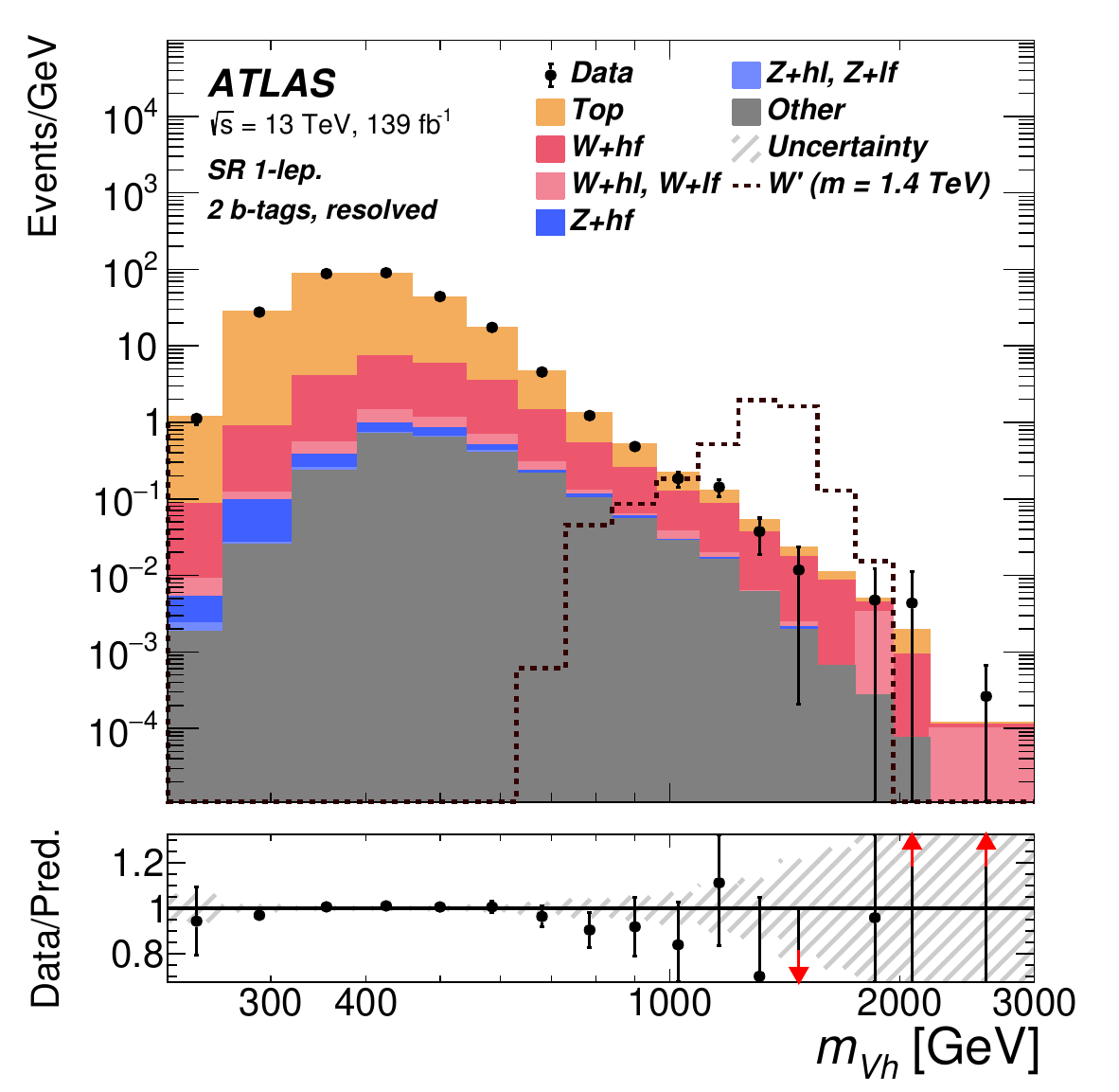}
}\\
\subfloat[]{
\includegraphics[width=0.50\textwidth]{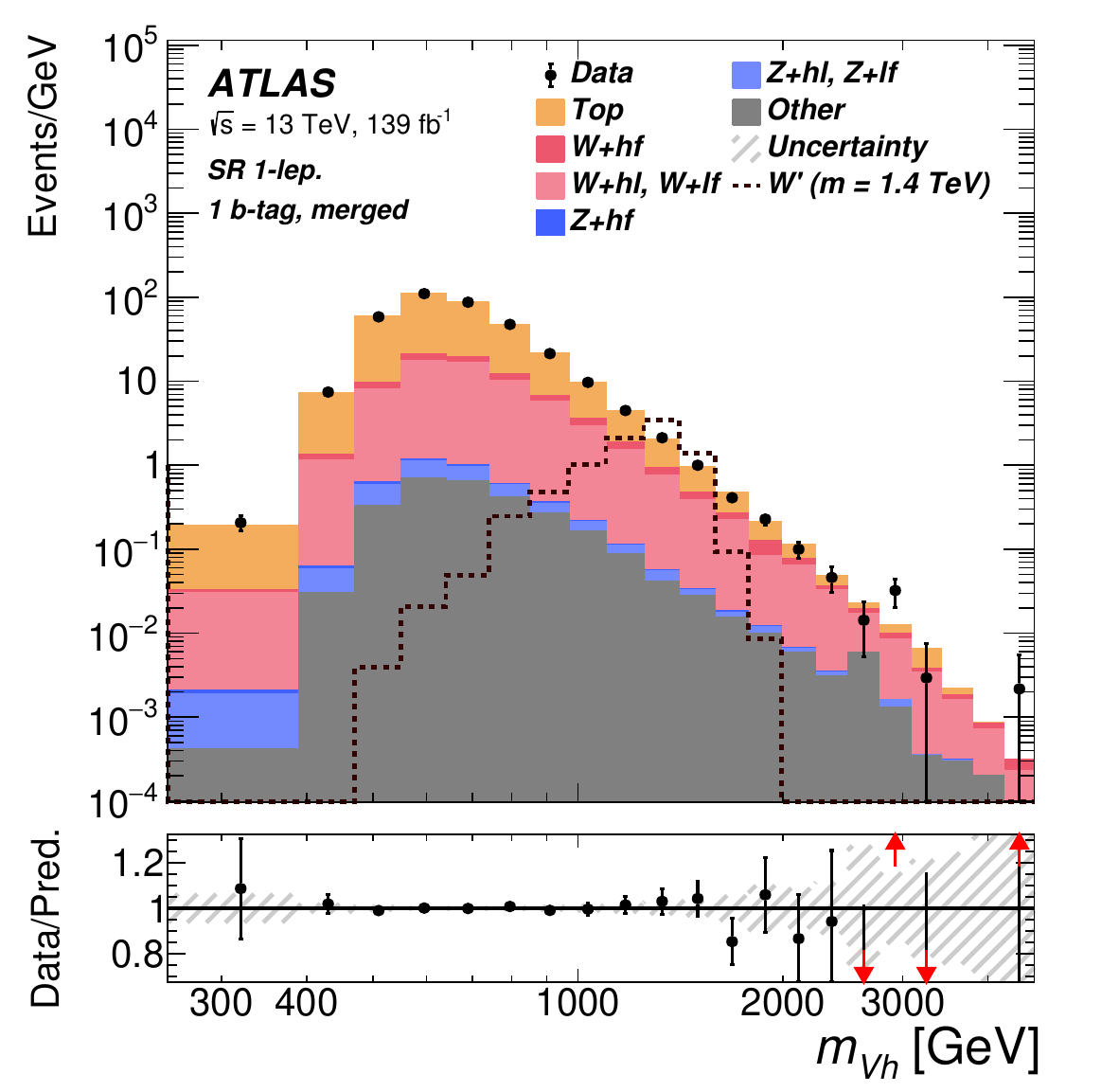}
}
\subfloat[]{
\includegraphics[width=0.50\textwidth]{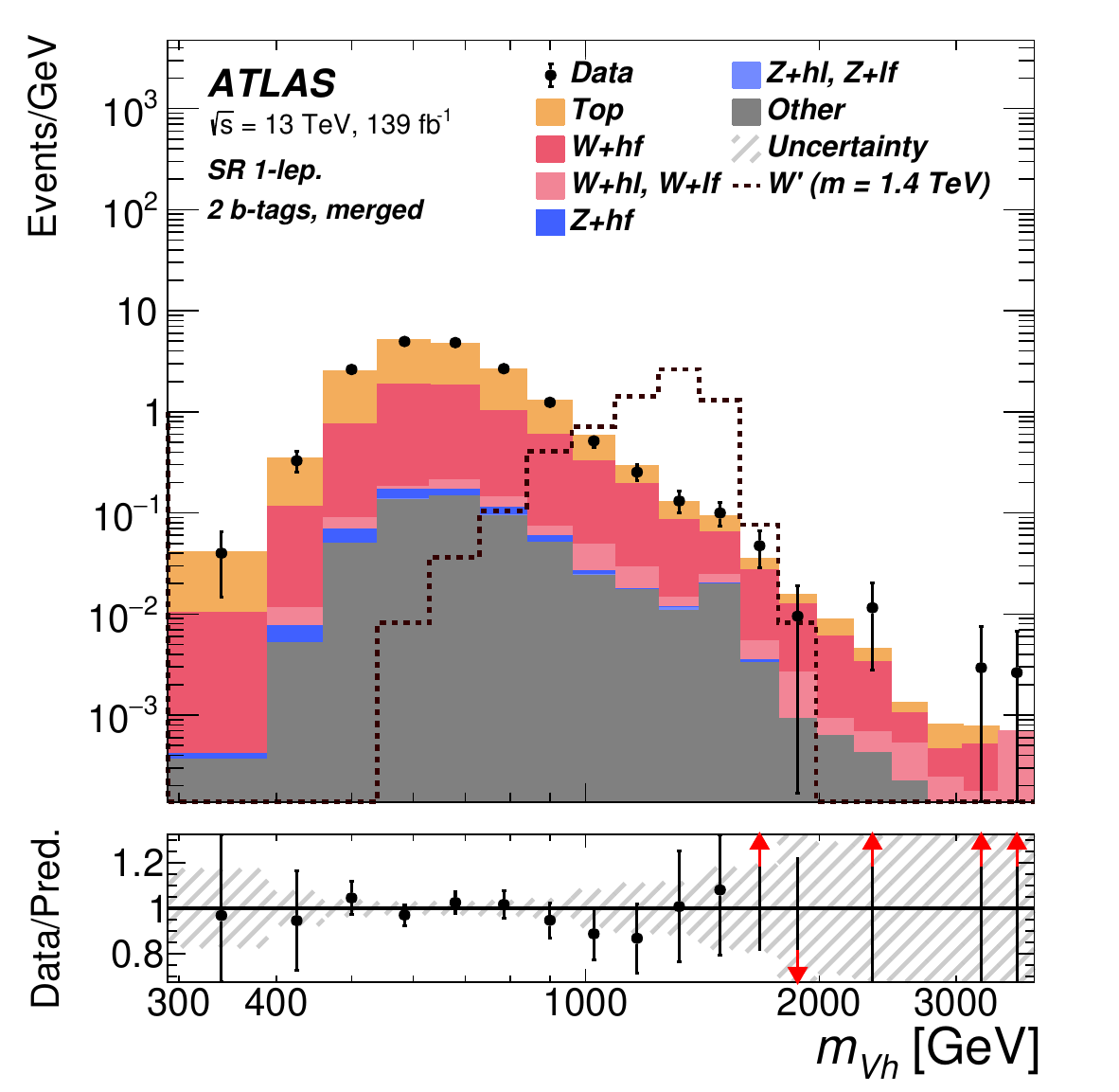}
}
\caption{
Event distributions of \mvh\   for the 1-lepton SR  in the resolved (a, b) and merged (c, d) categories of the \Wprime\ fit. The quantity on the vertical axis is the number of data events divided by the bin width in \GeV.
The term `Top' summarises events from \ttbar, single-top-quark, $\ttbar + h$  and $\ttbar + V$ contributions. In each plot, the last bin contains the overflow. The background prediction is shown after a background-only maximum-likelihood fit to the data.
The lower panels show the ratio of the observed to the estimated SM background. The background uncertainty band shows the post-fit statistical and systematic uncertainties added in quadrature.
The signal for the benchmark HVT \textit{Model A} with $m_{W'} = 1.4 $~\TeV\ is shown as a dashed line and normalised to a cross section times branching fraction of 0.1~pb.}
\label{fig:mVH_WPpostfit}
\end{figure}
\clearpage
 
\begin{figure}[h!]
\centering
\subfloat[]{
\includegraphics[width=0.50\textwidth]{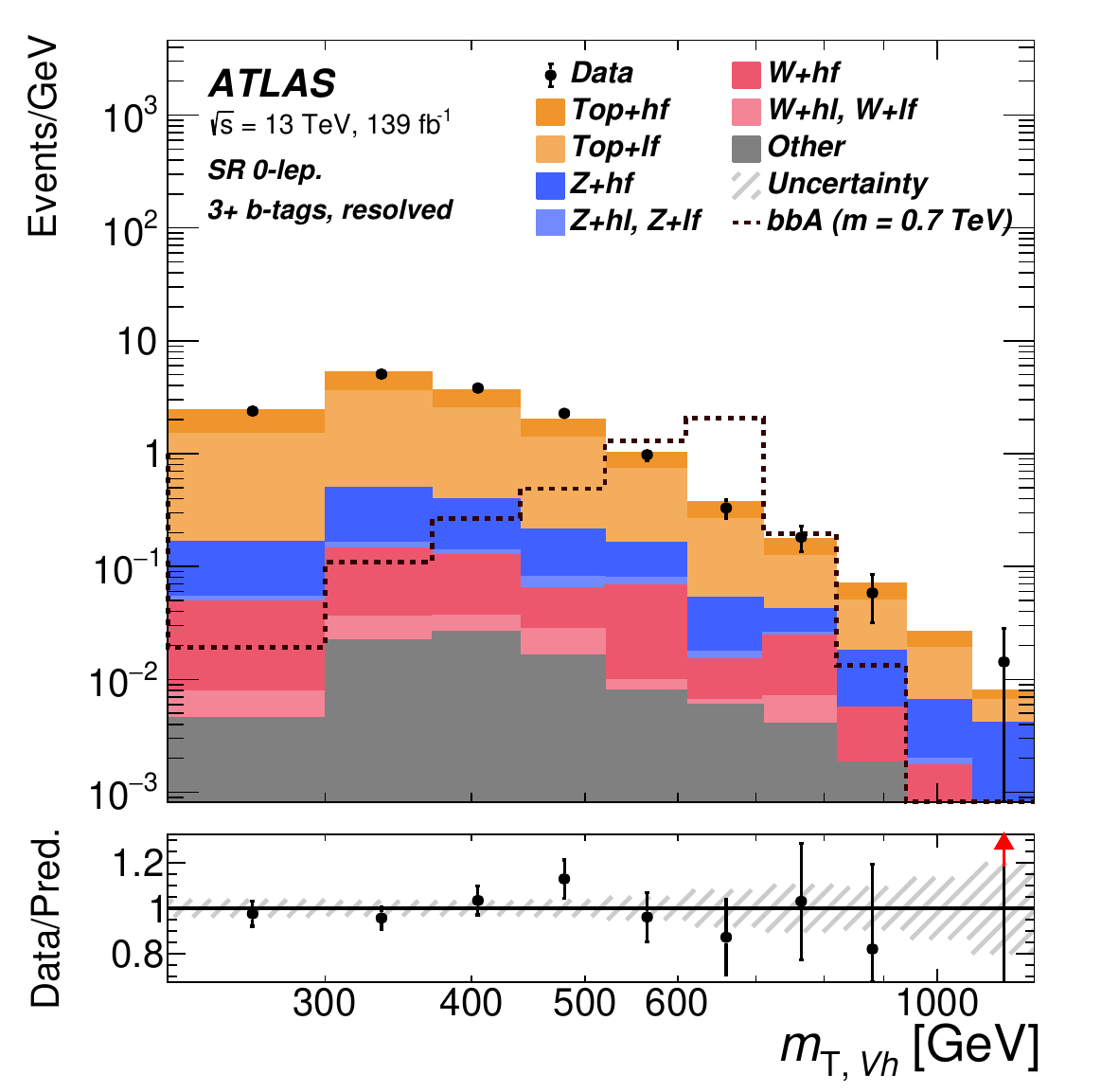}
}
\subfloat[]{
\includegraphics[width=0.50\textwidth]{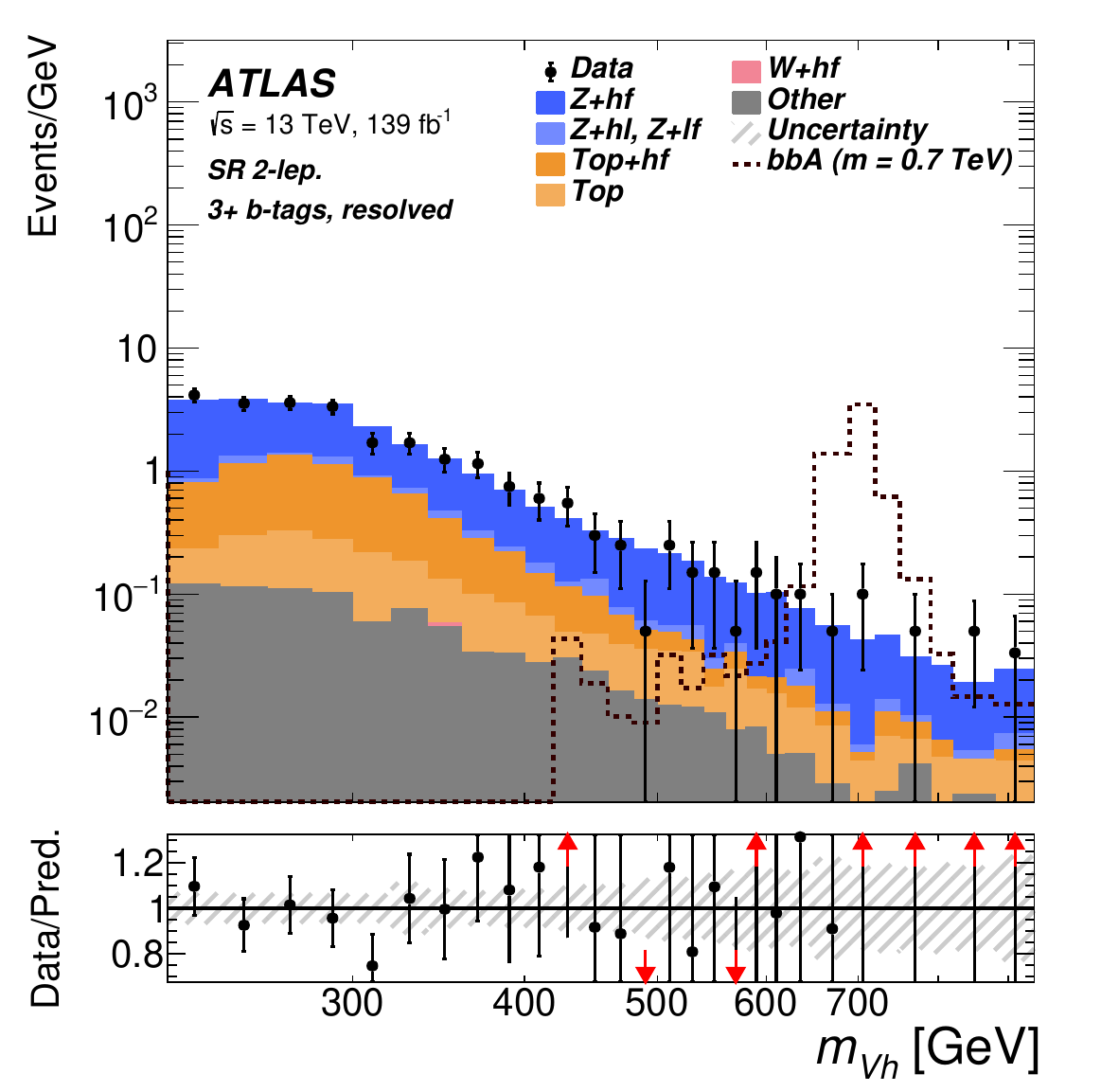}
}\\
\subfloat[]{
\includegraphics[width=0.50\textwidth]{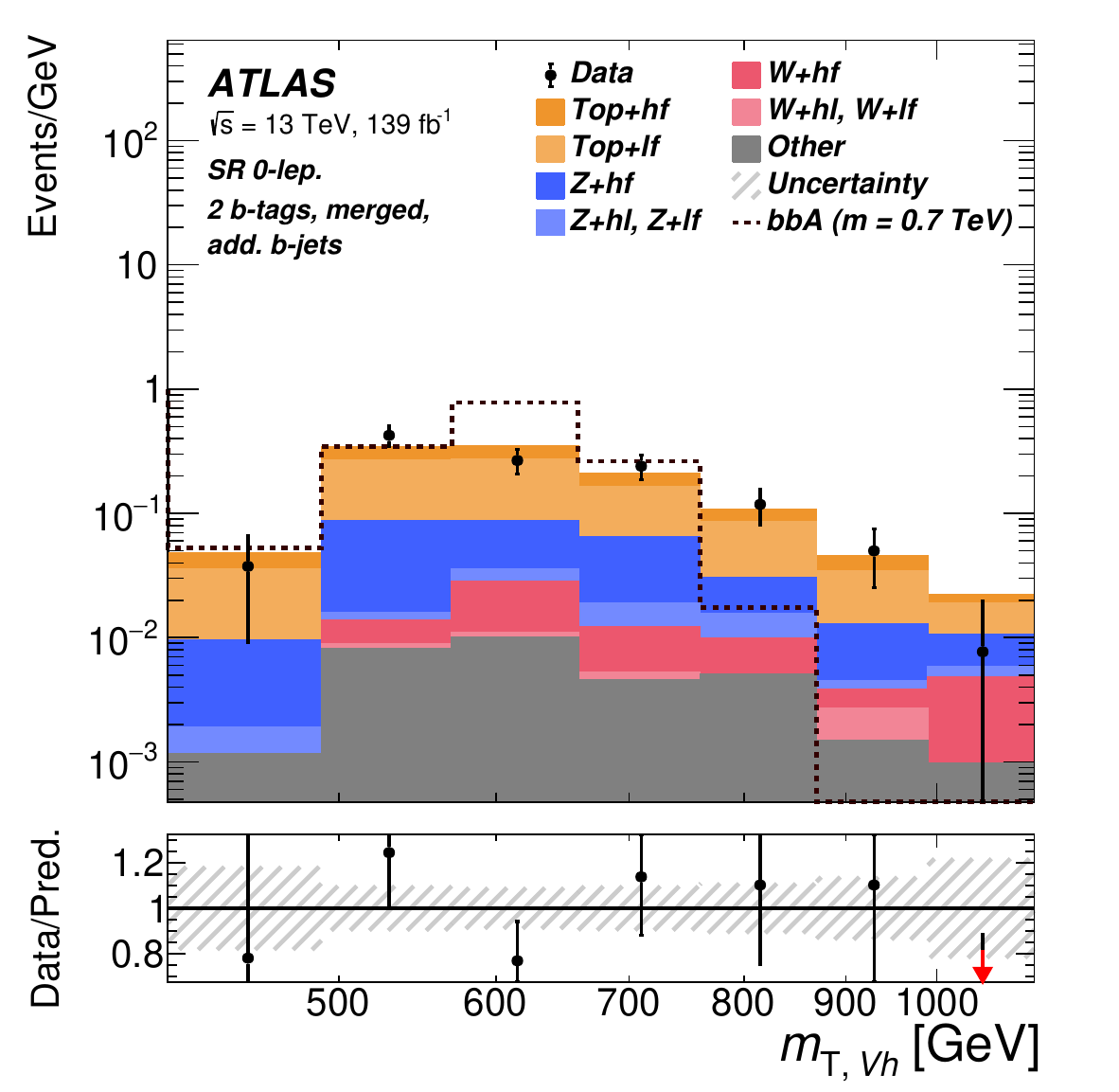}
}
\subfloat[]{
\includegraphics[width=0.50\textwidth]{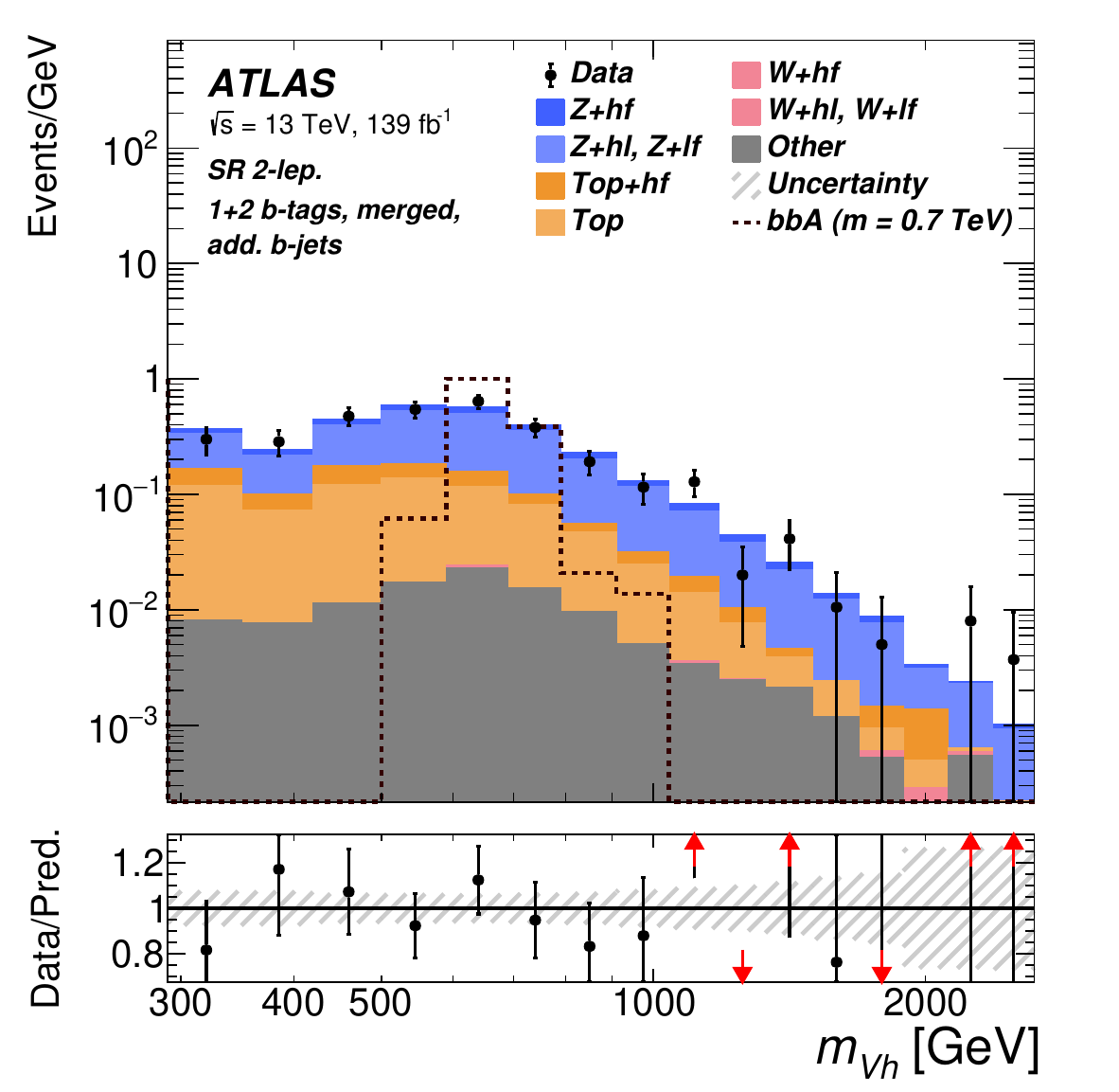}
}
\caption{
Event distributions of \mtvh\  for the 0-lepton channel (a, c) and of \mvh\ for the 2-lepton channels (b, d) in the merged and resolved 3+~tag SR categories of the \bbA\ fit.
The term `Top' summarises events from \ttbar, single-top-quark, $\ttbar + h$  and $\ttbar + V$ contributions.
The quantity on the vertical axis is the number of data events divided by the bin width in \GeV. In each plot, the last bin contains the overflow. The background prediction is shown after a background-only  maximum-likelihood fit to the data.
The signal for the benchmark 2HDM model with $m_{A} = 0.7 $~\TeV, normalised to $\sigma \times B(Zh) \times B(h\rightarrow b\bar{b}) = 0.1~\mathrm{pb}$,
is shown as a dashed line. The lower panels show the ratio of the observed to the estimated SM background. The background uncertainty
band shows the post-fit statistical and systematic components added in quadrature.}
\label{fig:mVH_bbApostfit1}
\end{figure}
\clearpage
 
\begin{figure}[h!]
\centering
\subfloat[]{\includegraphics[width=0.905\textwidth]{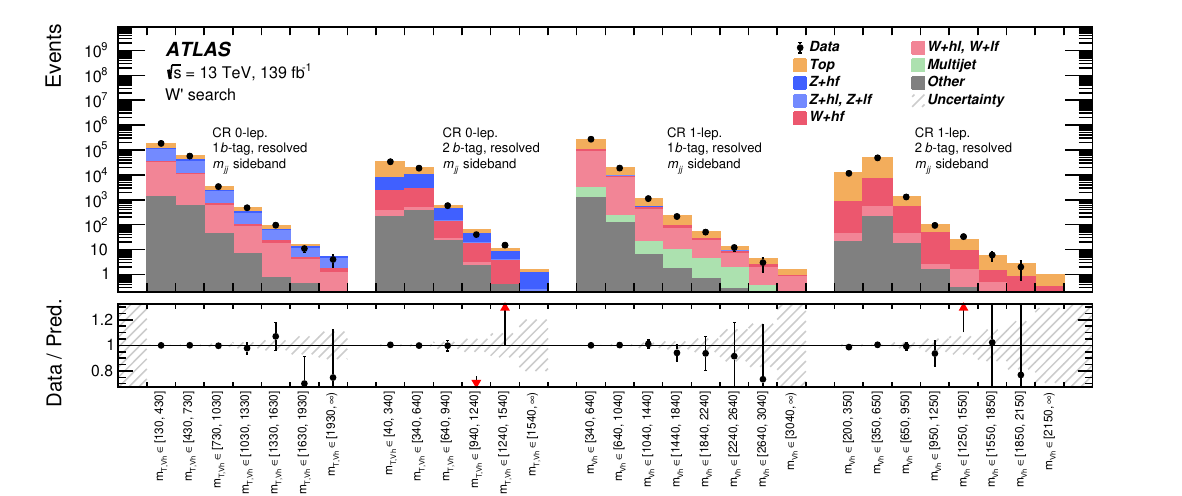}}\\
\subfloat[]{\includegraphics[width=0.905\textwidth]{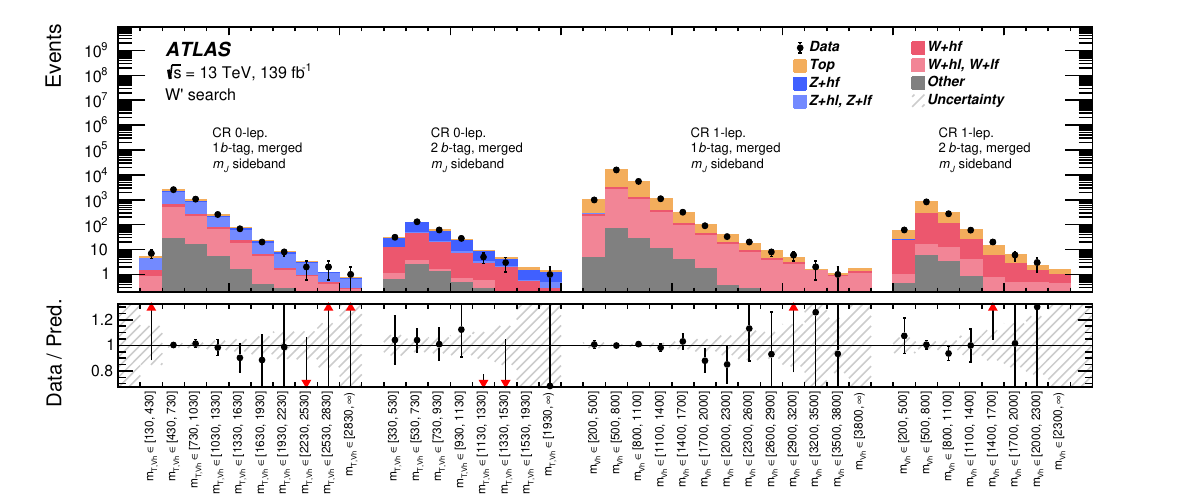}}\\
\subfloat[]{\includegraphics[width=0.905\textwidth]{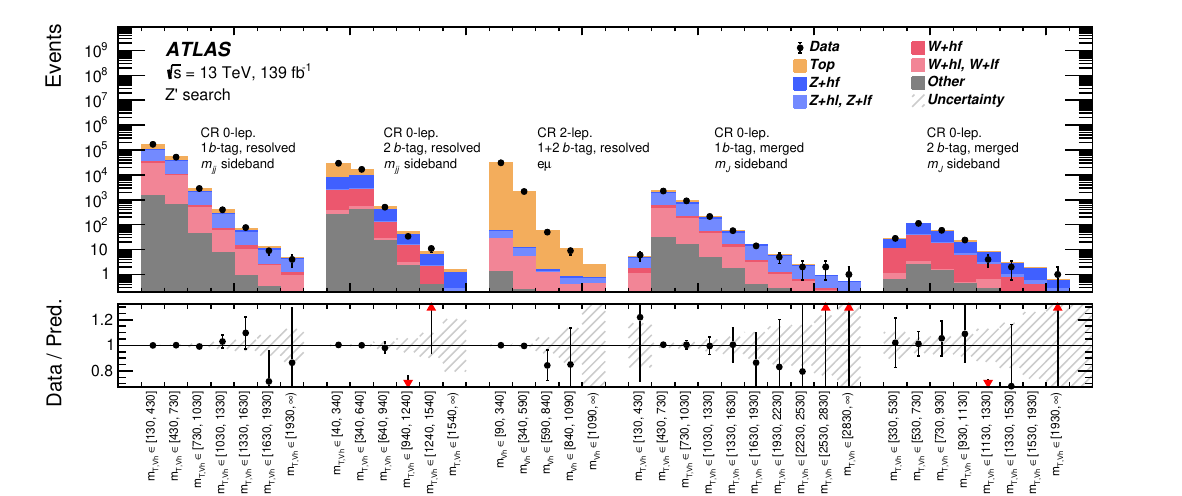}}
\caption{Distributions of the \mvh\ and \mtvh\ observables in the control regions of the (a) resolved and (b) merged event categories
used for the \Wprime\ search as well as (c) the control regions of the resolved and merged event categories used in the \Zprime\ search.
The term `Top' summarises events from \ttbar, single-top-quark, $\ttbar + h$  and $\ttbar + V$ contributions.
The distributions are presented after a background-only maximum-likelihood fit to data. The lower panels show the ratio of the observed to the estimated SM background.
The background uncertainty band shows the post-fit statistical and systematic uncertainties added in quadrature.}
\label{fig:CR_Summary_Vprime}
\end{figure}
\clearpage

% End of text imported from the .//texfiles/resultFiguresPostfit.tex input file

% The next lines are included from the .//texfiles/fitYieldsTable.tex input file
{
\def\arraystretch{1.15}
\begin{table}[h!]
\caption{\label{tab:postFitYields}
The observed event yields and background predictions in the various signal regions after fitting a signal and the various
background templates to the data. The yields in the 0- and 2-lepton channels correspond to the \bbA\ fit for a signal of mass 420~\GeV.  The yields in the 1-lepton channel are derived from the $W'$ fit for a signal
of mass $1400~\GeV$. The term Top summarises events from \ttbar, single-top-quark, \ttbar + $h$  and \ttbar +$V$ contributions. The quoted uncertainties are the
statistical and systematic uncertainties combined in quadrature after the fit. The uncertainties in the individual background predictions are larger than the total
background uncertainty due to correlations between the normalisation parameters in the fit.
}
 
\begin{center}
\begin{footnotesize}
\begin{tabular}{l| r @{$\pm$} l  r @{$\pm$} l r @{$\pm$} l |  r @{$\pm$} l  r @{$\pm$} l r @{$\pm$} l }
\hline\hline
& \multicolumn{6}{c|}{ Resolved }  & \multicolumn{6}{c}{ Merged }  \\
0-lepton  & \multicolumn{2}{c}{\textit{1 $b$-tag}} & \multicolumn{2}{c}{\textit{2 $b$-tags}} & \multicolumn{2}{c|}{\textit{3+ $b$-tags}} & \multicolumn{2}{c}{\textit{1 $b$-tag}} & \multicolumn{2}{c}{\textit{2 $b$-tags}} & \multicolumn{2}{c}{\textit{add. $b$-tags}} \\
\hline
Top    & 40400\phantom{0} & \phantom{0}600 & 10200\phantom{0} & \phantom{0}200 & 1140\phantom{0} & \phantom{0}31 & 2000\phantom{0} & \phantom{0}100 & 71.9\phantom{0} & \phantom{0}3.4 & 77\phantom{0} & \phantom{0}5 \\
\Zhf    & 3700\phantom{0} & \phantom{0}100 & 3000\phantom{0} & \phantom{0}100 & 82.1\phantom{0} & \phantom{0}3.3 & 651\phantom{0} & \phantom{0}32 & 298\phantom{0} & \phantom{0}12 & 19.2\phantom{0} & \phantom{0}1.4 \\
\Zhl    & 20700\phantom{0} & \phantom{0}400 & 106\phantom{0} & \phantom{0}13 & 5.5\phantom{0} & \phantom{0}0.8 & 2300\phantom{0} & \phantom{0}100 & 12.9\phantom{0} & \phantom{0}0.8 & 2.5\phantom{0} & \phantom{0}0.2 \\
\Zl     & 1600\phantom{0} & \phantom{0}200 & 5.0\phantom{0} & \phantom{0}2.1 & 0.1\phantom{0} & \phantom{0}0.1 & 396\phantom{0} & \phantom{0}34 & 0.8\phantom{0} & \phantom{0}0.3 & \multicolumn{2}{c}{< 0.1} \\
\Whf    & 1140\phantom{0} & \phantom{0}80 & 860\phantom{0} & \phantom{0}60 & 30.9\phantom{0} & \phantom{0}2.1 & 243\phantom{0} & \phantom{0}24 & 106\phantom{0} & \phantom{0}11 & 3.9\phantom{0} & \phantom{0}0.6 \\
\Whl    & 10000\phantom{0} & \phantom{0}400 & 90\phantom{0} & \phantom{0}10 & 3.6\phantom{0} & \phantom{0}0.5 & 1100\phantom{0} & \phantom{0}100 & 5.8\phantom{0} & \phantom{0}0.6 & 0.4\phantom{0} & \phantom{0}0.1 \\
\Wl        & 1915\phantom{0} & \phantom{0}270 & 13\phantom{0} & \phantom{0}4.4 & \multicolumn{2}{c|}{< 0.1} & 234\phantom{0} & \phantom{0}31 & 0.5\phantom{0} & \phantom{0}0.2 & \multicolumn{2}{c}{< 0.1} \\
Diboson    & 470\phantom{0} & \phantom{0}60   & 68\phantom{0} & \phantom{0}8 & 3.8\phantom{0} & \phantom{0}0.4 & 280\phantom{0} & \phantom{0}40 & 78\phantom{0} & \phantom{0}11 & 2.6\phantom{0} & \phantom{0}0.3 \\
SM $Vh$    & 223\phantom{0} & \phantom{0}13   & 292\phantom{0} & \phantom{0}16 & 3.7\phantom{0} & \phantom{0}0.2 & 24.2\phantom{0} & \phantom{0}1.7 & 15.2\phantom{0} & \phantom{0}1.0 & 0.5\phantom{0} & \phantom{0}0.1 \\
\hline
Total    & 80100\phantom{0} & \phantom{0}220 & 14650\phantom{0} & \phantom{0}90 & 1266\phantom{0} & \phantom{0}21 & 7260\phantom{0} & \phantom{0}60 & 587\phantom{0} & \phantom{0}13 & 106\phantom{0} & \phantom{0}5 \\
Data    & \multicolumn{2}{c}{80110} & \multicolumn{2}{c}{14681} & \multicolumn{2}{c|}{1265} & \multicolumn{2}{c}{7260} & \multicolumn{2}{c}{584} & \multicolumn{2}{c}{105} \\
\hline\hline
2-lepton  & \multicolumn{2}{c}{\textit{1 $b$-tag}} & \multicolumn{2}{c}{\textit{2 $b$-tags}} & \multicolumn{2}{c|}{\textit{3+ $b$-tags}} & \multicolumn{2}{c}{\textit{1 $b$-tag}} & \multicolumn{2}{c}{\textit{2 $b$-tags}} & \multicolumn{2}{c}{\textit{add. $b$-tags}} \\
\hline
Top        & 8100\phantom{0} & \phantom{0}200 & 6000\phantom{0} & \phantom{0}200 & 135\phantom{0} & \phantom{0}4 & 44.8\phantom{0} & \phantom{0}3.0 & 2.8\phantom{0} & \phantom{0}0.2 & 79.4\phantom{0} & \phantom{0}4.8 \\
\Zhf       & 23200\phantom{0} & \phantom{0}500 & 17100\phantom{0} & \phantom{0}100 & 327\phantom{0} & \phantom{0}9 & 208\phantom{0} & \phantom{0}7 & 98\phantom{0} & \phantom{0}4 & 33.2\phantom{0} & \phantom{0}2.2 \\
\Zhl       & 102500\phantom{0} & \phantom{0}1100 & 450\phantom{0} & \phantom{0}40 & 16.9\phantom{0} & \phantom{0}1.9 & 709\phantom{0} & \phantom{0}15 & 4.6\phantom{0} & \phantom{0}0.2 & 157\phantom{0} & \phantom{0}7 \\
\Zl        & 6800\phantom{0} & \phantom{0}1000 & 29\phantom{0} & \phantom{0}11 & \multicolumn{2}{c|}{< 0.1} & 154\phantom{0} & \phantom{0}11 & 0.3\phantom{0} & \phantom{0}0.1 & 18.1\phantom{0} & \phantom{0}1.7 \\
\Whf       & 4.9\phantom{0} & \phantom{0}1.2 & 1.5\phantom{0} & \phantom{0}0.1 & \multicolumn{2}{c|}{< 0.1}  & 0.2\phantom{0} & \phantom{0}0.0 & \multicolumn{2}{c}{< 0.1} & \multicolumn{2}{c}{< 0.1} \\
\Whl       & 25\phantom{0} & \phantom{0}4 & 0.2\phantom{0} & \phantom{0}0.1 & \multicolumn{2}{c|}{< 0.1} & 1.3\phantom{0} & \phantom{0}0.0 & \multicolumn{2}{c}{< 0.1} &  \multicolumn{2}{c}{< 0.1} \\
\Wl        & 0.7\phantom{0} & \phantom{0}0.1 & \multicolumn{2}{c}{< 0.1} & \multicolumn{2}{c|}{< 0.1} & 0.3\phantom{0} & \phantom{0}0.0  & \multicolumn{2}{c}{< 0.1} & \multicolumn{2}{c}{< 0.1} \\
Diboson    & 1660\phantom{0} & \phantom{0}80 & 480\phantom{0} & \phantom{0}20 & 13.1\phantom{0} & \phantom{0}0.6 & 85.5\phantom{0} & \phantom{0}3.0 & 20.5\phantom{0} & \phantom{0}0.9 & 10.1\phantom{0} & \phantom{0}0.4 \\
SM $Vh$    & 333\phantom{0} & \phantom{0}13 & 402\phantom{0} & \phantom{0}11 & 5.1\phantom{0} & \phantom{0}0.2 & 5.1\phantom{0} & \phantom{0}0.2 & 2.6\phantom{0} & \phantom{0}0.1 & 0.9\phantom{0} & \phantom{0}0.1 \\
\hline
Total    & 142700\phantom{0} & \phantom{0}300 & 24400\phantom{0} & \phantom{0}100 & 496\phantom{0} & \phantom{0}11 & 1208\phantom{0} & \phantom{0}16 & 129\phantom{0} & \phantom{0}4 & 298\phantom{0} & \phantom{0}7 \\
Data    & \multicolumn{2}{c}{142672} & \multicolumn{2}{c}{24371} & \multicolumn{2}{c|}{505} & \multicolumn{2}{c}{1220} & \multicolumn{2}{c}{133} & \multicolumn{2}{c}{311} \\
\hline\hline
1-lepton  & \multicolumn{2}{c}{\textit{1 $b$-tag}}   & \multicolumn{2}{c}{\textit{2 $b$-tags}}  & \multicolumn{2}{c|}{\textit{3+ $b$-tags}} & \multicolumn{2}{c}{\textit{1 $b$-tag}}  &\multicolumn{2}{c}{\textit{2 $b$-tags}} &\multicolumn{2}{c}{\textit{add. $b$-tags}}  \\
\hline
Top      & 86540\phantom{0} & \phantom{0}470 & 18174\phantom{0} & \phantom{0}88  & \multicolumn{2}{c|}{--} & 26070\phantom{0}& \phantom{0}200 & 1073\phantom{0} & \phantom{0}23 & \multicolumn{2}{c}{--}  \\
\Zl      & 108\phantom{0}   & \phantom{0}20  & 0.4\phantom{0}   & \phantom{0}0.1 & \multicolumn{2}{c|}{--}  & 22.6\phantom{0} & \phantom{0}5.9 & \multicolumn{2}{c}{< 0.1}  & \multicolumn{2}{c}{--}     \\
\Zhl     & 561\phantom{0}   & \phantom{0}83  & 5.2\phantom{0}   & \phantom{0}0.9 & \multicolumn{2}{c|}{--}  & 111\phantom{0}  & \phantom{0}14  & 0.9\phantom{0}  & \phantom{0}0.1 & \multicolumn{2}{c}{--}   \\
\Zhf     & 84\phantom{0}    & \phantom{0}20  & 59\phantom{0}    & \phantom{0}12  & \multicolumn{2}{c|}{--}  & 20.9\phantom{0} & \phantom{0}4.5 & 10.9\phantom{0} & \phantom{0}2.4 & \multicolumn{2}{c}{--}  \\
\Wl      & 2080\phantom{0}  & \phantom{0}230 & 0.7\phantom{0}   & \phantom{0}0.3 & \multicolumn{2}{c|}{--}  & 1011\phantom{0} & \phantom{0}81  & 1.8\phantom{0}  & \phantom{0}0.5 & \multicolumn{2}{c}{--}   \\
\Whl     &  22330\phantom{0}& \phantom{0}530 & 92.8\phantom{0}  & \phantom{0}6.6 & \multicolumn{2}{c|}{--}  & 5670\phantom{0} & \phantom{0}180 & 21.1\phantom{0} & \phantom{0}1.7 & \multicolumn{2}{c}{--}   \\
\Whf     & 2910\phantom{0}  & \phantom{0}110 & 1457\phantom{0}  & \phantom{0}54  & \multicolumn{2}{c|}{--}  & 1219\phantom{0} & \phantom{0}38  & 60\phantom{0}   & \phantom{0}20  & \multicolumn{2}{c}{--}   \\
Diboson  & 160\phantom{0}   & \phantom{0}15  & 16.3\phantom{0}  & \phantom{0}1.4 & \multicolumn{2}{c|}{--}  & 251\phantom{0}  & \phantom{0}19  & 32.9\phantom{0} & \phantom{0}2.7 & \multicolumn{2}{c}{--}   \\
SM \Vh   & 150\phantom{0}   & \phantom{0}14  & 190\phantom{0}   & \phantom{0}17  & \multicolumn{2}{c|}{--}  & 32.6\phantom{0} & \phantom{0}2.7 & 25.9\phantom{0} & \phantom{0}2.2 & \multicolumn{2}{c}{--}   \\
Multijet & 298\phantom{0}   & \phantom{0}58  & \multicolumn{2}{c}{< 0.1} & \multicolumn{2}{c|}{--} & \multicolumn{2}{c}{< 0.1} & \multicolumn{2}{c}{< 0.1} & \multicolumn{2}{c}{--}    \\
\hline
Total    & 115210\phantom{0}& \phantom{0}280 & 19995\phantom{0} & \phantom{0}80  & \multicolumn{2}{c|}{--}  & 34410\phantom{0} & \phantom{0}120 & 1769\phantom{0} & \phantom{0}27  & \multicolumn{2}{c}{--}   \\
Data     & \multicolumn{2}{c}{115145} & \multicolumn{2}{c}{20017} & \multicolumn{2}{c|}{--}  & \multicolumn{2}{c}{34403} & \multicolumn{2}{c}{1771} & \multicolumn{2}{c}{--}   \\
\hline\hline
\end{tabular}
 
\end{footnotesize}
\end{center}
\end{table}
}
\clearpage

% End of text imported from the .//texfiles/fitYieldsTable.tex input file

The largest deviation from the Standard Model expectations is found in the \Zprime\ and \ggA\ searches for a resonance mass of $500\;\GeV$ and corresponds to a
local significance of about 2.1 standard deviations (and a $p$-value, i.e.\ the probability that the background can produce a fluctuation greater than the excess
observed in data, of 0.017). The global significance of this excess corresponds to 1.1 standard deviations. The \bbA\ search displays a similar excess of 1.6
standard deviations ($p$-value of 0.059) around the same resonance mass. For all three fit models, the excess originates  mainly from the 2~$b$-tag category
of the 2-lepton channel, where small differences between the observations and SM predictions are found for $\mvh$ values around $500\;\GeV$. The local significance
in the \bbA\ search is lower than in the \Zprime and \ggA\ searches because the additional signal regions used in the search for the
\bbA\ process do not display any excess. Another excess is found in the \Zprime\ search at a resonance mass of $2.2\;\TeV$ and corresponds to a local
significance of about 2.0 standard deviations ($p$-value of 0.02). The \Wprime\ search finds mild excesses at resonance masses of $400\;\GeV$ (about 2.0 standard
deviations, $p$-value of 0.02) and $3.0\;\TeV$ (about 1.7 standard deviations, $p$-value of 0.05).
 
The 95\% CL upper limits on the production cross section for $Z^\prime\rightarrow Z\h$ are shown in Figure~\ref{fig:HVTlimits}(a) as a function of the resonance
mass. The observed limits range from $0.9~\pb$ for $m_{Z^{\prime}} = 300~\GeV$ to $0.3~\fb$ for $m_{Z^{\prime}} = 5~\TeV$. These limits exclude $Z^{\prime}$ masses below $2.8~\TeV$
for HVT benchmark \textit{Model A} with coupling constant $g_\text{V}=1$~\cite{Pappadopulo:2014qza}. For \textit{Model B} with coupling constant
$g_\text{V}=3$~\cite{Pappadopulo:2014qza}, $Z^{\prime}$ masses below $3.2~\TeV$ are excluded. The combined limits are dominated by the 0-lepton (2-lepton)
channel for resonance masses above (below) $800\;\GeV$. The limits on the $m_{Z^\prime} = 5~\TeV$ signal hypothesis were
also calculated using pseudo-experiments to validate the asymptotic approximation approach in a phase-space region strongly limited by a low number of data events.
It was found that the two statistical methods agree within $20\%$. \par
 
The dominant uncertainties for low resonance masses (i.e.\ $m_{Z^\prime}=500\;\GeV$) are due to the modelling of the parton shower in $t\bar{t}$ background events and
to the QCD scale uncertainties in the \Zhf\ background. At intermediate resonance masses (i.e.\ $m_{Z^\prime}=800\;\GeV$), the \largeR\ jet mass resolution
uncertainties become increasingly important and so do the uncertainties related to the matching of the matrix element to the parton shower for the various \Wjets\
background contributions. The latter uncertainties are also the dominant systematic uncertainties for $Z^{\prime}$ resonance masses above 2\;\TeV. The statistical
uncertainties of the data become dominant for resonance masses above $600\;\GeV$.
 
Figure~\ref{fig:HVTlimits}(b) shows the 95\% CL upper limits on the production cross section for $W^\prime\rightarrow W\h$ as a function of
the resonance mass. The limits range from $1.3~\pb$ for $m_{W^{\prime}} = 400~\GeV$ to $0.4~\fb$ for $m_{W^{\prime}} = 5~\TeV$. These limits exclude $W^{\prime}$
masses below  $2.95~\TeV$ for HVT benchmark \textit{Model A} with coupling constant $g_\text{V}=1$~\cite{Pappadopulo:2014qza}.
For \textit{Model B} with coupling constant $g_\text{V}=3$, $W^{\prime}$  masses below $3.3~\TeV$ are excluded. The combined limits are dominated
by the 1-lepton channel over the full range of tested mass hypotheses. The limits on the
$m_{W^\prime} = 5~\TeV$ signal hypothesis were also calculated using pseudo-experiments to validate the asymptotic approximation
approach in a phase-space region strongly limited by a low number of data events. It was found that the two statistical methods agree
within $10\%$. \par
 
Figure~\ref{fig:2HDMlimits}(a) shows the 95\% CL upper limits on the production cross section of \ggA\  multiplied by its branching
fraction to $Z\h$ as a function of the resonance mass. The observed limits range from $0.6~\pb$ for ${m_{A} = 220~\GeV}$ to $3~\fb$ for
$m_{A} = 2~\TeV$. Figure~\ref{fig:2HDMlimits}(b) shows the 95\% CL upper limits on the production cross section of \bbA\  multiplied by
its branching fraction to $Z\h$ as a function of the resonance mass. The observed limits range from $0.5~\pb$ for ${m_{A} = 220~\GeV}$ to
$2.5~\fb$ for $m_{A} = 2~\TeV$.

% The next lines are included from the .//texfiles/resultFiguresLimits.tex input file
\begin{figure*}[ht!]
\centering
\subfloat[]{
\includegraphics[width=0.475\textwidth]{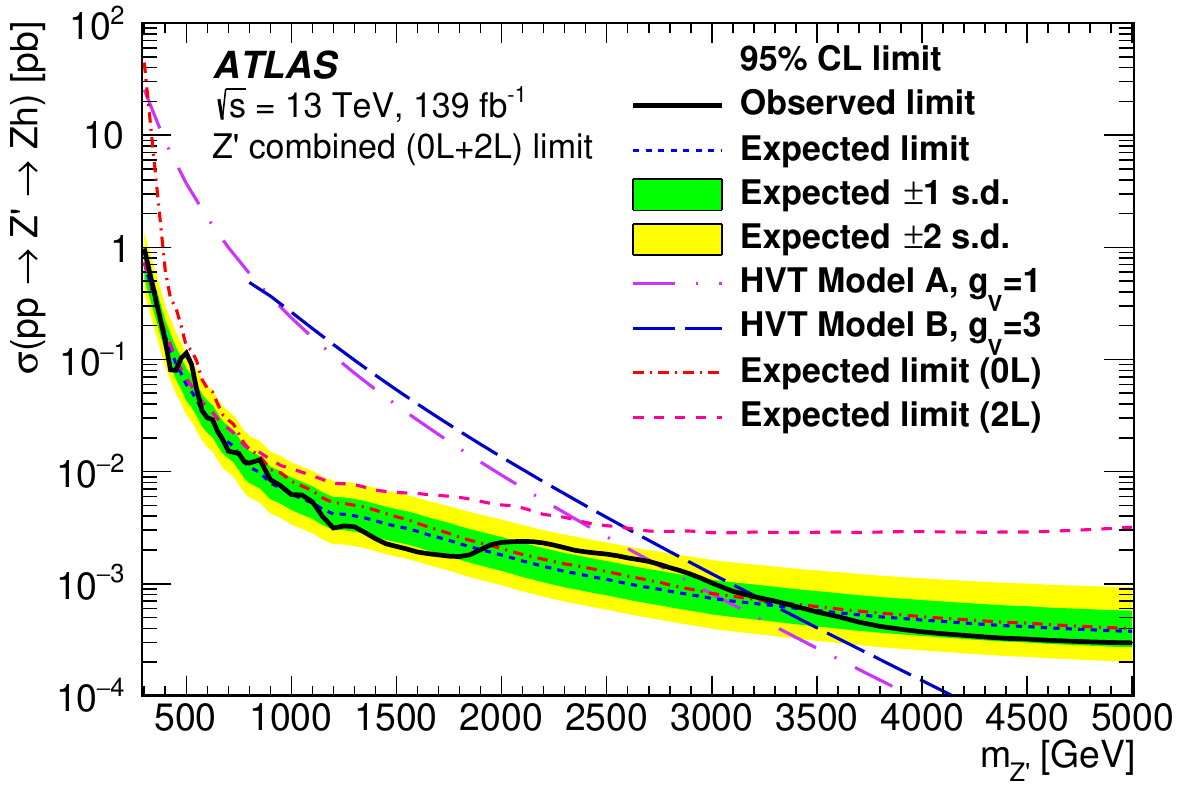}
}
\subfloat[]{
\includegraphics[width=0.475\textwidth]{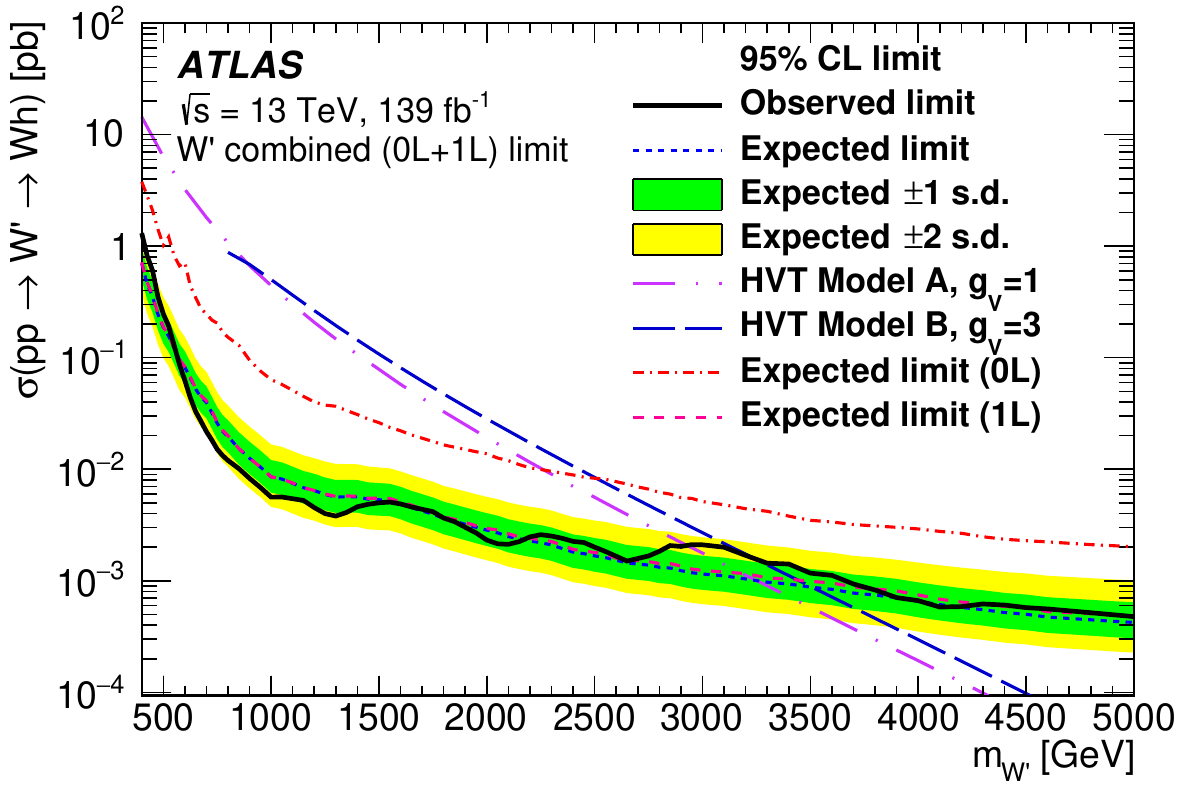}
}
\caption{
Upper limits at the 95\% CL on the product of the cross section for $\pp \to\ $\Zprime\ and the branching fraction to \Zh\ from the combination of the 0-lepton (0L)
and 2-lepton (2L) channels (a) and on the product of the cross section for $\pp \to\ $\Wprime\ and the branching fraction to \Wh\ from the combination of the
0-lepton and 1-lepton (1L) channels (b). For the $Z^\prime$ and $W^\prime$ search, the branching fraction of $h\rightarrow b\bar{b}, c\bar{c}$ is assumed to be 0.598.}
\label{fig:HVTlimits}
\end{figure*}

% End of text imported from the .//texfiles/resultFiguresLimits.tex input file

% The next lines are included from the .//texfiles/resultFiguresLimitsPScalar.tex input file
\begin{figure*}[ht!]
\centering
\subfloat[]{
\includegraphics[width=0.5\textwidth]{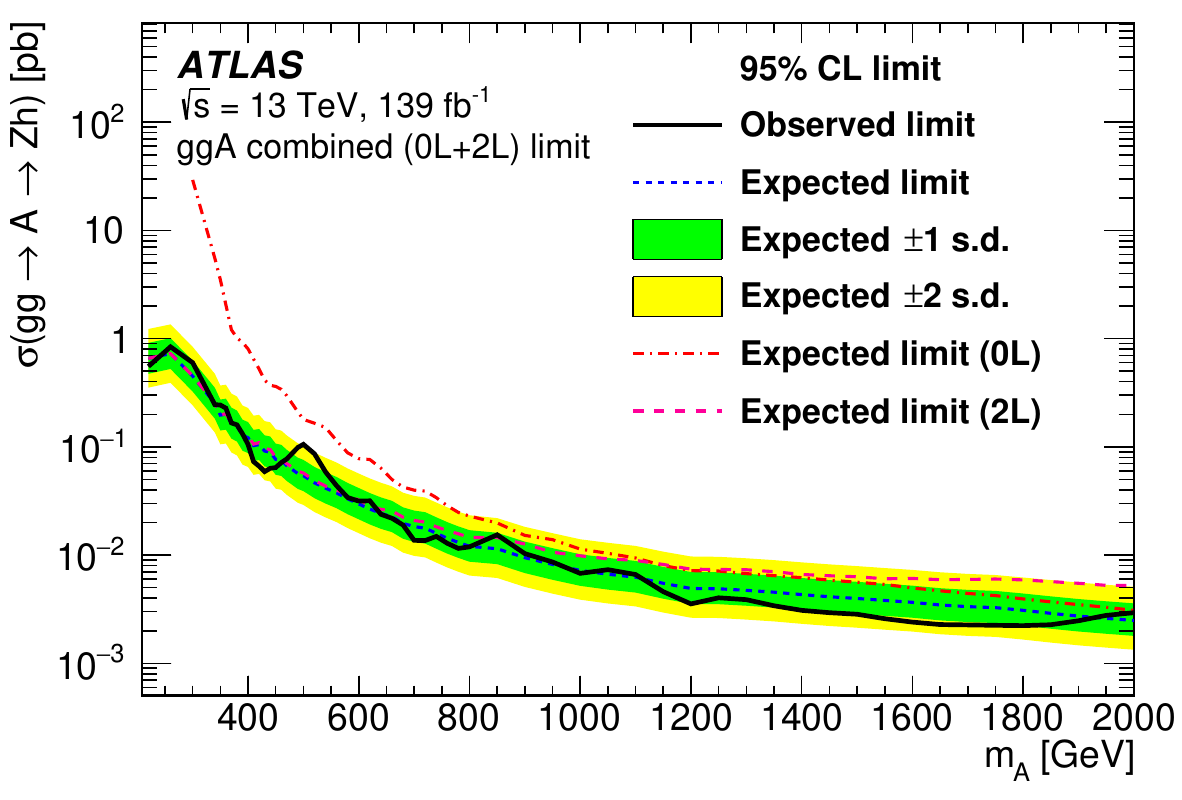}
}
\subfloat[]{
\includegraphics[width=0.5\textwidth]{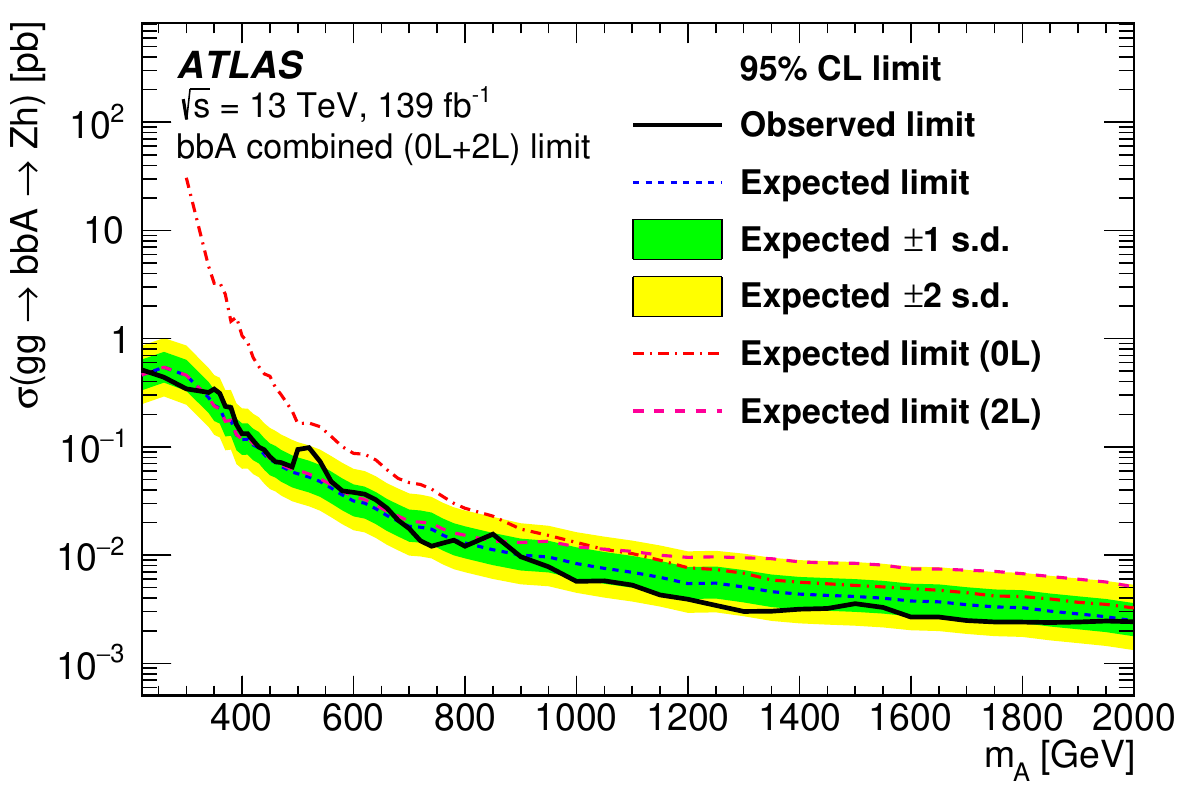}
}
\caption{
Upper limits at the 95\% CL on the product of the cross section for (a) $gg \to  \A$ and (b) $b\bar{b} \to  \A$ and their
respective branching fraction to \Zh\ from the combination of the 0-lepton (0L) and 2-lepton (2L) channels. For the $gg \to  \A$ search,
the possible signal components of the data are interpreted assuming pure gluon--gluon fusion production. For both searches, a branching
fraction of 0.569 for $h\rightarrow b\bar{b}$~\cite{deFlorian:2016spz} is assumed.}
\label{fig:2HDMlimits}
\end{figure*}

% End of text imported from the .//texfiles/resultFiguresLimitsPScalar.tex input file

The upper limits on the production cross section times branching ratio obtained from the \Zprime\ and \Wprime\ searches are used to set $95\%$ CL exclusion contours
in the HVT parameter plane \{$g_{\mathrm{F}}$, $g_{\mathrm{H}}$\}\footnote{The coupling constants $g_\mathrm{H}$ and $g_\mathrm{F}$ are related to those in Ref.~\cite{Pappadopulo:2014qza}
as follows: the Higgs boson coupling is $g_{\mathrm{H}} = g_{\mathrm{V}}c_{\mathrm{H}}$ and the universal fermion coupling is $g_{\mathrm{F}} = g^{2}c_{\mathrm{F}}/g_{\mathrm{V}}$,
where $g$ is the SM SU(2)$_\mathrm{L}$ gauge coupling.}~\cite{PhysRevD.98.052008}. Exclusion contours are shown in Figure~\ref{fig:HVTInterpret}
for resonance masses of $2$, $3$ and $4\,\TeV$. The constraints on $g_{\mathrm{F}}$ and $g_{\mathrm{H}}$ are stronger for large coupling parameter values
and become weaker as these coupling parameters approach zero. This is because the resonance couplings to $Vh$ vanish as the $g_{\mathrm{H}}$ parameter
reaches zero, while for $g_{\mathrm{F}} = 0$ the $\Zprime$ and $\Wprime$ production cross sections in the quark--antiquark annihilation mode become zero.

% The next lines are included from the .//texfiles/InterpretationFigures.tex input file
 
\begin{figure*}[h!]
\centering
\includegraphics[width=.445\textwidth]{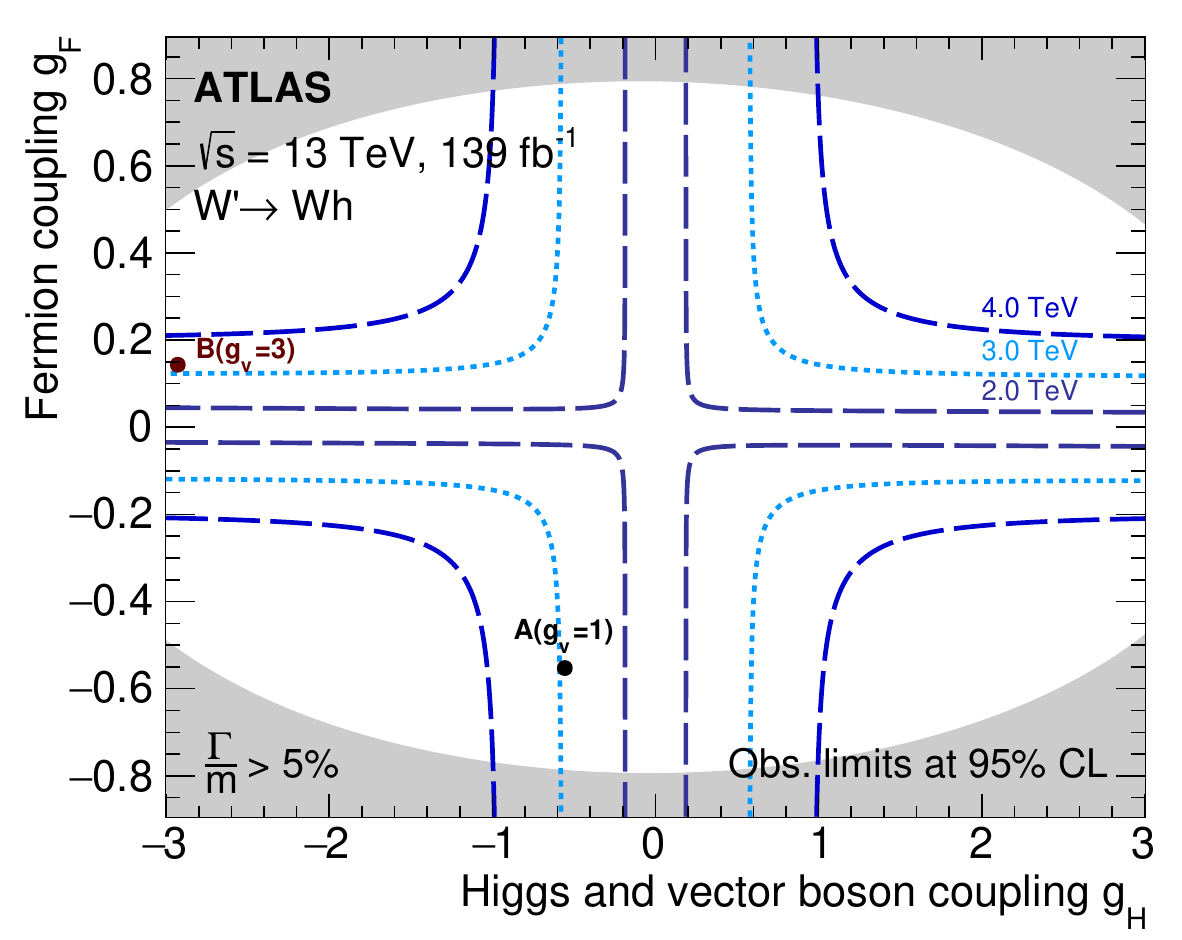}
\includegraphics[width=.445\textwidth]{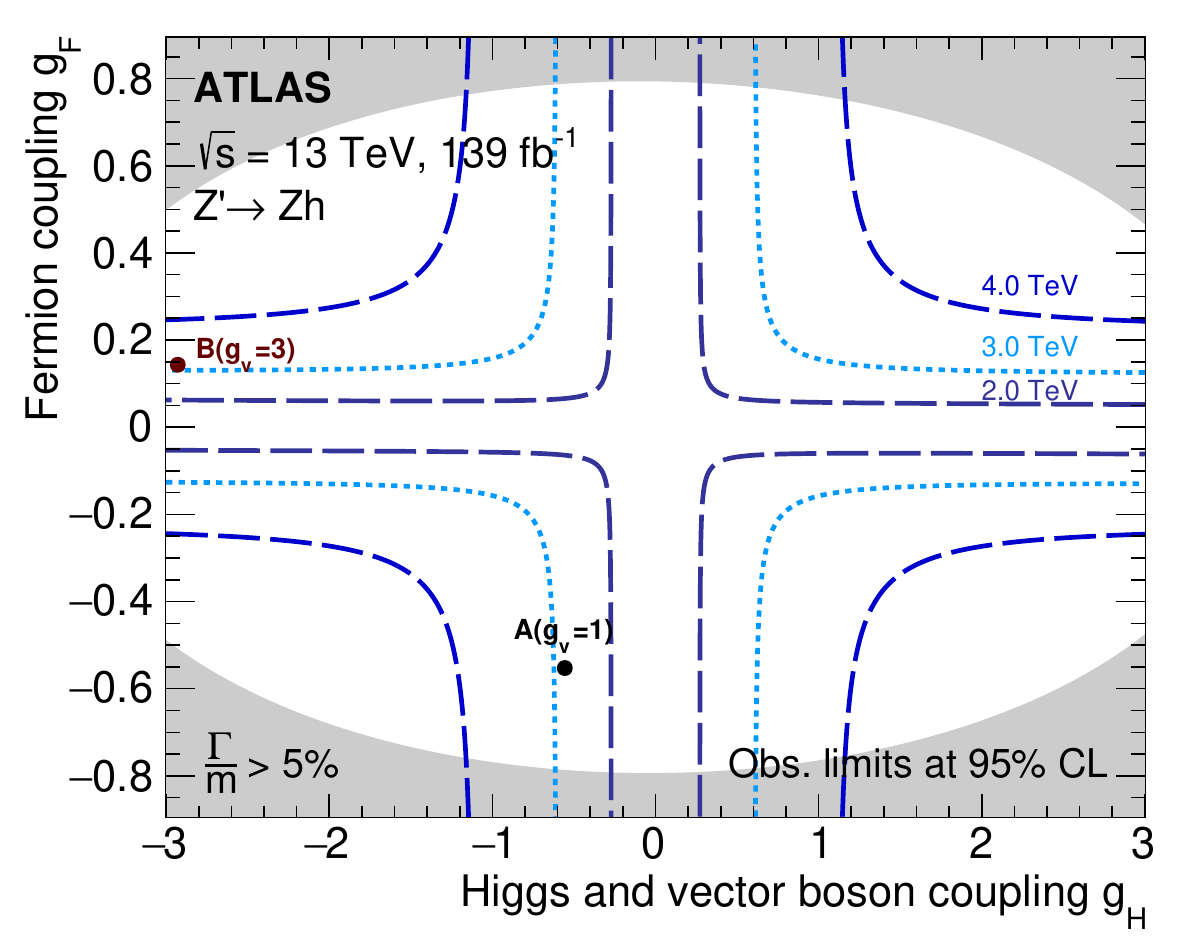}
\caption{
Observed limits on the HVT model at $95\%$ CL in the $g_{\mathrm{F}}$ vs $g_{\mathrm{H}}$ plane for resonance masses of $2$, $3$ and $4$~TeV in the \Wh\ channel (left) and in the \Zh\ channel (right).
The circles indicate the coupling values for models A and B and the grey region corresponds to the area of phase space where the decay width of the resonance is no longer negligible and the signal
$\mvh$ shape is no longer expected to be dominated by the experimental resolution.}
\label{fig:HVTInterpret}
\end{figure*}

% End of text imported from the .//texfiles/InterpretationFigures.tex input file

Figure~\ref{fig:2hdm-observed-exclusion-contours} shows the expected and observed two-dimensional likelihood scans of the $b$-associated
production cross section times branching fraction $B(A\rightarrow Zh)$ versus the gluon--gluon fusion production cross section times
branching fraction $B(A\rightarrow Zh)$ for various pseudoscalar masses $m_{A}$. For each mass hypothesis, the best-fit value is compatible
with the absence of a signal. The largest difference between the observed and expected best-fit values is found for a resonance mass of
${m_{A} = 500~\GeV}$.
 
To obtain the 95\% CL upper limits on the 2HDM parameters, 95\% CL upper limits on the production cross section for $A \rightarrow Z\h$ are calculated for
admixtures of \ggA\ and \bbA\ production modes and variations of the $A$ boson natural width up to $\Gamma_A/m_A = 20\%$. For this interpretation, $\mtvh$ and
$\mvh$ distributions of the simulated signal events are smeared according to a Breit–Wigner function with the width predicted by the parameters of the model.
It was verified that this procedure produces line-shapes that are the same as those predicted by simulation.
 
Figure~\ref{fig:2HDMinterpret1} shows the interpretation of the limits on \ggA\ and \bbA\ production in the Type-I, Type-II, Lepton-specific and Flipped 2HDM scenarios
as a function of the parameters $\tan\beta$ and $\cos(\beta - \alpha)$ for $m_A = 700~\GeV$. In the Type-I and Lepton-specific 2HDM models, a pseudoscalar with
mass $m_A = 700~\GeV$ is excluded at the 95\% CL for $\tan\beta$ values as large as 10. Greater sensitivity is observed at high $\tan\beta$ for
the Type-II and Flipped models, due to a larger cross section for $b$-quark associated production. In the alignment limit $\cos(\beta - \alpha)\rightarrow 0$,
the $A\rightarrow Zh$ branching fraction vanishes, thus the relatively low sensitivity around $\cos(\beta - \alpha)\approx 0$. Nevertheless,
$\cos(\beta - \alpha)$ values down to around 0.07 can be excluded for all tested model scenarios. The narrow inclined regions with no sensitivity at
low $\tan\beta$ are caused by a vanishing $h\rightarrow bb$ branching fraction.
 
Figure~\ref{fig:2HDMinterpret2} shows the interpretation of the cross-section limits as a function of the parameters $\tan\beta$ and $m_A$ for
$\cos(\beta - \alpha) = 0.1$. The shape of the expected exclusions is determined by the interplay of the expected cross-section limit,
which decreases as a function of $m_A$, and the signal production cross section times the $A \rightarrow Zh$ branching fraction at given $m_A$ and
$\tan\beta$ values. This branching fraction decreases significantly at $m_{A} = 350\,\GeV$ because the $A\rightarrow t\bar{t}$ decay becomes possible, but increases
again at higher $m_{A}$ values. The parameter $\tan\beta$ controls the admixture of the gluon--gluon fusion and $b$-quark associated production, and thus
affects the rate at which the signal cross section falls as a function of $m_{A}$, leading to a varying sensitivity as a function of $\tan\beta$.
The excesses or deficits in the data visible in Figure~\ref{fig:2HDMlimits} are also reflected in Figure~\ref{fig:2HDMinterpret2}.

% The next lines are included from the .//texfiles/InterpretationFigures_2hdm.tex input file
 
\begin{figure}[tpb]
\centering
\includegraphics[width=0.95\textwidth]{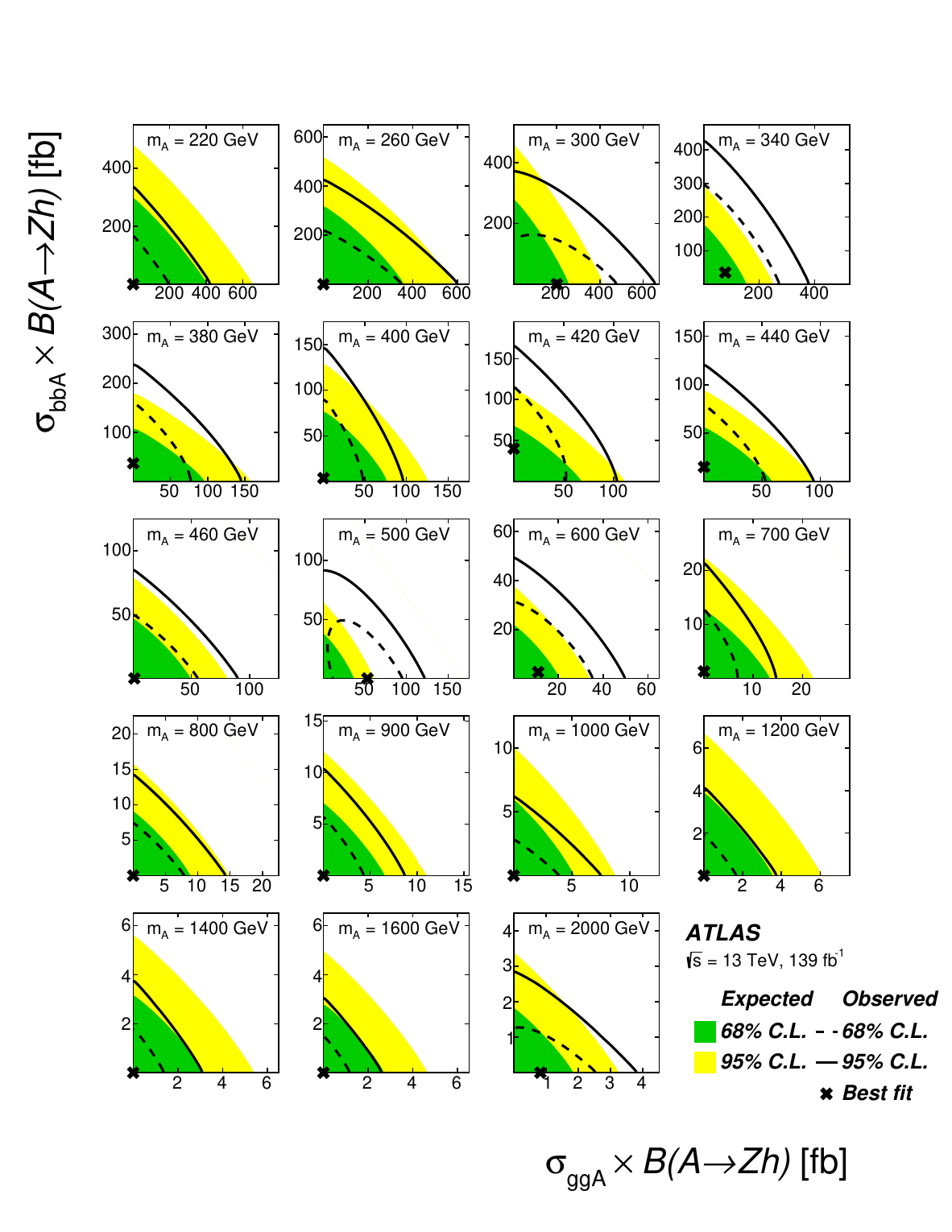}
\caption{Expected and observed two-dimensional likelihood scans of the \Pqb-associated production cross section times
branching fraction, $\sigma_{\bbA} \times B(A \to Zh)$, vs the gluon--gluon fusion cross section times branching fraction,
$\sigma_{\ggA} \times B(A \to Zh)$, for a given \PA boson mass ($m_{\PA}$).
For each mass, a variety of different cross-section hypotheses are scanned. At each point, $2\Delta(\text{NLL})$ is calculated, defined as the
negative-log-likelihood (NLL) of the conditional fit to the Asimov and observed datasets with $\sigma_{\bbA}$ and $\sigma_{\ggA}$
fixed to their values at that point and with the minimum NLL value at any point subtracted. The best-fit point
and the preferred \SI{68}{\percent} and \SI{95}{\percent} CL boundaries are found at $2\Delta(\text{NLL})$
values of 0.0, 2.30 and 5.99, respectively.}
\label{fig:2hdm-observed-exclusion-contours}
\end{figure}
 
\begin{figure*}[ht!]
\centering
\subfloat[][2HDM Type-I]{\includegraphics[height=0.4\textheight]{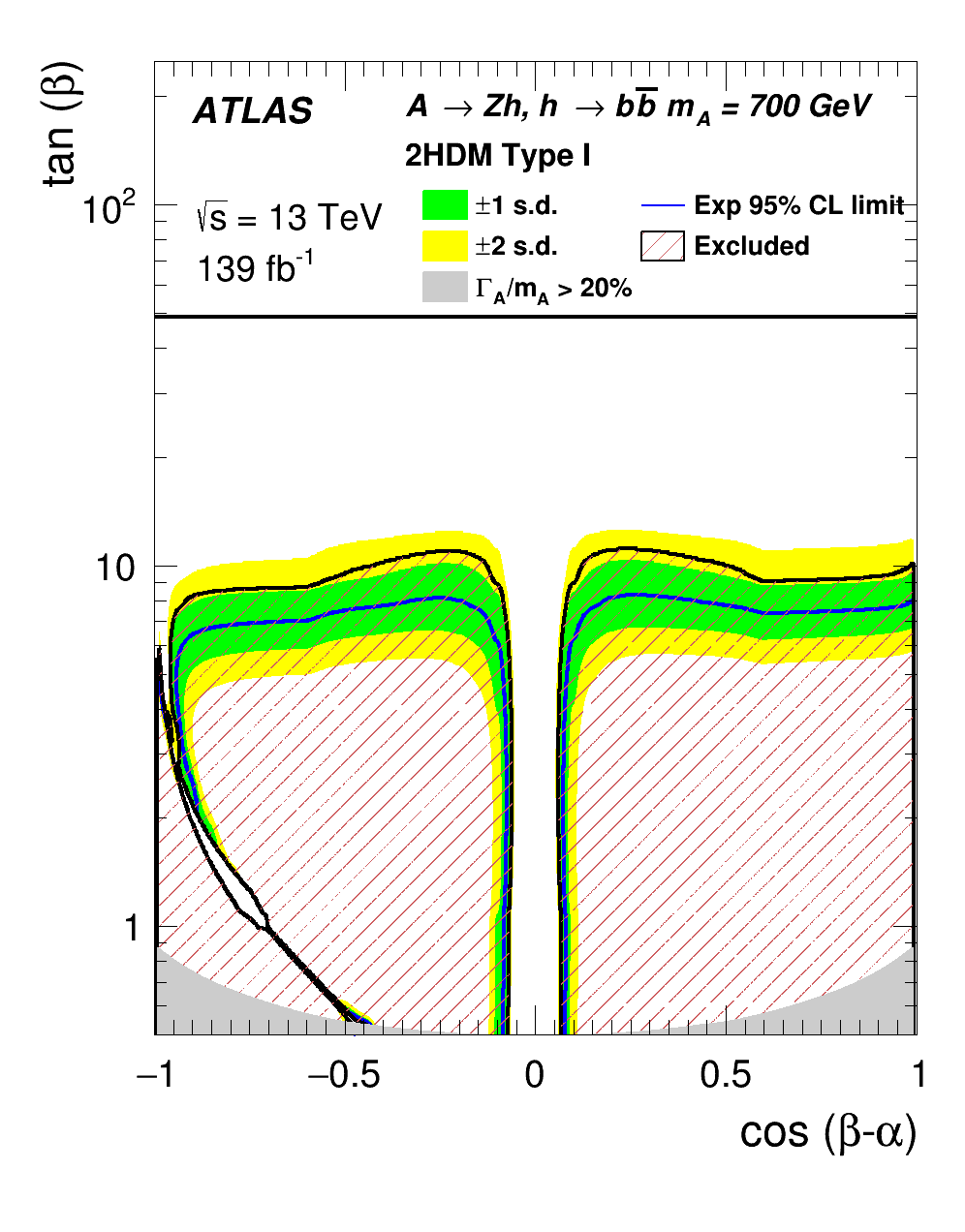}}
\subfloat[][2HDM Type-II]{\includegraphics[height=0.4\textheight]{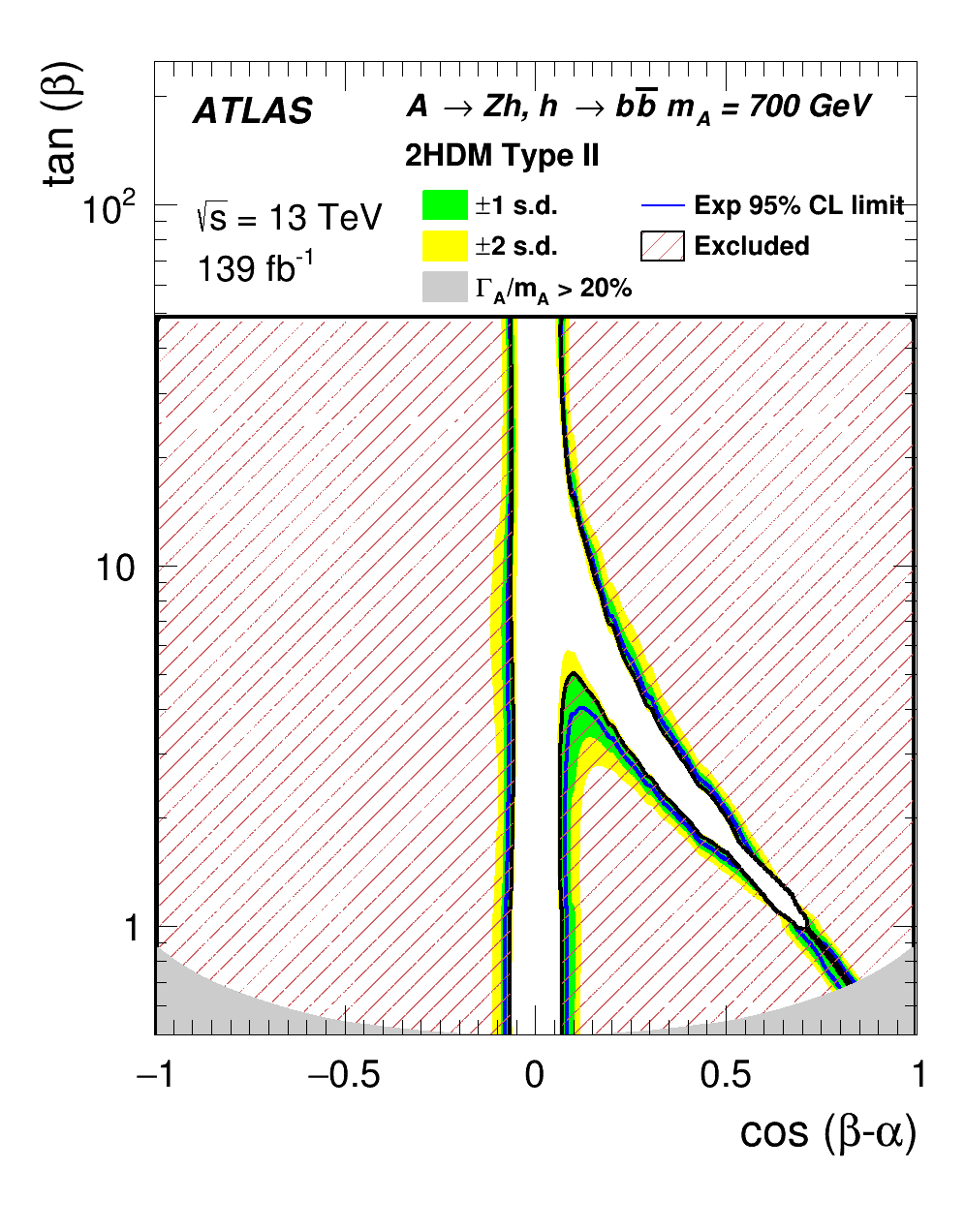}}\\
\subfloat[][2HDM Lepton-specific]{\includegraphics[height=0.4\textheight]{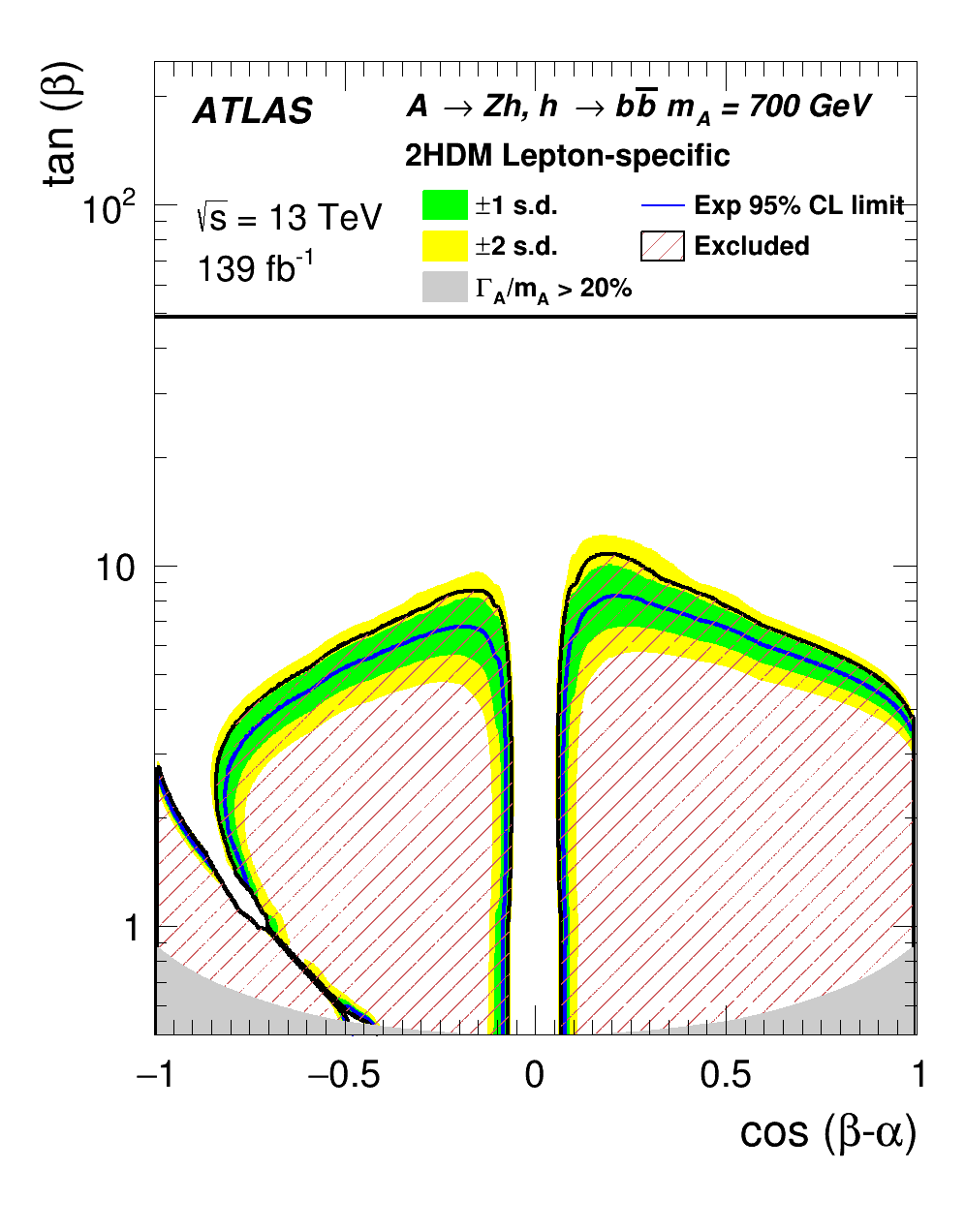}}
\subfloat[][2HDM Flipped]{\includegraphics[height=0.4\textheight]{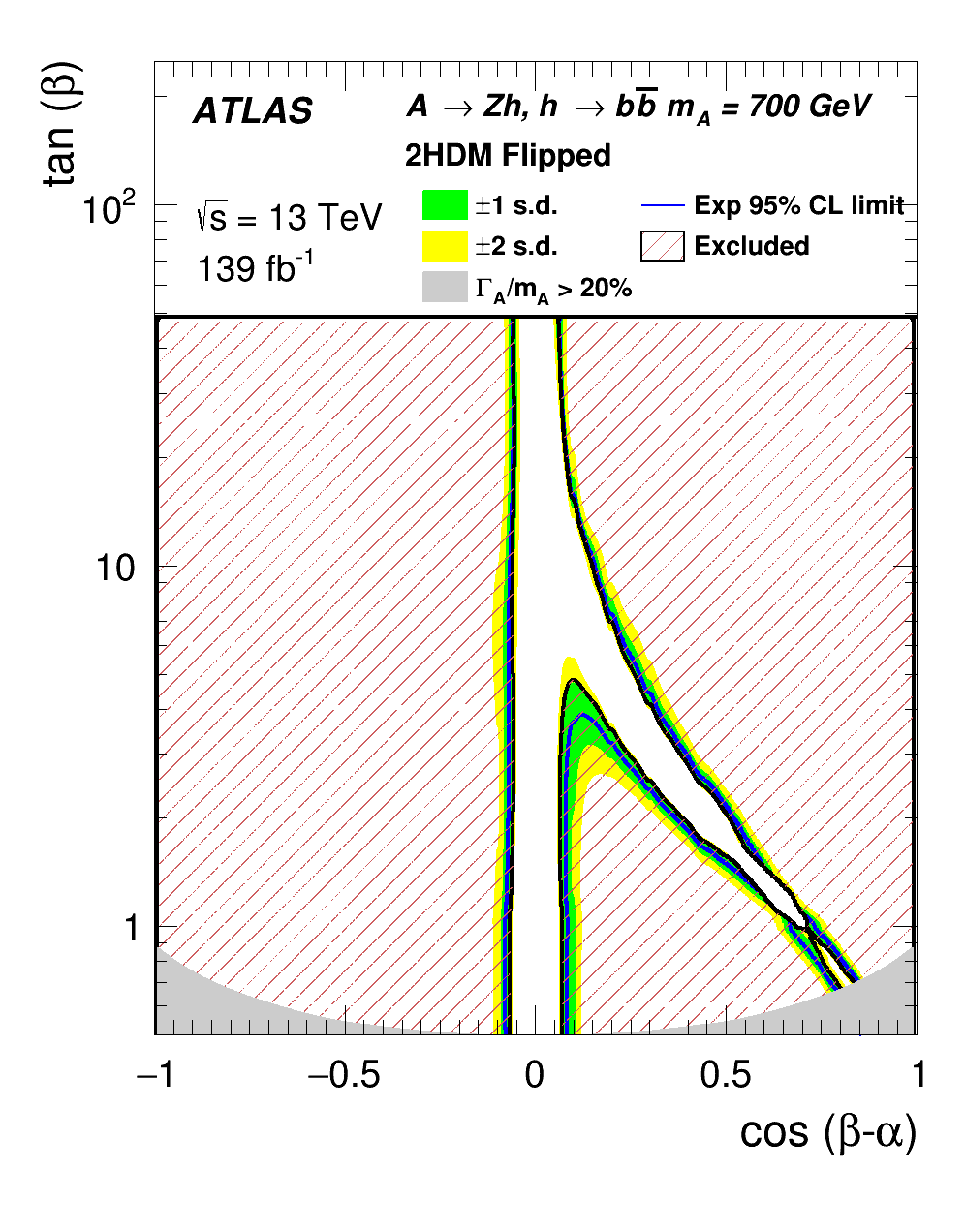}}
\caption{
The interpretation of the cross-section limits in the context of the various 2HDM types as a function of the parameters $\tan\beta$ and $\cos(\beta - \alpha)$ for
$m_A = 700$~\GeV: (a) Type-I , (b) Type-II, (c) Lepton-specific, and (d) Flipped. Variations of the natural width up to $\Gamma_A/m_A = 20\%$ have been taken into
account. For the interpretation in Type-II and Flipped 2HDM, the $b$-quark associated production is included in addition to the gluon--gluon fusion production.
The exclusion limits are only evaluated for $\tan\beta < 50$ because larger $\tan\beta$ values are disfavoured by most models (and cross-section
predictions become less reliable).}
\label{fig:2HDMinterpret1}
\end{figure*}

\begin{figure*}[ht!]
\centering
\subfloat[][2HDM Type-I]{\includegraphics[height=0.4\textheight]{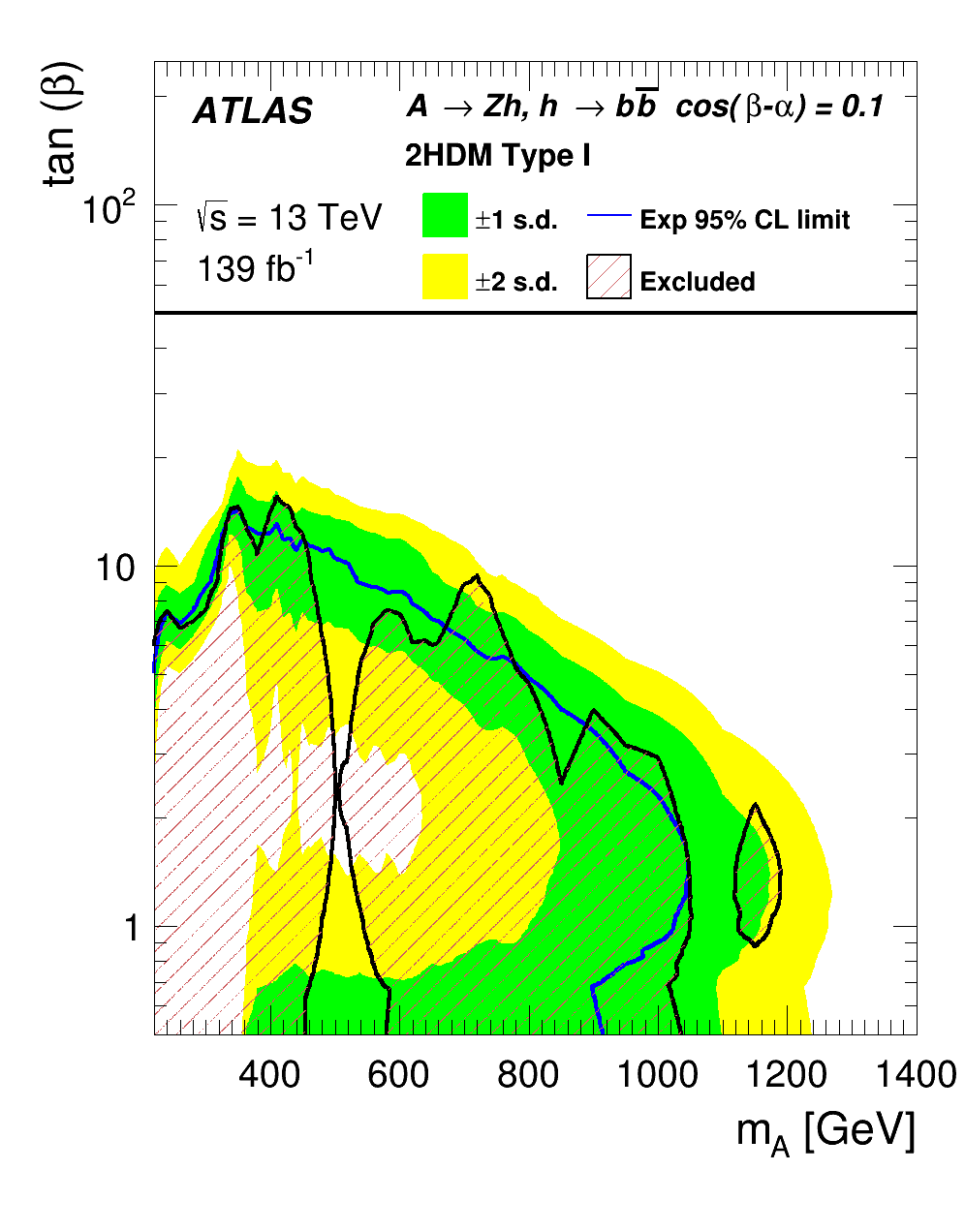}}
\subfloat[][2HDM Type-II]{\includegraphics[height=0.4\textheight]{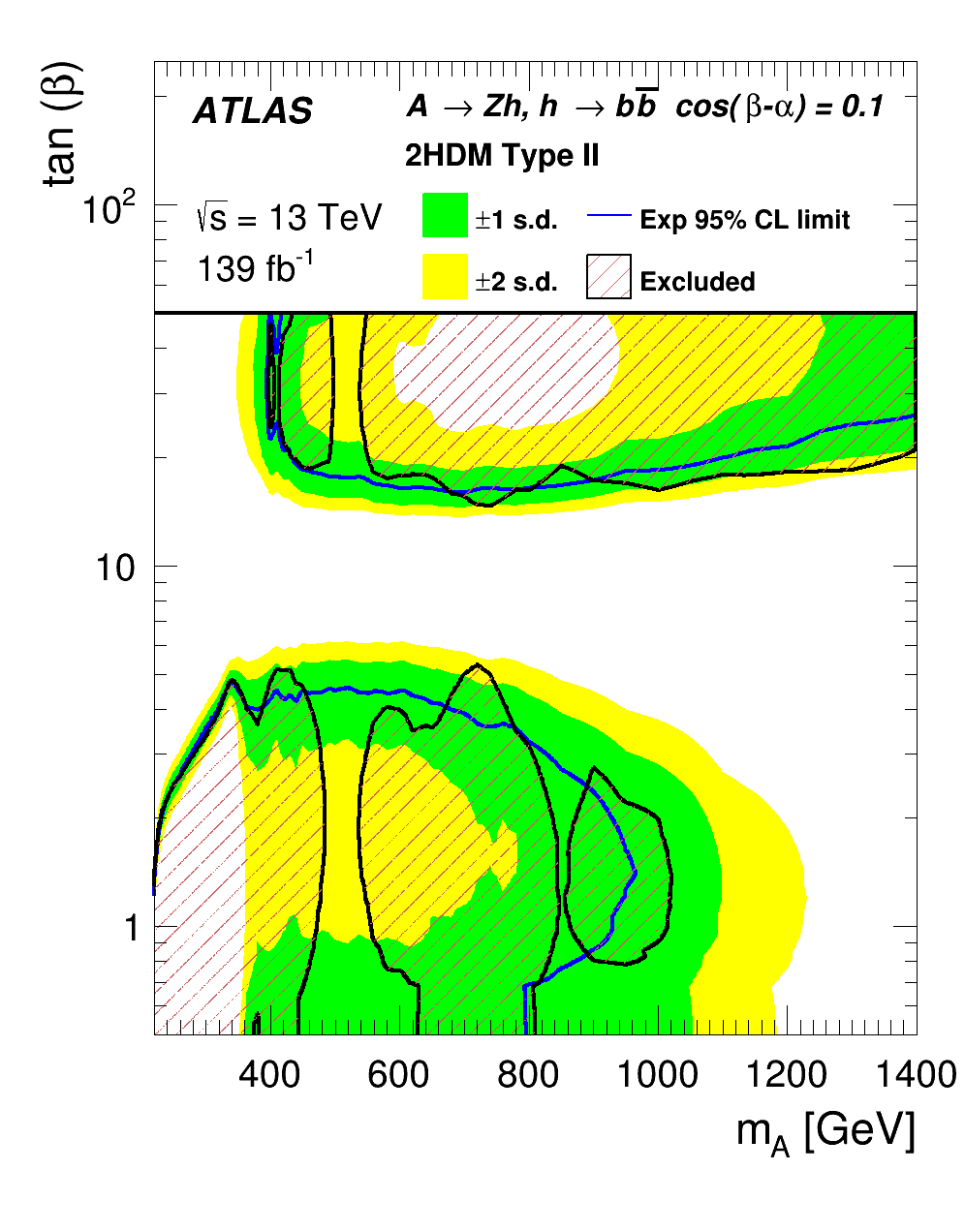}}\\
\subfloat[][2HDM Lepton-specific]{\includegraphics[height=0.4\textheight]{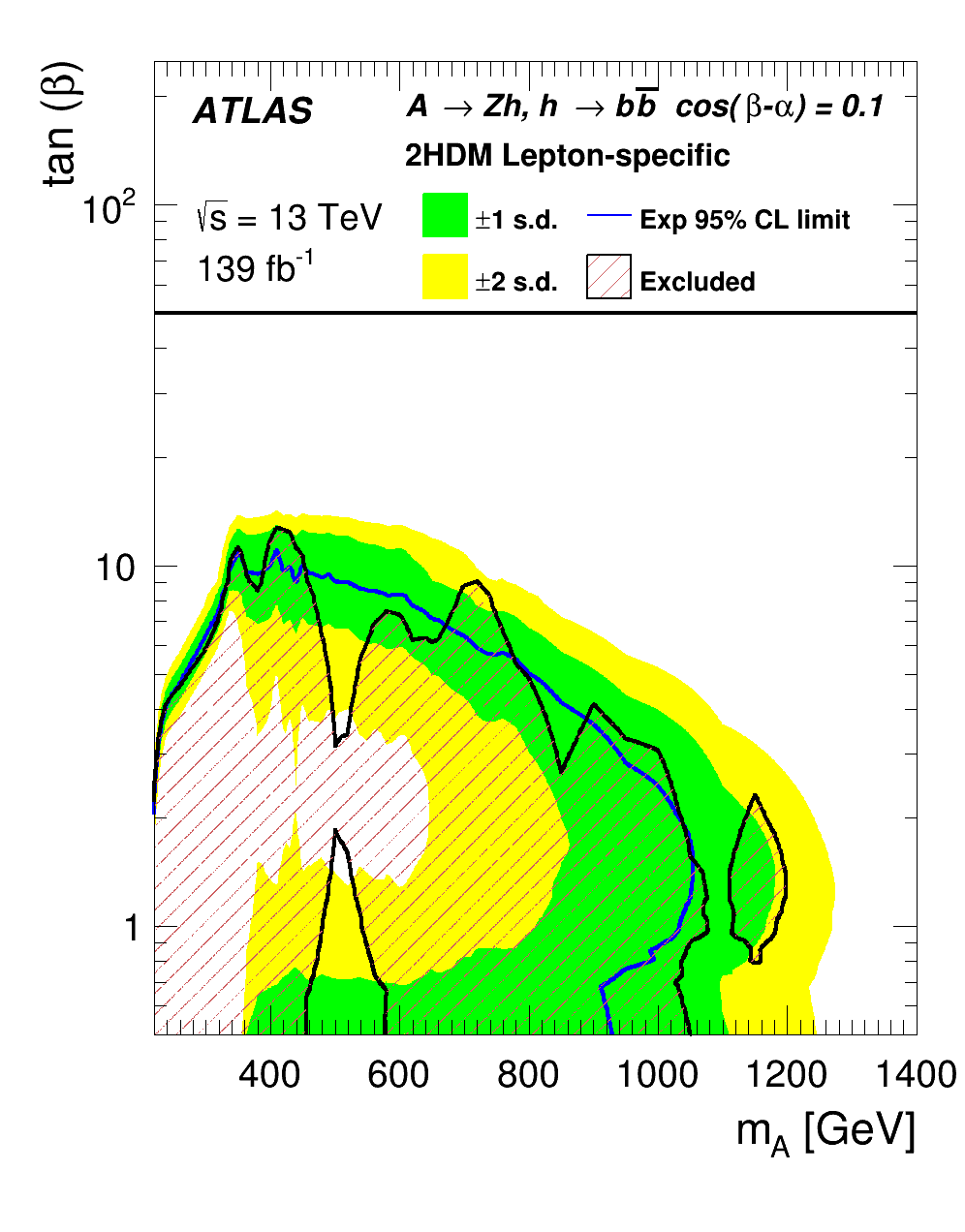}}
\subfloat[][2HDM Flipped]{\includegraphics[height=0.4\textheight]{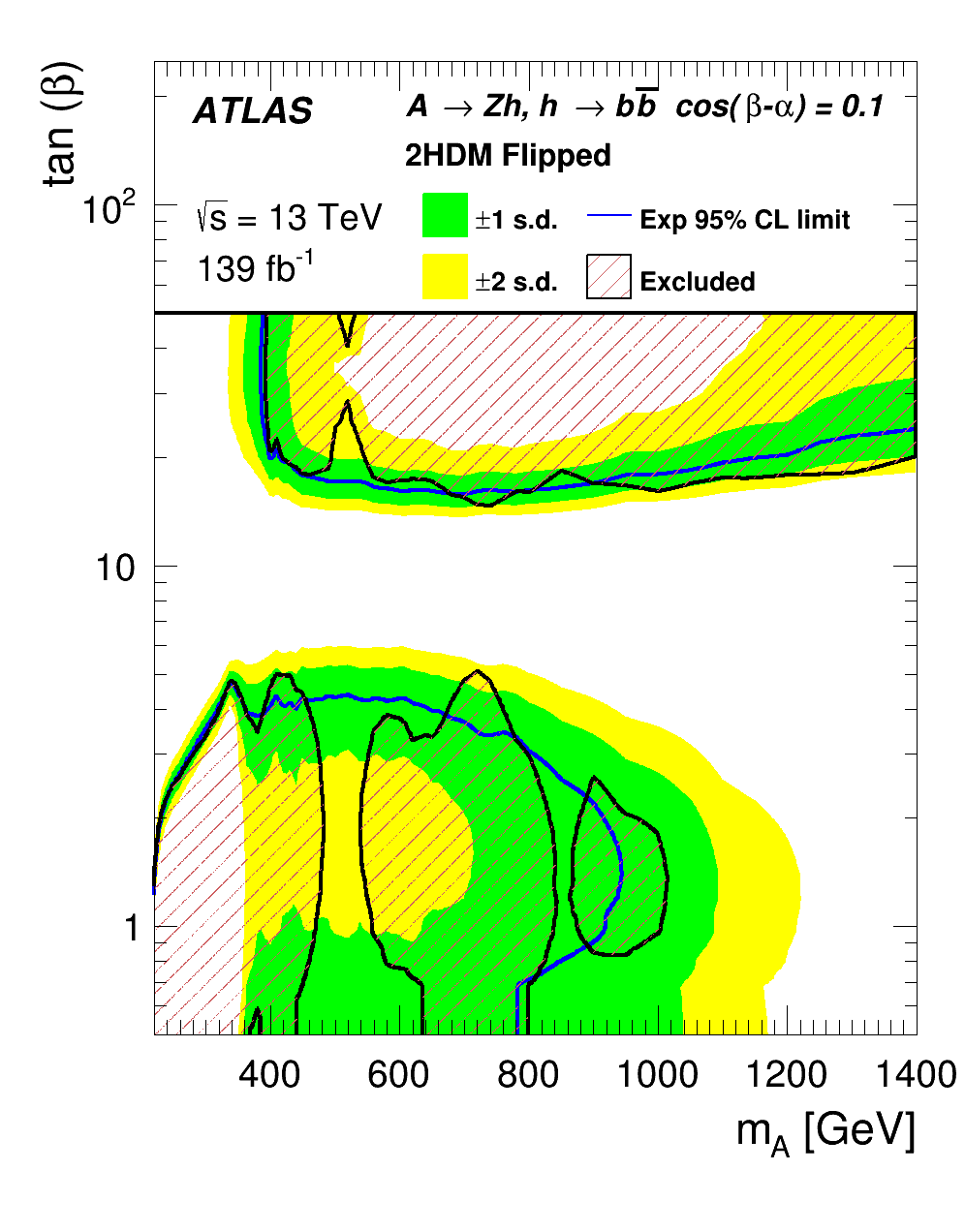}}
\caption{
The interpretation of the cross-section limits in the context of the various 2HDM types as a function of the parameters
$\tan\beta$ and $m_A$ for $\cos(\beta - \alpha) = 0.1$: (a) Type-I , (b) Type-II, (c) Lepton-specific, and (d) Flipped. Variations
of the natural width up to $\Gamma_A/m_A = 20\%$ have been taken into account. The exclusion limits are only evaluated for $\tan\beta < 50$
because larger $\tan\beta$ values are disfavoured by most models (and cross-section predictions become less reliable).}
\label{fig:2HDMinterpret2}
\end{figure*}

% End of text imported from the .//texfiles/InterpretationFigures_2hdm.tex input file

\clearpage

% End of text imported from the .//texfiles/Results.tex input file

% The next lines are included from the .//texfiles/Conclusion.tex input file
\section{Conclusion}\label{sec:Conclusion}
 
A search for \Zprime\ and \Wprime\ bosons and for a CP-odd Higgs boson $A$ in the \vvbb, \llbb\ and \lvbb\ final
states is performed using 139~fb$^{-1}$ of 13~\TeV\ $pp$ collision data collected with the
ATLAS detector at the LHC. No significant excess of events is observed above the SM
predictions in the three channels, and upper limits are set on the respective production
cross sections. \par
 
The 95\% CL upper limits on the production cross section for $Z^\prime\rightarrow Z\h$ range
from $0.9~\pb$ for $m_{Z^{\prime}} = 300~\GeV$ to $0.3~\fb$ for $m_{Z^{\prime}} = 5~\TeV$.
These limits exclude $Z^{\prime}$ masses below $2.8~\TeV$ for HVT benchmark \textit{Model A}
with coupling constant $g_\text{V}=1$. For \textit{Model B} with
coupling constant $g_\text{V}=3$, $Z^{\prime}$  masses below
$3.2~\TeV$ are excluded. The \Zprime\ cross-section limits are also converted into constraints on
the couplings in the $g_{\mathrm{F}}$ vs $g_{\mathrm{H}}$ plane for
resonance mass hypotheses of $2$, $3$ and $4$~TeV. \par
 
The 95\% CL upper limits on the production cross section for $\Wprime\rightarrow W\h$ range from around $1.3~\pb$ for ${m_{W^{\prime}} = 400~\GeV}$ to $0.4~\fb$
for $m_{W^{\prime}} = 5~\TeV$. These limits exclude $W^{\prime}$  masses below $2.95~\TeV$ for HVT benchmark \textit{Model A} with coupling constant
$g_\text{V}=1$. For \textit{Model B} with coupling constant $g_\text{V}=3$, $W^{\prime}$  masses below
$3.3~\TeV$ are excluded. The \Wprime\ cross-section limits are also converted into constraints on the couplings in the $g_{\mathrm{F}}$ vs $g_{\mathrm{H}}$
plane for resonance mass hypotheses of $2$, $3$ and $4$~TeV.\par
 
The 95\% CL upper limits on the cross section for \ggA\ production multiplied by the $A$ boson's
branching fraction to $Z\h$ range from $0.6~\pb$ for $m_{A} = 220~\GeV$ to $3~\fb$ for
$m_{A} = 2~\TeV$. The 95\% CL upper limits on the cross section for \bbA\ production multiplied by the $A$ boson's
branching fraction to $Z\h$ range from $0.5~\pb$ for $m_{A} = 220~\GeV$ to $2.5~\fb$ for $m_{A} = 2~\TeV$.
The limits on the \ggA\ and \bbA\ production cross sections are converted into constraints on the parameters $\tan\beta$,
$\cos(\beta - \alpha)$ and $m_A$ in the Type-I, Type-II, Lepton-specific and Flipped 2HDM scenarios.\par
 
The cross-section limits improve on the results from the ATLAS analysis of the partial
Run~2 dataset. The improvements range from about $50\%$
for a resonance mass of $220~\GeV$ to about $400\%$ for a mass of $5~\TeV$. The improvements at low
resonance masses are mainly due to the larger dataset, while at high resonance masses the improvements
are mostly due to refined $b$-tagging techniques, including the use of variable-radius track-jets instead of track-jets
clustered with a fixed radius parameter of $R=0.2$.

% End of text imported from the .//texfiles/Conclusion.tex input file

\section*{Acknowledgements}

% The next lines are included from the .//acknowledgements/Acknowledgements.tex input file

We thank CERN for the very successful operation of the LHC, as well as the
support staff from our institutions without whom ATLAS could not be
operated efficiently.
 
We acknowledge the support of
ANPCyT, Argentina;
YerPhI, Armenia;
ARC, Australia;
BMWFW and FWF, Austria;
ANAS, Azerbaijan;
CNPq and FAPESP, Brazil;
NSERC, NRC and CFI, Canada;
CERN;
ANID, Chile;
CAS, MOST and NSFC, China;
Minciencias, Colombia;
MEYS CR, Czech Republic;
DNRF and DNSRC, Denmark;
IN2P3-CNRS and CEA-DRF/IRFU, France;
SRNSFG, Georgia;
BMBF, HGF and MPG, Germany;
GSRI, Greece;
RGC and Hong Kong SAR, China;
ISF and Benoziyo Center, Israel;
INFN, Italy;
MEXT and JSPS, Japan;
CNRST, Morocco;
NWO, Netherlands;
RCN, Norway;
MEiN, Poland;
FCT, Portugal;
MNE/IFA, Romania;
MESTD, Serbia;
MSSR, Slovakia;
ARRS and MIZ\v{S}, Slovenia;
DSI/NRF, South Africa;
MICINN, Spain;
SRC and Wallenberg Foundation, Sweden;
SERI, SNSF and Cantons of Bern and Geneva, Switzerland;
MOST, Taiwan;
TENMAK, T\"urkiye;
STFC, United Kingdom;
DOE and NSF, United States of America.
In addition, individual groups and members have received support from
BCKDF, CANARIE, Compute Canada and CRC, Canada;
PRIMUS 21/SCI/017 and UNCE SCI/013, Czech Republic;
COST, ERC, ERDF, Horizon 2020 and Marie Sk{\l}odowska-Curie Actions, European Union;
Investissements d'Avenir Labex, Investissements d'Avenir Idex and ANR, France;
DFG and AvH Foundation, Germany;
Herakleitos, Thales and Aristeia programmes co-financed by EU-ESF and the Greek NSRF, Greece;
BSF-NSF and MINERVA, Israel;
Norwegian Financial Mechanism 2014-2021, Norway;
NCN and NAWA, Poland;
La Caixa Banking Foundation, CERCA Programme Generalitat de Catalunya and PROMETEO and GenT Programmes Generalitat Valenciana, Spain;
G\"{o}ran Gustafssons Stiftelse, Sweden;
The Royal Society and Leverhulme Trust, United Kingdom.
 
The crucial computing support from all WLCG partners is acknowledged gratefully, in particular from CERN, the ATLAS Tier-1 facilities at TRIUMF (Canada), NDGF (Denmark, Norway, Sweden), CC-IN2P3 (France), KIT/GridKA (Germany), INFN-CNAF (Italy), NL-T1 (Netherlands), PIC (Spain), ASGC (Taiwan), RAL (UK) and BNL (USA), the Tier-2 facilities worldwide and large non-WLCG resource providers. Major contributors of computing resources are listed in Ref.~\cite{ATL-SOFT-PUB-2021-003}.

% End of text imported from the .//acknowledgements/Acknowledgements.tex input file

\printbibliography

\clearpage
 
% ATLAS Collaboration author list
% Reference date of HDBS-2020-19 is 2022-03-25
% Author list last updated on date 11-AUG-23
% Data extracted on 11-Aug-2023 for paper reference HDBS-2020-19
% at 2:46pm
 
\begin{flushleft}
\hypersetup{urlcolor=black}
{\Large The ATLAS Collaboration}

\bigskip

\AtlasOrcid[0000-0002-6665-4934]{G.~Aad}$^\textrm{\scriptsize 101}$,
\AtlasOrcid[0000-0002-5888-2734]{B.~Abbott}$^\textrm{\scriptsize 119}$,
\AtlasOrcid[0000-0002-7248-3203]{D.C.~Abbott}$^\textrm{\scriptsize 102}$,
\AtlasOrcid[0000-0002-1002-1652]{K.~Abeling}$^\textrm{\scriptsize 55}$,
\AtlasOrcid[0000-0002-8496-9294]{S.H.~Abidi}$^\textrm{\scriptsize 29}$,
\AtlasOrcid[0000-0002-9987-2292]{A.~Aboulhorma}$^\textrm{\scriptsize 35e}$,
\AtlasOrcid[0000-0001-5329-6640]{H.~Abramowicz}$^\textrm{\scriptsize 150}$,
\AtlasOrcid[0000-0002-1599-2896]{H.~Abreu}$^\textrm{\scriptsize 149}$,
\AtlasOrcid[0000-0003-0403-3697]{Y.~Abulaiti}$^\textrm{\scriptsize 116}$,
\AtlasOrcid[0000-0003-0762-7204]{A.C.~Abusleme~Hoffman}$^\textrm{\scriptsize 136a}$
% ...
\bigskip
\\

% ...
$^{29}$Physics Department, Brookhaven National Laboratory, Upton NY; United States of America.\\
% ...
$^{35}$$^{(a)}$Facult\'e des Sciences Ain Chock, R\'eseau Universitaire de Physique des Hautes Energies - Universit\'e Hassan II, Casablanca;$^{(b)}$Facult\'{e} des Sciences, Universit\'{e} Ibn-Tofail, K\'{e}nitra;$^{(c)}$Facult\'e des Sciences Semlalia, Universit\'e Cadi Ayyad, LPHEA-Marrakech;$^{(d)}$LPMR, Facult\'e des Sciences, Universit\'e Mohamed Premier, Oujda;$^{(e)}$Facult\'e des sciences, Universit\'e Mohammed V, Rabat;$^{(f)}$Institute of Applied Physics, Mohammed VI Polytechnic University, Ben Guerir; Morocco.\\
% ...
$^{55}$II. Physikalisches Institut, Georg-August-Universit\"{a}t G\"ottingen, G\"ottingen; Germany.\\
% ...
$^{101}$CPPM, Aix-Marseille Universit\'e, CNRS/IN2P3, Marseille; France.\\
$^{102}$Department of Physics, University of Massachusetts, Amherst MA; United States of America.\\
% ...
$^{116}$Department of Physics, New York University, New York NY; United States of America.\\
% ...
$^{119}$Homer L. Dodge Department of Physics and Astronomy, University of Oklahoma, Norman OK; United States of America.\\
% ...
$^{136}$$^{(a)}$Departamento de F\'isica, Pontificia Universidad Cat\'olica de Chile, Santiago;$^{(b)}$Millennium Institute for Subatomic physics at high energy frontier (SAPHIR), Santiago;$^{(c)}$Instituto de Investigaci\'on Multidisciplinario en Ciencia y Tecnolog\'ia, y Departamento de F\'isica, Universidad de La Serena;$^{(d)}$Universidad Andres Bello, Department of Physics, Santiago;$^{(e)}$Instituto de Alta Investigaci\'on, Universidad de Tarapac\'a, Arica;$^{(f)}$Departamento de F\'isica, Universidad T\'ecnica Federico Santa Mar\'ia, Valpara\'iso; Chile.\\
% ...
$^{149}$Department of Physics, Technion, Israel Institute of Technology, Haifa; Israel.\\
$^{150}$Raymond and Beverly Sackler School of Physics and Astronomy, Tel Aviv University, Tel Aviv; Israel.\\
% ...
$^{a}$ Also Affiliated with an institute covered by a cooperation agreement with CERN.\\
$^{b}$ Also at An-Najah National University, Nablus; Palestine.\\
$^{c}$ Also at Borough of Manhattan Community College, City University of New York, New York NY; United States of America.\\
$^{d}$ Also at Bruno Kessler Foundation, Trento; Italy.\\
$^{e}$ Also at Center for High Energy Physics, Peking University; China.\\
$^{f}$ Also at Centro Studi e Ricerche Enrico Fermi; Italy.\\
% ...

\end{flushleft}

% Created with Glance <Atlas.Glance@cern.ch>

\end{document}